\documentclass[12pt]{article}

\usepackage[dvips]{graphicx}
\usepackage{epsfig}
\usepackage{amsmath,amsfonts,amssymb,amsthm}
\usepackage{mathrsfs,mathtools}
\usepackage{verbatim}
\usepackage{psfrag}
\usepackage{bm}
\usepackage{bbm}
\usepackage[utf8]{inputenc}
\usepackage[square,comma,sort&compress,numbers]{natbib}
\usepackage[dvipsnames]{xcolor}
\usepackage{slashed}
\usepackage{upgreek}
\usepackage{enumitem}
\usepackage{float}
\usepackage[T1]{fontenc}
\usepackage{soul}
\usepackage{pgfplots}
\usepackage{extarrows}
\pgfplotsset{compat=1.18}
\usetikzlibrary{external,decorations.pathmorphing}
\usepackage{cases}
\usepackage{soul}
\usepackage{cancel}
\usepackage[normalem]{ulem}
\usepackage{lmodern}
\usepackage{tikz}

\usepackage[T1]{fontenc}

\tikzset{
  external/only named=true,
  thick/.style={line width=.5pt},
  approximation/.style={line width=1.2pt},
  numerics/.style={black, dotted, line width=.8pt},
  amplitude/.style={dashed},
  estimate/.style={dashed, line width=.8pt},
  normal plot/.style={line width=.8pt},
}

\usetikzlibrary{shapes.misc}
\tikzset{cross/.style={cross out, draw=black, minimum size=2*(#1-\pgflinewidth), inner sep=0pt, outer sep=0pt},
cross/.default={1pt}}

\tikzset{snake it/.style={decorate, decoration=snake}}

\usepackage[colorlinks=true,urlcolor=blue,anchorcolor=blue,citecolor=blue,filecolor=blue
,linkcolor=blue,menucolor=blue,linktocpage=true,pdfproducer=medialab,pdfa=true
]{hyperref}

\usepackage{epsf,epsfig}
\usepackage{graphics}
\usepackage{subcaption}

\newcommand\blfootnote[1]{%
  \begingroup
  \renewcommand\thefootnote{}\footnote{#1}%
  \addtocounter{footnote}{-1}%
  \endgroup
}

\def\d{\mathrm{d}}

\def\L{\mathcal{L}}

\def\C{\mathcal{C}}
\def\O{\mathcal{O}}
\def\vec{\mathbf}
\def\i{\mathrm{i}}
\def\e{\mathrm{e}}
\def\P{\mathcal{P}}

\def\M{\mathcal{M}}
\def\veck{\vec{k}}
\def\vecp{\vec{p}}
\def\vecq{\vec{q}}
\def\vecx{\vec{x}}

\usepackage[errorstop]{feynmp}
\newcommand{\lsim}
{\;\raisebox{-.3em}{$\stackrel{\displaystyle <}{\sim}$}\;}

\newcommand{\rom}[1]{\uppercase\expandafter{\romannumeral #1\relax}}
\newcommand{\overbar}[1]{\mkern 1.5mu\overline{\mkern-1.5mu#1\mkern-1.5mu}\mkern 1.5mu}

\begin{document}

\thispagestyle{empty}

\begin{flushright}
{\small
MITP-25-029
\\
TUM-HEP-1562/25}
\end{flushright}

\vspace{-0.5cm}

\begin{center}
\Large\bf\boldmath
Bubble wall dynamics from \\ nonequilibrium quantum field theory
\unboldmath
\end{center}

\vspace{-0.2cm}

\begin{center}
Wen-Yuan Ai,$^{1,*}$\blfootnote{$^*$ wenyuan.ai@oeaw.ac.at} Matthias Carosi,$^{2,\dagger}$\blfootnote{$^\dagger$ matthias.carosi@tum.de} Björn Garbrecht,$^{2,\ddagger}$\blfootnote{$^\ddagger$ garbrecht@tum.de} Carlos Tamarit$^{3,\S}$\blfootnote{$^\S$ ctamarit@uni-mainz.de}\\ and  Miguel Vanvlasselaer$^{4,\parallel}$\blfootnote{$^\parallel$ miguel.vanvlasselaer@vub.be} \\
\vskip0.4cm

{\it $^1$Institute of High Energy Physics, Austrian Academy of Sciences,\\ Dominikanerbastei 16, 1010 Vienna, Austria}

{\it  $^2$Physik Department T70, Technische Universit\"{a}t M\"{u}nchen,\\ James-Franck-Stra\ss e, 85748 Garching, Germany
}

{\it $^3$PRISMA+ Cluster of Excellence \& Mainz Institute for Theoretical Physics,\\
Johannes Gutenberg-Universität Mainz, 55099 Mainz, Germany}

{\it $^4$
Theoretische Natuurkunde and IIHE/ELEM, Vrije Universiteit Brussel,\\
\& The International Solvay Institutes, Pleinlaan 2, B-1050 Brussels, Belgium
}
\vskip1.cm
\end{center}

\begin{abstract}
We derive the coupled dynamics between the bubble wall and the plasma from first principles using nonequilibrium quantum field theory. The commonly used equation of motion of the bubble wall in the kinetic approach is shown to be incomplete. In the language of the two-particle-irreducible effective action, the conventional equation misses higher-loop terms generated by the condensate-particle type vertices (e.g.,~$\varphi \phi\chi^2$, where $\varphi$ is the background field describing the bubble wall, $\phi$ the corresponding particle excitation and $\chi$ another particle species in the plasma). From the missing terms, we identify an additional dissipative friction which is contributed by particle production processes from the condensate-particle type vertices. We also show how other transmission processes beyond the 1-to-1 elementary transmission studied in the literature for ultrarelativistic bubble walls, e.g., 1-to-1 mixing and 1-to-2 transition radiation, can be understood from the kinetic approach. 
\end{abstract}

\newpage

\hrule
\tableofcontents
\vskip.8cm
\hrule

\section{Introduction}
\label{sec:Intro}

First-order phase transitions (FOPTs) in the early universe have been receiving growing attention. First and foremost, they have attracted interest because they are powerful sources of observable gravitational waves (GWs)\,\cite{Witten:1984rs,Hogan_GW_1986,Kosowsky:1992vn,Kosowsky:1992rz,Kamionkowski:1993fg}, making them a unique experimental probe of the early universe physics. In addition to those observational hopes, they also have far-reaching phenomenological consequences. They offer a natural avenue for baryogenesis\,\cite{Kuzmin:1985mm, Shaposhnikov:1986jp,Nelson:1991ab,Carena:1996wj,Long:2017rdo,Bruggisser:2018mrt,Bruggisser:2018mus,Morrissey:2012db,Azatov:2021irb, Baldes:2021vyz, Huang:2022vkf,Chun:2023ezg,Dichtl:2023xqd,Cataldi:2024pgt}, production of primordial magnetic fields\,\cite{Vachaspati:1991nm,Ahonen:1997wh,Vachaspati:2001nb,Ellis:2019tjf,Di:2020kbw,Olea-Romacho:2023rhh,Balaji:2024rvo},  production of dark matter\,\cite{Falkowski:2012fb, Baldes:2020kam,Hong:2020est, Azatov:2021ifm,Baldes:2021aph,Asadi:2021pwo,Baldes:2022oev,Azatov:2022tii,Baldes:2023cih,Kierkla:2022odc, Gehrman:2023qjn,Giudice:2024tcp,Ai:2024ikj,Cembranos:2024pvy,Benso:2025vgm}, formation of primordial black holes\,\cite{Kodama:1982sf,Kawana:2021tde,Liu:2021svg,Jung:2021mku,Gouttenoire:2023naa,Lewicki:2024ghw,Ai:2024cka,Murai:2025hse}. From a model-building perspective, FOPTs are a unique signature of beyond the Standard Model (BSM) physics and occur naturally in a large variety of BSM models.

In most of the phenomena mentioned above, a particularly important quantity is the bubble wall velocity $v_w$, which is determined by the interactions between the bubble wall and the plasma. Specifically, it was shown recently that the bubble wall velocity has a large impact on the amplitude and spectrum of the gravitational wave signal\,\cite{Gowling:2021gcy}. In the literature, there are two main but different approaches to dealing with bubble wall dynamics.\footnote{Holography is also used to deal with bubble wall dynamics in strongly coupled theories\,\cite{Bea:2021zsu,Bigazzi:2021ucw,Janik:2022wsx,Li:2023xto,Bea:2024bls}.}

In the first approach, to which we refer as the {\it kinetic approach}, one solves the coupled equations of motion (EoMs) for the background scalar field (condensate) and the plasma\,\cite{Moore:1995ua,Moore:1995si}
\begin{subequations}
\label{eq:conventional-starting-point}
\begin{align}
\label{eq:condensate-eom}
&\Box\varphi+V'(\varphi)+\sum_i\frac{\d m^2_i(\varphi)}{\d\varphi}\int \frac{\d^3{\bf k}}{(2\pi)^32E_i}\,f_i(\veck,x)=0\,,\\
\label{eq:Boltzmann_eq}
&\frac{\d f_i}{\d t}=-\C[f]\,,  
\end{align}
\end{subequations}
where $\Box=\partial_\mu \partial^\mu$, $E_i=\sqrt{\vec{k}^2_i+m_i^2}$. Above, $f_i(k,x)$ are the particle distribution functions and\,$V$ is the zero-temperature renormalized effective potential, typically considered up to one loop. 
The index\,$i$ runs over the particle species in the plasma. In solving the above equations, one needs to carefully take into account the inhomogeneous plasma temperature and velocity distributions across the wall. This leads to a standard hydrodynamic classification of the bubble expansion modes\,\cite{Espinosa:2010hh}. This is the most used approach in the literature, see e.g.\,\cite{Liu:1992tn,Konstandin:2010dm,Huber:2011aa,Kozaczuk:2015owa,Dorsch:2018pat,Friedlander:2020tnq,Balaji:2020yrx,Cline:2021iff,Ai:2021kak,Lewicki:2021pgr,Dorsch:2021nje,Jiang:2022btc,Laurent:2022jrs,Wang:2022txy,Ai:2023see,Krajewski:2023clt,Wang:2023kux,Dorsch:2023tss,Kang:2024xqk,Ai:2024shx,Krajewski:2024gma,Wang:2024wcs,Barni:2024lkj,Ekstedt:2024fyq,Ai:2024btx,Krajewski:2024xuz,Dorsch:2024jjl,Carena:2025flp}.\footnote{In the local thermal equilibrium limit, the Boltzmann equations essentially decouple. We consider this case still to belong to the kinetic approach.}

In the second approach, which is valid for ultrarelativistic bubble walls, one uses a more microscopic picture where one counts the particle processes that have momentum exchange between the wall and the particles passing through the wall\,\cite{Dine:1992wr,Bodeker:2009qy, Bodeker:2017cim}. The microscopic particle process can be visualised as a particle kicking the wall when passing through the latter. Hence, we will refer to this approach as the {\it kick approach}. This approach is typically associated with the ballistic approximation;\footnote{The ballistic approximation can also be incorporated into the kinetic approach\,\cite{Liu:1992tn,BarrosoMancha:2020fay,Wang:2024wcs,Ai:2024btx}. } particles pass through the wall so quickly that there is no time for them to collide with each other during their passage through the wall. The latter approximation is, of course, justified by the assumption that the bubble wall under consideration is ultrarelativistic. Often, this also means that the bubble expansion is assumed to be in the so-called detonation mode, such that the fluid in front of the wall is not perturbed. Due to the wall being ultrarelativistic, particles are much more likely to enter the bubble rather than exit it. In this approach, the friction simplifies to\,\cite{Bodeker:2009qy,Bodeker:2017cim}
\begin{equation}
\label{eq:kick_picture}
    \mathcal{P}_{\rm kick} = \sum_{a, X} \int \frac{\d^3 \vecp}{(2\pi)^3 } \frac{p^z}{p^0}\, \d \mathbb{P}_{a\to X}(\vecp)\, f_a(\vecp)\, \Delta p^z_{a \to X} \,,  \qquad \text{(friction in the kick picture)} \, ,
\end{equation}
where $p^0$ is the on-shell energy
and $\d\mathbb{P}_{A\rightarrow X}$ is the differential probability for the transition process $a(p)\rightarrow X$. This formula has a very clear interpretation: $\Delta p^z$ is the momentum exchange between the wall and the particles assuming a given process, while $1/(2\pi)^3\int\d^3\vecp\, p^z/p^0 f_a(\vecp)$ is the flux of $a$ impinging on the wall. Finally, the sum runs over all processes for which momentum is exchanged between the particles and the wall. So one can understand the expression in Eq.\,\eqref{eq:kick_picture} as the sum over all the ``kicks'' that the wall receives from the plasma.

The kick approach has been widely followed in the recent literature, see \cite{Bodeker:2009qy, Bodeker:2017cim,Hoche:2020ysm, Azatov:2020ufh,Gouttenoire:2021kjv,GarciaGarcia:2022yqb,Ai:2023suz,Azatov:2023xem,Baldes:2024wuz,Azatov:2024auq} and, despite being introduced to simplify the calculation of the wall velocity, it has brought several unexpected results. In the evaluation of Eq.\,\eqref{eq:kick_picture}, the following processes have been discovered and discussed:
\begin{itemize}
    \item[(1)] 1-to-1 elementary transmission\,\cite{Dine:1992wr,Bodeker:2009qy}, $a\rightarrow a$: a particle enters into the bubble and its mass changes,
    \item[(2)]  1-to-1 mixing transmission\,\cite{Azatov:2020ufh}, $a\rightarrow b$ : a particle enters into the bubble and transits to another, typically heavier, state via mixing,
    \item[(3)] 1-to-2 transition radiation\,\cite{Bodeker:2017cim,Hoche:2020ysm,Gouttenoire:2021kjv, Azatov:2023xem}, $a\rightarrow b + c$: a particle $a$ enter into the bubble and radiate a particle $c$ with a $\varphi$-dependent mass $m_c(\varphi)$ due to a trilinear interaction term of the form, e.g., $\bar{\psi}\gamma^\mu A_\mu \psi$ or $(\partial_\mu \phi) A^\mu \phi^\dagger$,
    \item[(4)] 1-to-2 particle production\,\cite{Azatov:2021ifm,Ai:2023suz}, $a(+\varphi)\rightarrow b+c$: a particle $a$ enter into the bubble and pair produces $b$ and $c$ due to a $\varphi$-dependent vertex, e.g., $\varphi \phi\chi^2$.
\end{itemize}

\begin{figure}[ht]
    \centering
    \includegraphics[scale=1]{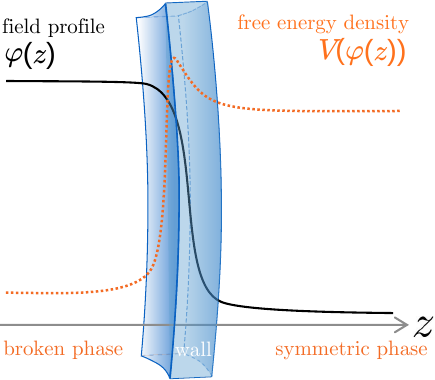}
    \caption{Rest frame of the bubble wall (indicated by the blue region) used for defining the frictional pressures. }
    \label{fig:wall}
\end{figure}

In terms of Feynman diagrams, the four processes (1) - (4) can be represented as  
\begin{align}
\begin{tikzpicture}
[baseline={-0.025cm*height("$=$")}]
\draw[thick,double] (-1.5,0) -- (0,0);
    \draw[thick,double] (0,0) -- (1.5,0);
    \fill (0,0) circle (0.06);
    \fill (-1,0.2)   node[left] {$\scriptstyle a$};
    \fill (1.5,0.2)   node[left] {$\scriptstyle a$};
\end{tikzpicture}\,, 
\qquad
\begin{tikzpicture}
[baseline={-0.025cm*height("$=$")}]
\draw[thick,double] (-1.5,0) -- (0,0);
    \draw[dashed,double] (0,0) -- (1.5,0);
    \fill (0,0) circle (0.06);
    \fill (-1,0.2)   node[left] {$\scriptstyle a$};
    \fill (1.5,0.2)   node[left] {$\scriptstyle b$};
\end{tikzpicture}\,, 
\qquad 
\begin{tikzpicture}
[baseline={-0.025cm*height("$=$")}]
\draw[thick,double] (-1.5,0) -- (0,0);
    \draw[dashed,double] (0,0) -- (1.5,0);
    \fill (0,0) circle (0.06);
    \draw[line width=0.3mm, decorate,decoration={snake,amplitude=.5mm,segment length=1.5mm},double] (0,0) -- (1.2,0.8);
    \fill (-1,0.2)   node[left] {$\scriptstyle a$};
    \fill (1.5,0.2)   node[left] {$\scriptstyle b$};
     \fill (1.2,0.9)   node[left] {$\scriptstyle c$};
\end{tikzpicture}\,,
\qquad
\begin{tikzpicture}
[baseline={-0.025cm*height("$=$")}]
    \draw[thick, double] (-1.5,0) -- (0,0);
    \draw[dashed,double] (0,0) -- (1.2,1);
    \draw[dotted,double] (0,0) -- (1.2,-1);
    \fill (0,0) circle (0.06);
    \fill (-1,0.2)   node[left] {$\scriptstyle a$};
    \fill (1.2,1)   node[left] {$\scriptstyle b$};
    \fill (1.2,-1)   node[left] {$\scriptstyle c$};
    \draw[thick] (0,0) -- (0,-0.8);
    \draw (0,-0.95) circle (0.15);
    \draw[thick] (0-0.1,-0.95-0.1) -- (0+0.1,-0.95+0.1);
    \draw[thick] (0-0.1,-0.95+0.1) -- (0+0.1,-0.95-0.1);
\end{tikzpicture}
 \,,
\end{align}
where a line ended with a wheel cross indicates a power of the background field $\varphi$ attached to the vertex. We have used double lines to indicate that particles may have $\varphi$-dependent masses and thermally corrected dispersion relations. A crucial difference between type\,(3) and type\,(4) is that for the latter, there is a background field\,$\varphi$ insertion in the vertex, e.g., the $\varphi$ in $\varphi\phi\chi^2$, while for the former, there is none (no $\varphi$ appearing in e.g. $\bar{\psi}\gamma^\mu A_\mu \psi$). We refer to such a vertex for type (4) processes as {\it condensate-particle vertex} or $\varphi$-dependent vertex.\footnote{In this paper, a vertex term refers to any term that is higher than quadratic in the {\it fluctuation fields}, such as $\phi\chi^2$ or $\varphi\phi\chi^2$. In contrast, a term that is purely quadratic in the fluctuation fields, such as $\varphi^2\phi^2$ or $\varphi^2 A_\mu A^\mu$, is referred to as a mass term.} 
Assuming a planar wall moving in the $z$ direction, as shown in Fig.\,\ref{fig:wall}, all the above interactions violate the conservation of $z$-momentum among the particles. 
The $z$-momentum gain in the particles $\Delta p^z_{a \to X}=(p^z_X-p_a^z)>0$ is transmitted from the bubble wall,\footnote{Here take the convention that the wall moves along the positive $z$ direction. If the wall moves in the negative $z$-direction, then there would be a total $z$-momentum loss in the particles.}
creating friction on the wall from the process under study.

However and perhaps surprisingly, from previous studies, it is not clear how such contributions can emerge from the kinetic approach. Although it has been shown in\,\cite{Ai:2024shx,Ai:2024btx} that the friction from the 1-to-1 elementary transmission can be recovered from Eq.\,\eqref{eq:condensate-eom} in the ultrarelativistic limit\,$\gamma_w\equiv 1/\sqrt{1-v_w^2}\rightarrow\infty$, the contribution of the other three types of processes (2)-(4) has not been understood well within the kinetic approach. Reconciling the kick and kinetic approaches and recovering the above-mentioned contributions to the friction within the kinetic approach are the main goals and motivations of the current work.

To meet such an end, we shall use the Closed-Time-Path (CTP)\,\cite{Schwinger:1960qe,Keldysh:1964ud,Chou:1984es} formalism (also called Schwin\-ger--Keldysh or in-in formalism) to give a first-principle derivation of the condensate EoM. In this formalism, the condensate is described by the {\it one-point function}, $\varphi\equiv \langle \Phi\rangle$, where the brackets denote the ``vacuum'' expectation value (VEV)\footnote{We are not necessarily studying vacuum state, but still call one-point functions as vacuum expectation values.}, while the connected {\it two-point functions} describe particles. For example, for the scalar field $\Phi$, the fluctuation field upon the background, $\phi \equiv  \Phi-\langle \Phi\rangle$, takes care of the $\phi$ particle excitations. Clearly, $\phi$ has a vanishing VEV, and hence only its two-point function, $\Delta_\phi\equiv \langle\phi\phi\rangle= \langle \Phi\Phi\rangle_c$ where the subscript ``c'' stands for ``connected'', describes particles.

We will show that the conventionally used condensate EoM\,\eqref{eq:condensate-eom} misses higher loop non-local self-energy terms and, therefore, is incomplete. 
In the language of the two-particle-irreducible (2PI) effective action\,\cite{Cornwall:1974vz}, the last term in Eq.\,\eqref{eq:condensate-eom} is derived from the one-loop term of the 2PI effective action\,$\Gamma_{\rm 2PI}[\varphi,\Delta]$, where here $\Delta$ denotes general two-point functions (including $\Delta_\phi$). Higher-loop diagrams with an explicit dependence on the condensate\,$\varphi$ then generate non-local self-energy terms in the condensate EoM.  Yet, as we will show, the processes we are after are all captured by the leading order local and non-local contributions, while sub-leading local terms only yield higher-order corrections. 
We will also show that non-local self-energy terms capture the particle production processes induced by the condensate-particle type vertices.

Since type\,(2) and\,(3) processes are not from vertices directly involving the condensate, they are in principle captured already by Eq.\,\eqref{eq:condensate-eom}. However, to see the contribution from the transition radiation process in the conventional kinetic approach, one has to take into account the $\varphi$-dependent mass terms in solving the Boltzmann equations. This is usually not done. Also, computing the collision terms with either $\varphi$-dependent mass terms or $\varphi$-dependent vertices taken into account is challenging. Therefore, in this work to decouple from the Boltzmann equations and to make the kick picture explicit in the kinetic approach, we will perform an expansion in\,$\varphi$, i.e., VEV insertion approximation (VIA), in the condensate EoM. This expansion generates new non-local self-energy-like terms in the condensate EoM. We shall show that type\,(2) and\,(3) processes are captured by these new terms after the field expansion. We verify in Appendix \ref{app:expansion_Boltz} that this expansion is consistent with the Dyson-Schwinger equations for the two-point functions.

In summary, our derivation allows us to embed the kinetic approach in the framework of out-of-equilibrium quantum field theory. Thus, we provide a scheme that is {\it systematically improvable} loop by loop.
We show that in this framework, the kinetic approach is capable of capturing all of the relevant processes. In particular, non-local terms corresponding to two-loop self-energies, usually absent in the conventional kinetic picture, are shown to describe particle production processes from $\varphi$-dependent vertices. Last but not least, this framework does not rely on ultrarelativistic wall expansion and can be employed in full generality. Solving the coupled system of equations would yield the full real-time evolution of the system. This is only feasible numerically, and even then rather challenging, as it requires solving a coupled system of integrodifferential equations (see\,\cite{Carosi:2024lop} for an application to vacuum decay). This is left for future work.

In Table~\ref{tab:different_pictures}, we classify different particle processes contributing to the friction on the bubble wall. The momentum-conserving processes are decays and scatterings that conserve total momentum among the particles. In the kick approach, these processes do not contribute to the friction on the wall since $\Delta p^z_{a\rightarrow X}=0$ in Eq.\,\eqref{eq:kick_picture}. However, they 
do indirectly contribute to friction; they enter the collision terms in the Boltzmann equations and modify the particle distributions $f_i$, impacting on the wall motion through the last term in Eq.\,\eqref{eq:condensate-eom}. On the other hand, the contribution from hydrodynamic obstruction originates from the inhomogeneity in the temperature across the bubble wall, reflecting heating effects in the shock wave. 
\begin{table}[ht]
    \centering
    \begin{tabular}{|c||c|c|c}
        \hline
        \textit{Processes contributing to the friction} & \textit{kinetic} & \textit{kick} 
        \\
         \hline\hline
         Momentum-conserving processes &   \cite{Moore:1995ua,Moore:1995si} & Not included \\
         Particles gaining mass & \cite{Moore:1995ua,Moore:1995si} &  \cite{Dine:1992wr, Bodeker:2009qy} \\
         Particle production and mixing & This work &  \cite{Azatov:2020ufh}\\
         Transition radiation & This work &  \cite{Bodeker:2017cim,Hoche:2020ysm, Gouttenoire:2021kjv}  \\
         Hydrodynamic obstruction &  \cite{Ignatius:1993qn,Konstandin:2010dm,BarrosoMancha:2020fay,Balaji:2020yrx,Ai:2021kak,Ai:2023see} & 
         Not included \\
         \hline
         Valid for all velocities ?  & Yes & No (only ultrarelativistic) \\
         \hline
    \end{tabular}
    \caption{Different processes that affect the bubble motion and where they have been discussed. We only name a few references in this table; for a more complete list, see the main text. All such processes are captured in the CTP formalism, wherein finite-temperature and nonequilibrium effects can be systematically accounted for. }
    \label{tab:different_pictures}
\end{table}

Before closing the Introduction, we make a specific comment on Ref.\,\cite{Konstandin:2014zta}, which argues that the condensate EoM\,\eqref{eq:condensate-eom} is complete—contrary to our claim above. Their argument relies on the assumption that all particle collisions—including decays and scatterings—conserve the total energy-momentum {\it within} the plasma when summed over all species. Equivalently, it assumes there is no net energy-momentum exchange between the scalar condensate and the plasma via such collisions. Under this assumption, the only source of energy-momentum transfer between the condensate and the plasma is the variation in particle masses across the wall, captured by the last term in Eq.\,\eqref{eq:condensate-eom}. However, this assumption does not hold in the presence of $\varphi$-dependent interaction vertices, such as the $\varphi(z)\phi\chi^2$ term discussed earlier. Such vertices lead to collisions that {\it do} violate energy-momentum conservation within the plasma, even after summing over all species.

The remainder of this article is organised as follows. In Section\,\ref{sec:incompleteness}, we explain why the conventional condensate EoM\,\eqref{eq:condensate-eom} is incomplete. In Section\,\ref{sec:2PI}, we then derive the complete condensate EoM using the CTP formalism within the framework of the 2PI effective action. From the condensate EoM, we identify and classify friction contributions on the wall. To evaluate the expression for the different contributions to the friction,  in Section\,\ref{sec:small_field}, we apply the VIA and recover several results present in the literature on ultrarelativistic bubble walls. 
To make our condensate EoM amenable to applications, in Section\,\ref{sec:localisation}, we make the new term local using a localisation procedure. Finally, we conclude in Section\,\ref{sec:Conc}. \\

{\it Note added:} While we were finalising this paper, we learned that another group (Michael J. Ramsey-Musolf and Jiang Zhu) has been working on similar topics. Their paper\,\cite{Ramsey-Musolf:2025jyk} is posted jointly with ours.

\section{Incompleteness of the conventional bubble equation of motion}
\label{sec:incompleteness}

In this section, we show that Eq.\,\eqref{eq:condensate-eom} is not complete. We shall first illustrate this incompleteness by looking at the operator EoM. In the next section, we derive the complete bubble EoM by including higher-order corrections using the CTP formalism and the 2PI effective action.

Let us start our analysis with the theory of two interacting scalars, described by the following Lagrangian
\begin{align}
\label{eq:toy-model}
    \L=\frac{1}{2}(\partial_\mu\Phi)\partial^\mu\Phi-V_0(\Phi)+\frac{1}{2}(\partial_\mu\chi)\partial^\mu\chi-\frac{1}{2}\mu^2_\chi\chi^2-\frac{\lambda_\chi}{4!}\chi^4-\frac{g}{4}\Phi^2\chi^2 \,,
\end{align}
where
\begin{align}
\label{eq:toy-potential}
V_0(\Phi)=-\frac{1}{2}\mu_\phi^2\Phi^2+\frac{\lambda_\phi}{4!}\Phi^4  \,.
\end{align}
All parameters are assumed to be positive. Finite-temperature corrections to the potential could allow for a FOPT for the field $\Phi$ from zero to a symmetry-broken value. The field\,$\Phi$ thus has a non-vanishing one-point function, $\langle\Phi\rangle\equiv \varphi$, which describes the background bubble configuration, also called condensate in the literature. In this background, the zero-temperature masses of~$\phi$ and~$\chi$ are given by 
\begin{align}
m^2_\phi(\varphi)=V_0''(\varphi)=-\mu_\phi^2+\lambda\varphi^2/2\,,\qquad m^2_\chi(\varphi)=\mu^2_\chi+g\varphi^2/2\,, \label{eqn: zero temp masses}
\end{align}
which are $\varphi$-dependent. The EoM of $\Phi$ is
\begin{align}
\label{eq:EoM-operator}
    \Box\Phi+V_0'(\Phi)+\frac{g}{2}\Phi\chi^2=0\,.
\end{align}
In the above equation, one can do the expansion $\Phi=\varphi+\phi$ where $\phi$ is the fluctuation field on top of the background $\varphi$. Taking further the expectation value of Eq.\,\eqref{eq:EoM-operator} and using $\langle \phi\rangle =\langle \Phi\rangle-\varphi = 0$,  one obtains
\begin{align}
\label{eq:full-eom}
\Box\varphi+V_0'(\varphi)+\frac{\lambda_\phi}{2}\varphi\langle\phi^2\rangle+\frac{g}{2}\varphi\langle\chi^2\rangle+ \frac{\lambda_\phi}{6} \langle\phi^{3}\rangle+\frac{g}{2}\langle\phi \chi^2\rangle=0\,.
\end{align}
Note that due to  $\langle\phi\rangle =\langle\chi\rangle=0$, $\langle\phi\phi\rangle$ and $\langle\chi\chi\rangle$ receive contributions from only the connected two-point functions. Importantly, Eq.\,\eqref{eq:full-eom} is exact, but not closed. However, Eq.\,\eqref{eq:condensate-eom} does not fully follow this exact EoM. Rather, it is derived from Eq.\,\eqref{eq:full-eom} by simply discarding the last two terms\footnote{Below we will show in greater detail how Eq.\,\eqref{eq:condensate-eom} is equivalent to Eq.\,\eqref{eq:eom-1} with mild approximations.}
\begin{align}
   \label{eq:eom-1}
   {\rm \bf Truncated \ condensate\ EoM:\ }
     \Box\varphi+V_0'(\varphi)+\frac{\lambda_\phi}{2}\varphi\langle\phi^2\rangle+\frac{g}{2}\varphi\langle\chi^2\rangle=0\,. 
\end{align}
At this point, we would like to emphasise that Eq.\,\eqref{eq:eom-1} misses additional contributions, namely the term $\langle \phi^3\rangle$ and $\langle \phi\chi^2\rangle$. As we shall see below, these two additional terms describe the direct interaction between the bubble wall and particles via $\varphi$-dependent vertices. Of course, this does not say anything about the EoMs for the two-point functions, which, for the present purposes, can be approximated by the Boltzmann equations in Eq.\,\eqref{eq:Boltzmann_eq}.

\section{Complete bubble equation of motion and most general frictional pressure}
\label{sec:2PI}

For the reader who is not familiar with the CTP formalism and 2PI effective action techniques, we begin this section with a brief motivation and overview. First, the CTP formalism is designed to calculate expectation values of observables at a given time. This allows one to follow their time evolution, which is crucially important for non-equilibrium processes such as those taking place in the background of an expanding bubble, as considered in this paper. This possibility to follow the time evolution of expectation values is in stark contrast to the usual $S$-matrix formalism in QFT, which is concerned with transition amplitudes between asymptotic states after an infinite time evolution. While the $S$-matrix formalism allows to write equations for time-dependent processes in a plasma (such as Boltzmann equations) involving reaction rates computed from $S$-matrix elements, these equations are not derived from first principles and rely on the assumption that the time-scales involved in the individual particle interactions are much shorter than the timescales involved in the evolution of hydrodynamic quantities. 

In contrast to this, the CTP formalism enables the derivation of evolution equations from first principles, and for this reason, we will rely on it for the rest of the paper. When studying the time-evolution of the expectation value of a scalar field undergoing a phase transition in a plasma, one expects that the dynamics will be affected by quantities characterising the medium, such as the number densities and the dispersion relation of excitations of the plasma. Such quantities are encoded in the two-point functions of the different fields, evaluated in the finite-density state as opposed to the vacuum. For example, the number current for a fermion field is of the form $j^\mu_\psi=\bar\psi\gamma^\mu\psi$, and its expectation values are given by fermionic two-point functions.

Hence, to study the phase transition dynamics, it is useful to work with the one-point function $\langle \Phi\rangle$ of the scalar field undergoing the transition, as well as the two-point functions of all the fields that couple to the scalar, and which characterise the properties of the medium. The goal is to derive general equations for the one and two-point functions, which will determine the dynamics of the phase transition and how it affects the medium. Ideally, these equations should be derived by extremizing an appropriate functional of the one and two-point functions. Such a functional is in fact known under the name of 2PI effective action \cite{Cornwall:1974vz,Chou:1984es,Calzetta:1986cq}, which is a generalisation of the more familiar 1PI effective action, which depends only on one-point functions.

As in the 1PI case, extremizing the 2PI action with respect to the one-point functions gives rise to quantum equations of motion for the scalar condensate. On the other hand, extremizing with respect to the two-point functions give rise to Schwinger--Dyson-type equations. Remarkably, as the two-point functions encode the number densities of the different species, the Schwinger--Dyson equations can be used to derive Boltzmann equations. For a review of the CTP formalism and the use of the 2PI effective action, we refer the reader to  Ref.\,\cite{Berges:2004yj}.

\subsection{Complete bubble equation of motion from the CTP formalism}

\begin{figure}[H]
    \centering
    \includegraphics[scale=0.3]{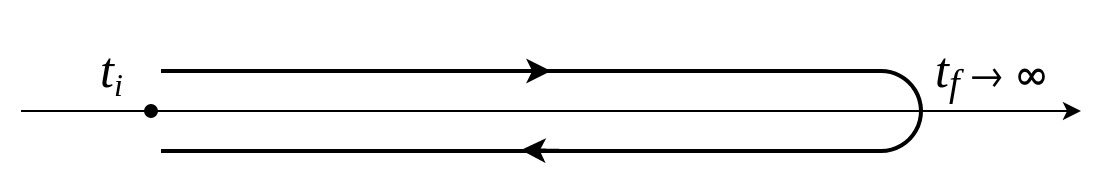}
    \caption{The Keldysh contour $\mathcal{C}$ for the generating functional in the CTP formalism. Here, the forward and backward time contours are slightly shifted off the real line only for the purpose of illustration; both contours should be understood as lying exactly on the real line.}
    \label{fig:keldyshcontour}
\end{figure}
In the CTP formalism, the generating functional is formulated on a closed time path as shown in Fig.\,\ref{fig:keldyshcontour}. To distinguish the fields on this contour, one denotes fields on the forward and backward branches with superscripts ``+'' and ``-'' (Schwinger--Keldysh indices), respectively. First of all, within this formalism, one can define the one-point functions, e.g. $\varphi^a\equiv \langle \Phi^a\rangle$ where $a=\pm$. Thus, for each field, there are two different one-point functions. 

The same reasoning applies to two-point functions. The closed-time contour defines the time-ordering $T_\C$, therefore, for each field there are four propagators, e.g. for~$\Phi$ we have
\begin{subequations}
\begin{align}
\label{eq:4_wightman}
    &\Delta^{<}_\phi(x,y)\equiv \Delta_\phi^{+-}(x,y) =\langle T_\C \Phi^+(x)\Phi^{-}(y)\rangle_c=\langle \Phi(y)\Phi(x)\rangle_c\,,\\
     & \Delta^{>}_\phi(x,y)\equiv \Delta^{-+}_\phi(x,y)=\langle T_\C \Phi^-(x)\Phi^+(y)\rangle_c=\langle \Phi(x)\Phi(y)\rangle_c\,,\\
    &\Delta_\phi^T(x,y)\equiv \Delta_\phi^{++}(x,y) = \langle T_\C \Phi^+(x)\Phi^+(y)\rangle_c=\langle T \Phi(x)\Phi(y)\rangle_c\,,\\
    &\Delta_\phi^{\bar{T}}(x,y)\equiv \Delta_\phi^{--}(x,y) =\langle T_\C\Phi^-(x)\Phi^-(y)\rangle_c=\langle\overline{T}\Phi(x)\Phi(y)\rangle_c\,,
\end{align}
\end{subequations}
where~$T$~($\overline{T}$) is the (anti-)time-ordering operator and~$\Delta_\phi^{\gtrless}$ are called {\it Wightman functions}. From the previous expressions and the definitions of the time-ordering operations $T,\bar T$, the following symmetry property is immediate
\begin{align}
    \label{eq:symprop}
    \Delta_\phi^{ab}(x,x')=\Delta_\phi^{ba}(x',x).
\end{align}

One now needs to determine the dynamics of these one- and two-point functions. In this respect, we make use of the aforementioned 2PI effective action, which is a functional of the one- and two-point functions,~${\Gamma_{\rm 2PI}[\varphi,\Delta_\phi,\Delta_\chi]}$\footnote{For the simplicity of the following discussion, we enforce that  $\langle\chi\rangle=0$.}. For the model with two scalar fields in Eq.\,\eqref{eq:toy-model}, this functional takes the general form (see, e.g., Ref.\,\cite{Berges:2004yj})
\begin{align}
\label{eq:2PI-general-form}
    \Gamma_{\rm 2PI}[\varphi,\Delta_\phi,\Delta_\chi]=& \underbrace{S[\varphi^+]-S[\varphi^-]}_{\rm classical}+\underbrace{\frac{\i}{2}{\rm Tr}\ln \Delta_\phi^{-1}+\frac{\i}{2}{\rm Tr} \left[G_\phi^{-1}(\varphi)\Delta_\phi\right]+\frac{\i}{2}{\rm Tr}\ln \Delta_\chi^{-1}+\frac{\i}{2}{\rm Tr} \left[G_\chi^{-1}(\varphi)\Delta_\chi\right]}_{\rm one-loop}\notag\\
    &+\underbrace{\Gamma_2[\varphi,\Delta_\phi,\Delta_\chi]}_{\rm two-loop\ and\ higher}\,,
\end{align}
where $S[\varphi]$ is the classical action, and~$G_{\phi}^{-1}$ and~$G_\chi^{-1}$ are the kinetic operators for fields $\phi$ and $\chi$ respectively, computed in the $\varphi$ background. They read
\begin{align}
\label{eq:G-inverse-A}
   G_\phi^{ab,-1}(\varphi) =\i c^{ab}\left[\Box+m_\phi^2(\varphi^a)\right]\, , \qquad \qquad
   G_{\chi}^{ab,-1}(\varphi) =\i  c^{ab}\left[\Box+m_\chi^2(\varphi^a)\right]\,,
\end{align}
where $m_\phi^2$ and~$m_\chi^2$ are defined in Eq.\,\eqref{eqn: zero temp masses} and~$c^{ab}={\rm diag}(+1,-1)$.
The trace is taken over the position space as well as the Schwinger--Keldysh indices~${a,b}$. For instance,
\begin{align}
    {\rm Tr} \left[G^{-1}_\phi(\varphi)\Delta_\phi\right]=\sum_{a, \, b} \int\d^4 x\, \left.\left[G_\phi^{ab,-1}(\varphi(x)) \Delta_\phi^{ba}(x,y)\right]\right|_{y \, = \, x}\,.
\end{align}
Finally for the remaining term in the 2PI effective action
\begin{align}
   \Gamma_2[\varphi,\Delta_\phi,\Delta_\chi]= -\i\times {\rm the\ sum\ of\ 2PI\ vacuum\ diagrams}\,,
\end{align}
where 2PI means that the diagrams cannot be separated into two disconnected parts by cutting two propagators. Note that, in the 2PI effective action formalism, all lines appearing in the diagrammatic expansion are exact propagators~$\Delta_\phi$ and~$\Delta_\chi$. In practice, some truncation will be necessary. In the present work, in view of the scattering processes of particles from the bubble wall that we aim to describe, we will frequently make use of the quasi-particle approximation.

 In the 2PI formalism, the EoMs for the one- and two-point functions take the compact form
\begin{subequations}
\begin{align}
    \label{eq:eoms-2PI-CTP}
    &\left.\frac{\delta \Gamma_{\rm 2PI}[\varphi,\Delta_\phi,\Delta_\chi]}{\delta\varphi^+(x)}\right|_{\varphi^+=\varphi^-=\varphi}=0\,,\\[1.5ex]
    &\left.\frac{\delta \Gamma_{\rm 2PI}[\varphi,\Delta_\phi,\Delta_\chi]}{\delta\Delta^{ab}_\phi(x,x')}\right|_{\varphi^+=\varphi^-=\varphi}=0\,, \qquad \left.\frac{\delta \Gamma_{\rm 2PI}[\varphi,\Delta_\phi,\Delta_\chi]}{\delta\Delta^{ab}_\chi(x,x')}\right|_{\varphi^+=\varphi^-=\varphi}=0\,.
\end{align}
\end{subequations}

These EoMs set the dynamics of the theory. In this section, we are interested in the EoM for the one-point function, namely EoM for the condensate (background field), which drives the expansion of the bubble wall. The EoMs for the two-point functions can be reduced to the Boltzmann equations\,\cite{Chou:1984es,Calzetta:1986cq}, see \cite{Ai:2023qnr} for a pedagogical derivation.

To write the condensate EoM compactly, one defines the {\it condensate self-energies},
\begin{align}
\label{eq:Piphiab}
    \Pi^{ab}_\varphi(x,x')\equiv -(ab)\frac{\delta^2\Gamma_2}{\delta\varphi^a(x)\delta\varphi^b(x')}\,. 
\end{align}
This form of self-energy terms, descending from $\Gamma_2$, can be identified with the proper self-energy. 
We shall also introduce the notation $\Pi_\varphi^{>}\equiv \Pi_\varphi^{-+}$ and $\Pi_\varphi^{<}\equiv \Pi_\varphi^{+-}$, analogous to the Wightman functions. With this definition for the self-energies $\Pi_\varphi^{ab}(x,x')$, the functional $\Gamma_2$ can be written as
\begin{align}
\label{eq:Gamma2-self-energy}
    \Gamma_2 \supset -\frac{1}{2}\,\sum_{a,b}(ab)\int\d^4 x\,\d^4 x' \,\varphi^a(x) \Pi_\varphi^{ab}(x,x')\varphi^b(x')\,,  
\end{align}
where we only kept terms containing two insertions of the condensate. We emphasize that terms that do not depend on~$\varphi$ also appear in~$\Gamma_2$, however, these terms do not enter the condensate EoM, and therefore, they are not of interest to us in this discussion.
Then, varying with respect to the condensate, we have 
\begin{align}
    &-\frac{\delta \Gamma_2}{\delta\varphi^+(x)}=\int \d^4 x'\, \Pi_\varphi^{++}(x,x')\varphi^+(x')-\int\d^4 x'\, \Pi_\varphi^{+-}(x,x')\varphi^-(x')\notag\\[2ex]
    &\qquad\quad\Rightarrow -\left.\frac{\delta \Gamma_2}{\delta\varphi^+(x)}\right|_{\varphi^+=\varphi^-=\varphi}=\int \d^4 x'\, \Pi_\varphi^{\rm R}(x,x')\varphi(x')\,,
\end{align}
where $\Pi_{\varphi}^{\rm R}(x,x')\equiv \Pi_\varphi^{++}(x,x')-\Pi_\varphi^{+-}(x,x')
$ is the retarded condensate self-energy.
From Eq.\,\eqref{eq:eoms-2PI-CTP}, the condensate EoM can thus be written as
\begin{align}
\label{eq:eom-condensate-CTP}
    &\Box\varphi(x)+V_0'(\varphi(x))+\frac{\lambda_\phi}{2}\varphi(x)\Delta^{++}_\phi(x,x)+\frac{g}{2} \varphi(x) \Delta^{++}_\chi(x,x)+ \int \d^4 x'\, \Pi^{\rm R}_\varphi(x,x')\varphi(x')=0\,,
\end{align}
Again, Eq.\,\eqref{eq:condensate-eom} corresponds to Eq.\,\eqref{eq:eom-condensate-CTP} with the last term discarded. In this paper, we turn to the study of the impact of NLO corrections, contained in the~$\Pi_\varphi^{\rm R}$ term, on the bubble wall dynamics. This term has been studied for particle production from an oscillating background field\,\cite{Cheung:2015iqa,Ai:2021gtg,Wang:2022mvv,Ai:2023ahr}, where it was shown to have a relevant impact.

To leading order, the terms in~$\Gamma_2$ that contribute to the condensate retarded self-energy are the two-loop 2PI vacuum diagrams that have an explicit dependence on $\varphi$. For our model described by the Lagrangian in Eq.\,\eqref{eq:toy-model}, these are
\begin{align}
\Gamma_2\supset \: & -\i\left(
\begin{tikzpicture}[baseline={-0.025cm*height("$=$")}]
\draw[draw=black,thick] (-0.5,0) circle (0.15) ;
\draw[draw=black,thick] (-0.5-0.11,0+0.11) -- (-0.5+0.11,0-0.11);
\draw[draw=black,thick] (-0.5-0.11,0-0.11) -- (-0.5+0.11,0+0.11);
\draw[draw=black,dashed] (-0.35,0) -- (0.25,0);
\draw[draw=black,dashed] (0.75,0) circle (0.5);
\draw[dashed] (0.25,0) -- (1.25,0);
\draw[draw=black,dashed] (1.25,0) -- (1.85,0);
\draw[draw=black,thick] (2,0) circle (0.15) ;
\draw[draw=black,thick] (2-0.11,0+0.11) -- (2+0.11,0-0.11);
\draw[draw=black,thick] (2-0.11,0-0.11) -- (2+0.11,0+0.11);
\end{tikzpicture}
\, +\,
\begin{tikzpicture}[baseline={-0.025cm*height("$=$")}]
\draw[draw=black,thick] (-0.5,0) circle (0.15) ;
\draw[draw=black,thick] (-0.5-0.11,0+0.11) -- (-0.5+0.11,0-0.11);
\draw[draw=black,thick] (-0.5-0.11,0-0.11) -- (-0.5+0.11,0+0.11);
\draw[draw=black,dashed] (-0.35,0) -- (0.25,0);
\draw[draw=black,thick] (0.75,0) circle (0.5);
\draw[dashed] (0.25,0) -- (1.25,0);
\draw[draw=black,dashed] (1.25,0) -- (1.85,0);
\draw[draw=black,thick] (2,0) circle (0.15) ;
\draw[draw=black,thick] (2-0.11,0+0.11) -- (2+0.11,0-0.11);
\draw[draw=black,thick] (2-0.11,0-0.11) -- (2+0.11,0+0.11);
\end{tikzpicture}
    \right) \notag \\[2ex]
     &= \:   \frac{1}{2}\frac{\i\lambda_\phi^2}{3!}  \sum_{a,b} \int \d^4x\d^4x' \, (ab)\,\varphi^a (x) \left( \Delta_\phi^{ab}(x,x')\right)^3 \varphi^b(x')  \notag \\
    &  + \frac{1}{2} \frac{\i g^2}{2} \sum_{a,b} \int \d^4x\d^4x'  (ab)\, \varphi^a (x) \left( \Delta_\chi^{ab}(x,x')\right)^2 \Delta_\phi^{ab}(x,x') \varphi^b(x') \,,\label{eq:diagram-2PI-2loop}
\end{align}
where in the first line we suppressed symmetry factors. Above, a dashed line denotes the $\phi$ propagator~$\Delta_\phi$, a solid line the $\chi$ propagator~$\Delta_\chi$, and a dashed line ending with a crossed circle indicates an insertion of the background~$\varphi$ at the corresponding vertex. Taking the variation of Eq.\,\eqref{eq:diagram-2PI-2loop} with respect to the condensate, we find the two-loop retarded self-energy entering the condensate EoM~\eqref{eq:eom-condensate-CTP}. It reads
\begin{align}
    \Pi^{\rm R}_\varphi(x,x')=&-\frac{\i \lambda_\phi^2}{3!} \left[(\Delta_\phi^{++}(x,x'))^3-(\Delta_\phi^{+-}(x,x'))^3\right]  \notag\\
    &-\frac{\i g^2}{2} \left[\Delta_\phi^{++}(x,x') (\Delta_\chi^{++}(x,x'))^2-\Delta_\phi^{+-}(x,x')(\Delta_\chi^{+-}(x,x'))^2\right]\,,
    \label{Eq:Pi}
\end{align}
as follows from Eqs.\,\eqref{eq:Piphiab}, \eqref{eq:diagram-2PI-2loop} and  Eq.\,\eqref{eq:symprop}. 
Dissipative processes described by the diagrams in Eq.\,\eqref{eq:diagram-2PI-2loop} can be interpreted via the cutting rules at finite temperature\,\cite{Cutkosky:1960sp,Weldon:1983jn,Kobes:1985kc,Kobes:1986za,Landshoff:1996ta,Gelis:1997zv,Bedaque:1996af} and for bubble wall dynamics these are scattering processes between particles and the background field, i.e. the bubble. For an oscillating condensate, they can also describe its decay\,\cite{Cheung:2015iqa,Ai:2021gtg,Wang:2022mvv,Ai:2023ahr,Ai:2023qnr}. 

A distinct feature of these processes is that they involve the condensate-particle vertices. In our toy model, these vertices are $\varphi\phi^3$ and $\varphi \phi \chi^2$. In the conventional bubble EoM\,\eqref{eq:condensate-eom}, the bubble wall interacts with particles only through the modified dispersion relation of the fluctuation fields, i.e. via $\varphi$-dependent mass terms of the fluctuation fields. The condensate-particle vertices also give rise to new collision terms in the Boltzmann equations and could also be important there.

Equation \eqref{eq:eom-condensate-CTP} is non-local and consequently it is extremely difficult to solve exactly and analytically. The standard procedure for localisation is to do the Wigner transform for the correlation functions and then perform the gradient expansion. A detailed discussion of this procedure can be found in, e.g., Ref.\,\cite{Prokopec:2003pj} (See also\,\cite{Ai:2023qnr} for a pedagogical introduction).

The Wigner transform of a two-point function $G(x_1,x_2)$ is defined as
\begin{align}\label{eq:Wigner}
    \overbar{G}(k,x)=\int\d^4 r \, \e^{\i k \cdot r} G\left(x+\frac{r}{2},x-\frac{r}{2}\right)\,,
\end{align}
where $x=(x_1+x_2)/2$ and $r=x_1-x_2$. The inverse Wigner transform reads
\begin{align}
    G(x,x')=\int \frac{\d^4 k}{(2\pi)^4} \, \e^{-\i k(x-x')}\overbar{G}\left(k,\frac{x+x'}{2}\right)\,.
\end{align}
In the quasi-particle limit and at leading order in the gradient expansion, the propagators in Wigner space~$\{k,x\}$ read\,\cite{Garbrecht:2018mrp}\footnote{Note the difference between our notation and that in Ref.\,\cite{Garbrecht:2018mrp}; Our $\Delta$ is the $\i\Delta$ of Ref.\,\cite{Garbrecht:2018mrp} }
\begin{subequations}
\label{eq:on-shell-limit}
\begin{align} \overbar{\Delta}^<(k,x)&=
2\pi \delta(k^2-m^2)\left[
\vartheta(k^0) f(\mathbf k,x)
+\vartheta(-k^0) (1+ f(-\mathbf k,x))\right]
\,,
\\
\overbar\Delta^>(k,x)&=
2\pi \delta(k^2-m^2)\left[
\vartheta(k^0) (1+f(\mathbf k,x))
+\vartheta(-k^0)  f(-\mathbf k,x)\right]
\,,
\\
\overbar{\Delta}^T(k,x)&=
\frac{\rm i}{k^2-m^2+{\rm i}\varepsilon}+
2\pi \delta(k^2-m^2)\left[
\vartheta(k^0) f(\mathbf k,x)
+\vartheta(-k^0)  f(-\mathbf k,x)\right]
\,, \label{eqn:barDelta++}
\\
\overbar{\Delta}^{\bar T}(k,x)&=
-\frac{\rm i}{k^2-m^2-{\rm i}\varepsilon}+
2\pi \delta(k^2-m^2)\left[
\vartheta(k^0) f(\mathbf k,x)
+\vartheta(-k^0)  f(-\mathbf k,x)\right]
\,,
\end{align}
\end{subequations}
where the particle density~$f$ is a function of the space-time coordinate~$x$, the {\it three}-momentum~$\vec k$ and $\vartheta$ designates the Heaviside theta-function.
Note that in these expressions only the tree-level, $\varphi$-dependent mass $m^2$ appears. Our procedure, however, can be easily generalised to the case when we include corrections to the mass by replacing~$m^2\to M^2$, where~$M^2$ now receives both thermal and quantum corrections.

Fourier transforming Eq.\,\eqref{eqn:barDelta++}, one can write 
\begin{subequations}
 \label{eq:modified_potential}   
 \begin{align}
   & \frac{\lambda_\phi}{2}\varphi(x)\Delta^{++}_\phi(x,x)\equiv \frac{\partial\delta V_\phi(\varphi)}{\partial\varphi}+ \frac{\d m^2_\phi(\varphi)}{\d\varphi} \int_{\vec k,\phi}\,{f_\phi(\vec k,x)}\,,\\
    &\frac{g}{2}\varphi(x)\Delta^{++}_\chi(x,x)\equiv \frac{\partial\delta V_\chi(\varphi) }{\partial\varphi} + \frac{\d m^2_\chi(\varphi)}{\d\varphi}\int_{\vec k,\chi}\,{f_\chi(\vec k,x)}\,,
\end{align}
\end{subequations}
where $E_{\phi/\chi}=\sqrt{m_{\phi/\chi}^2+\veck^2}$. In Eq.\,\eqref{eq:modified_potential} we introduced the following short-hand notation for phase space integrals with the Lorentz invariant measure for particle species~$i$
\begin{equation}
    \int_{\vec k,i} = \: \int \frac{\d^3 \vec k}{(2\pi)^3 2E_i} \,,
\end{equation}
which we will use in the remainder of this work. Wherever there is no room for confusion, we will drop the index~$i$ identifying the particle species.
We have identified the divergent zero-temperature one-loop correction to the classical potential
\begin{equation}
    \frac{\partial\delta V_{\phi}(\varphi)}{\partial\varphi} = \: \frac{\lambda_\phi}{2} \varphi(x) \int \frac{\d^4 k }{(2\pi)^4} \frac{\i}{k^2-m_\phi^2+\i\varepsilon}\,,
\end{equation}
and analogously for~$\chi$. These divergent corrections will be made finite following standard regularisation and renormalisation procedures and consequently, below, we should understand both $V_0$ and $\delta V$ as renormalised potentials.
Summing up, we find the zero-temperature one-loop effective potential
\begin{align}
    V_0(\varphi)+\delta V_\phi(\varphi) +\delta V_\chi(\varphi) = \: V_0(\varphi) + \delta V_{T=0} ( \varphi) = \: V(\varphi)\,.
\end{align}
Then, we can write the condensate EoM as
\begin{align}
\label{eq:eom-final form}
    \Box\varphi + V' (\varphi)+\sum_{i=
    \phi,
    \chi}\frac{\d m^2_i(\varphi)}{\d\varphi} \int_{\vec k,i} \,{f_i(\vec k,x)}+\hskip-1.7cm\underbrace{\int \d^4 x'\, \Pi^{\rm R}_\varphi (x,x') \varphi(x')}_{\rm {not\  considered}\ in\ earlier\ studies\ on \ the\ kinetic\ approach}\hskip-1.5cm=0\,.
\end{align}
We take $i=\phi,\chi$ for the specific model we are considering. In general models, $i$ runs over every field that obtains a $\varphi$-dependent mass term. Above, we have highlighted that the last term has been overlooked in the literature on bubble wall dynamics.

{Before we proceed further, we make a comment here. Usually, it is thought that a degeneracy in the effective potential between the phases leads to a static bubble wall if one uses Eq.\,\eqref{eq:condensate-eom}. And one may think that the new term above contradicts this statement. However, this statement itself is ambiguous as it depends on which loop order of the effective potential one refers to. Also, due to the inhomogeneous background field, there are gradient effects that are not captured by the effective potential but affect the wall profile and motion. Actually, for a static wall, the new term in Eq.\,\eqref{eq:eom-final form} can be understood as higher-order and non-local corrections to the effective potential.}

To understand how friction on the bubble wall enters the EoM, we consider a stationary wall and work in its rest frame. If the bubble radius is large compared to other length scales involved in the process, we can approximate the bubble to a planar wall.
We then assume that the wall moves in the $z$-direction, as shown in Fig.\,\ref{fig:wall}. In this frame, due to the translation invariance in the time and in the spatial directions tangential to the wall, denoted as $\vec{x}_\perp$, it is convenient to define
\begin{align}
\label{eq:piR}
    \pi^{\rm R}_\varphi(z,z')=\int\d t'\d^2 \vec{x}'_\perp\, \Pi^{\rm R}_\varphi(t-t',\vec{x}_\perp-\vec{x}'_\perp; z,z')\,, 
\end{align}
so that
\begin{align}
\label{eq:self-energy-term-wall-frame}
    \int \d^4 x'\, \Pi^{\rm R}_\varphi (x,x') \varphi(x')=\int \d z' \, {\pi}^{\rm R}_\varphi(z,z')\varphi(z')\, .
\end{align}
Multiplying Eq.\,\eqref{eq:eom-final form} by $\d\varphi/\d z$ and integrating over $z$, we find
\begin{align}
\label{eq:friction}
  \int_{-\delta}^\delta \d z\,\frac{\d\varphi}{\d z}\left[\Box\varphi + V^\prime(\varphi) + \sum_{i=\phi,\chi}\frac{\d m^2_i(\varphi)}{\d\varphi} \int_{\vec k,i} \,f_i(\veck,z)+ \int \d z' \, {\pi}^{\rm R}_\varphi(z,z')\varphi(z')\right]=0\,,
\end{align}
where~$\delta$ is a length scale larger than the wall width but still much smaller than the bubble radius, $L_w <  \delta\ll R_w$.
The first term is zero because it is the integral of a total derivative of a quantity that vanishes for large $|z|$ where the scalar field is constant. 
The second term is also the integral of a total derivative, that, however, does not vanish at the boundaries. It yields the \textit{zero-temperature} effective potential difference between the two phases, $\Delta V \equiv V(\delta)-V(-\delta)$. Finally, we find
\begin{align}
\label{eq:DeltaV}
  \Delta V = -\sum_{i=\phi,\chi} \int_{-\delta}^\delta \d z \frac{\d m_i^2(\varphi(z))}{\d z}\,\int_{\vec k,i}\,f_i(\veck,z) -\int_{-\delta}^\delta \d z\, \frac{\d \varphi}{\d z}\int \d z' \pi^{\rm R}_\varphi(z,z')\varphi(z')\,.
 \end{align}
 We identify the driving and the frictional pressures 
\begin{subequations}
\label{eq:pressures}
    \begin{align}
\label{eq:friction-formula}
&\P_{\rm driving} = \,  \Delta V\,,\\
\label{eq:P-friction}
&\P_{\rm friction}=  \P_{\rm mass} + \P_{\rm vertex}  \,.
\end{align}
\end{subequations}
We distinguish between the friction induced by the condensate-dependent mass term
\begin{align}
\label{eq:P_mass}
  \P_{\rm mass} \equiv \hskip-0.2cm\underbrace{-\sum_i \int_{-\delta}^\delta \d z \frac{\d m_i^2(\varphi(z))}{\d z}\,\int_{\vec k,i}\, f_i(\veck,z)}_{\mathrm{{included\ in\ earlier\ studies\ on\ the\ kinetic\ approach}}}\,\hskip-0.3cm,  
\end{align}
and the friction induced by condensate-dependent interaction vertices
\begin{align}
\label{eq:P_vertex}
\P_{\rm vertex} \equiv \hskip-0.5cm\underbrace{-\int_{-\delta}^{\delta}\d z\, \frac{\d \varphi}{\d z}\int \d z'  \pi^{\rm R}_\varphi(z,z')\varphi(z')}_{\mathrm{{not\ considered}\ in\ {earlier\ studies\ of\ the\ kinetic\ approach}}}\,\hskip-0.7cm.
\end{align}

\subsection{Friction induced by condensate-dependent masses: process (1)}
\label{sec:Pmass}

\noindent
Let us first focus on the pressure induced by the $\varphi$-dependent mass term, Eq.\,\eqref{eq:P_mass}. Here we discuss two typical cases: near local thermal equilibrium and the ultrarelativistic regime.

\paragraph{Near local thermal equilibrium.} Close to equilibrium, we expand the density functions as ${f_i=f_i^{\rm eq}+\delta f_i}$, so that 
\begin{align}
\label{eq:thermal-potential-correction}
    \sum_{i=
    \phi,
    \chi}\frac{\d m^2_i(\varphi)}{\d\varphi} \int_{\vec k,i} \, f_i(\veck,z)
    =\frac{\partial\delta V_{T\neq 0}(\varphi)}{\partial\varphi}+\sum_{i=
    \phi,
    \chi}\frac{\d m^2_i(\varphi)}{\d\varphi} \int_{\vec k,i} \,\delta f_i(\veck,z)\,.
\end{align}
Above, we have introduced the thermal correction to the potential\footnote{Identifying the RHS as the derivative of a potential correction with respect to $\varphi$ is only valid at the leading-order in the derivative expansion due to $\varphi$-dependent masses. See Ref.\,\cite{Dashko:2020qwy} for a discussion beyond the derivative expansion.}
\begin{equation}
    \frac{\partial \delta V_{T\neq 0}(\varphi)}{\partial\varphi} = \: \sum_{i=
    \phi,
    \chi}\frac{\d m^2_i(\varphi)}{\d\varphi} \int_{\vec k,i}\, f_i^{\rm eq}(\veck,z)\,.
\end{equation}
Combining $\delta V_{T\neq 0}$ with $V$ defines the effective potential $V_{\rm eff}$ at finite temperature,
\begin{align}
    V_{\rm eff}(\varphi,T) \equiv V(\varphi)+\delta V_{T\neq 0}(\varphi,T)\,.
\end{align}
Then Eq.\,\eqref{eq:P_mass} can be written as
\begin{align}
 \P_{\rm mass} & =-\int_{-\delta}^\delta \d z\, \frac{\d\varphi}{\d z}\frac{\partial \delta V_{T\neq 0}(\varphi,T)}{\partial\varphi} -\sum_i \int_{-\delta}^\delta \d z \frac{\d m_i^2(\varphi(z))}{\d z}\,\int_{\vec k,i}\, \delta f_i(\veck,z)\notag\\[1.5ex]
 & = -\int_{-\delta}^\delta \d z \left(\frac{\d \delta V_{T\neq 0}}{\d z}-\frac{\partial \delta V_{T\neq 0}(\varphi,T)}{\partial T}\frac{\d T}{\d z}\right)-\sum_i \int_{-\delta}^\delta \d z \frac{\d m_i^2(\varphi(z))}{\d z}\,\int_{\vec k,i}\delta f_i(\veck,z)\notag\\[1.5ex]
 & = -\Delta\delta V_{T\neq 0}+\P_{\rm LTE}+\P_{\rm dissipative}\,,
\end{align} 
where 
\begin{subequations}
\begin{align}
 \P_{\rm LTE} &=\int_{-\delta}^\delta \d z\, \frac{\partial \delta V_{T\neq 0}}{\partial T}\frac{\d T}{\d z} \,,\\
 \P_{\rm dissipative} &=
 \underbrace{-\sum_i \int_{-\delta}^\delta \d z \frac{\d m_i^2(\varphi(z))}{\d z}\,\int_{\vec k,i}\delta f_i(\veck,z)}_{\text{out-of-equilibrium\ effects}} \,.
\end{align}
\end{subequations}
$\P_{\rm LTE}$ is the friction that persists in local thermal equilibrium and is due to the inhomogeneous temperature distribution of the plasma across the wall\,\cite{Ai:2021kak} while $\P_{\rm dissipative}$ is the force due to out-of-equilibrium effects. Normally, one needs to solve the Boltzmann equations to obtain $\delta f_i$, see Refs.\,\cite{DeCurtis:2022hlx,DeCurtis:2023hil,DeCurtis:2024hvh} for some recent numerical calculations.

In Ref.\,\cite{Ai:2021kak}, $\Delta \delta V_{T\neq 0}$ is identified as a part of the driving pressure such that there~${\P_{\rm driving}=\Delta V_{\rm eff}}$. Since only the difference ${\P_{\rm driving}-\P_{\rm friction}}$ is physical, the identification of driving and frictional pressures is not unique. The identification in Eqs.\,\eqref{eq:pressures} agrees with the convention used in studies of friction on ultrarelativistic bubble walls\,\cite{Bodeker:2009qy,Bodeker:2017cim}. An advantage of this identification is that the driving force is a constant for a given particle physics model, i.e., independent of the fluid state or microscopic particle processes.

\paragraph{The ultrarelativistic regime.}

For $\gamma_w\gg 1$, the wall is thin enough in the plasma frame such that one can neglect the collisions between the particles when they cross the wall, this approximation is also called ballistic approximation\,\cite{Dine:1992wr,Moore:1995si}. Mathematically, the ballistic condition is given by  
\begin{align}
L_w \ll \gamma_w  L_{\rm MFP} \qquad \text{(ballistic condition)}\,,
\end{align}
where $L_w$ is the width of the wall in the wall frame, $L_{\rm MFP}$ is the mean free path of the incoming particles in the plasma frame. In the wall frame, the plasma moves too fast such that the particles do not have time to collide with each other during the passage of the plasma through the wall. If we do not consider particle-production processes for now, the distribution functions of the plasma remain unchanged even when the latter enters the wall. At this leading-order
\begin{align}
    f^{(0)}_i(\veck)= f_i^{\rm out}(\veck,T_+)\,,
\end{align}
where $f_i^{\rm out}$ and $T_+$ are the distribution function and temperature for the plasma in front of the wall, respectively.\footnote{In Ref.\,\cite{Ai:2024shx}, a different notation $f_i^{\rm in}(\veck,T_+)$ is used to denote the distribution of the incoming particles from the outside of the wall. There, the superscript ``in'' stands for ``incoming''. } 
Furthermore, for ultrarelativistic walls, it is typically assumed that the bubble expansion is in the detonation mode such the particles in front of the wall are in thermal equilibrium at the nucleation temperature, $T_+=T_n$ and $f_{i}^{\rm out}=f_i^{\rm out;eq}$.

Therefore, in the ultrarelativistic regime, a first contribution from $\mathcal{P}_{\rm mass}$ comes from the particles gaining a mass while crossing the wall. Inserting the distribution function of the incoming particles impinging on the wall into Eq.\,\eqref{eq:P_mass}, we obtain 
\begin{align}
\label{eq:pres_BM}
   \mathcal{P}_{\rm mass} \supset -\sum_i \Delta m_i^2 \int_{\vec k,i}\, f_i^{\rm out;eq}(\veck;T_n) \equiv \P_{\rm BM} \, , 
\end{align}
which is nothing but the B\"{o}deker-Moore thermal friction\,\cite{Bodeker:2009qy}, which corresponds to the transition processes we called (1) in the Introduction. Notice that in the ballistic regime, the reflection of the particles off the bubble wall can also be accounted for\,\cite{Dine:1992wr,Azatov:2020ufh,Ai:2024btx}. 

It is known that $\P_{\rm BM}$ is not the full friction induced by the $\varphi$-dependent mass terms in the ultrarelativistic regime. In fact, due to particle-production processes, e.g. transition radiation, decays and collisions, the distribution functions $f_i$ are not exactly $f^{(0)}=f_i^{\rm out; eq}$. The corrections in principle need to be obtained by solving the EoMs for the two-point functions or, as a pertinent approximation, the Boltzmann equations in the quasi-particle limit. In this paper, we omit this analysis but still reveal the said additional friction contained in $\P_{\rm mass}$ using some approximations. This will be done in Section\,\ref{sec:small_field}.

\subsection{Friction induced by condensate-dependent vertices: process (4)}\label{subsec: friction from pair production}

We now turn to the analysis of the contribution to the pressure due to $\varphi$-dependent vertices, as defined in Eq.\,\eqref{eq:P_vertex}. To be specific, let us look at one of the diagrams contributing to the self-energy
\begin{align}
\label{eq:self-energy-phichichi}
    \Pi^{ab}_\varphi\supset -\frac{\i g^2}{2} \Delta_\phi^{ab}(x,x') (\Delta_\chi^{ab}(x,x'))^2\,.
\end{align}
Since $\P_{\rm vertex}$ contains an integral over $z'$ and thus is non-local, computing it is generally difficult. In Section\,\ref{sec:localisation}, we will discuss a localisation procedure for it. In this subsection, we consider a simple situation where the self-energy $\pi^{\rm R}_\varphi(z,z')$ is $z$-translation invariant, i.e.,~${\pi^{\rm R}_\varphi(z,z')=\pi^{\rm R}_\varphi(z-z')}$. This greatly simplifies the analysis and allows for a clear interpretation of~$\P_{\rm vertex}$. As we will see shortly,~$\P_{\rm vertex}$ is the friction induced by particle production processes via condensate-particle vertices, e.g. the vertex~$g\varphi \phi\chi^2$ for the diagram in Eq.\,\eqref{eq:self-energy-phichichi}.

Translational invariance in the~$z$ direction is restored as the wall reaches ultrarelativistic velocities, namely as~$\gamma_w\gg1$. In this regime, in fact, the following applies:
\begin{itemize}
    \item Any particle impinging the bubble wall, say a $\phi$-particle, would have a large~$z$-momentum ($\sim \gamma_w T$) in the rest frame of the wall, as enforced by the particle's distribution function. As a result, its $\varphi$-dependent mass, and even its temperature-dependent thermal mass, can be neglected. Consequently, the two-point function $\Delta_\phi$ approximately exhibits translation invariance in the $z$-direction, i.e., $\Delta_\phi(z, z') \approx \Delta_\phi(z - z')$. 
    \item In the wall frame, any transition process conserves energy. Therefore, the outgoing states, say two $\chi$-particles, must also have large energy $\sim \gamma_w T/2$. Therefore, just like the incoming particles, their~$\varphi$-dependent and thermal masses can be neglected, and we have ${\Delta_\chi (z,z')\approx \Delta_\chi(z-z')}$.
\end{itemize}
The above implies that we are also allowed to ignore the coordinate dependence of the distribution functions~$f_{\phi/\chi}$, which we will take as functions solely of the three-momentum.
Furthermore, we assume $\chi$-particles to be very heavy
\begin{align}
\label{eq:heavy_condition}
    m_\chi  \gg v_b\sim T\,,
\end{align}
where $v_b$ is the value of the field~$\Phi$ in the symmetry-broken phase. 
This additional assumption implies that~$\chi$-particles are Boltzmann suppressed and thus selects the production process~$\phi\rightarrow\chi\chi$ over the emission and absorption processes~$\phi\chi\to\chi$ and~$\chi\to\chi\phi$. In the rest frame of the plasma, this process can be described as the bubble wall hitting a $\phi$ particle that thus changes its momentum and transits into two $\chi$ particles.
The production process has been used as a mechanism of heavy dark matter production, originally proposed in Ref.\,\cite{Azatov:2021ifm}, and later generalised to cases of fermion and vector dark matter\,\cite{Azatov:2024crd,Ai:2024ikj}. 
Our analysis below, however, can be easily generalised to the case when $\chi$ particles are not heavy, as we will comment later in this section (see Eq.\,\eqref{eq:P-phi-phi-phi} below).
Using the approximate translational invariance along the~$z$-direction,
\begin{equation}
\label{eq:pi_z_translation}
{\pi^{\rm R}_\varphi(z,z')\approx \pi^{\rm R}_\varphi(z-z')} \, ,
\end{equation}
we can Fourier transform $\pi^{\rm R}_\varphi(z-z')$ with respect to $z-z'$. Using Eq.\,\eqref{eq:piR}, we have 
\begin{align}
\label{eq:relation_tildepiR_tildePiR}
    \widetilde{\pi}_\varphi^{\rm R}(q^z)= \int \d(z-z') \, \e^{{-\i q^z (z-z')}} \pi^{\rm R}_\varphi (z-z')= \int \d^4 (x-x') \, \e^{{\i \hat{q} (x-x')} } \Pi_\varphi^{\rm R} (x-x')=\widetilde{\Pi}^{\rm R}_\varphi (\hat{q})\,,  
\end{align}
where $\hat{q}=(0,0,0,q^z)$. Note that although $\pi^{\rm R}_\varphi$ and $\Pi^{\rm R}_\varphi$ have different dimensions, their Fourier transforms share the same dimension. Taking the Fourier transform of the wall as well
\begin{align}
\label{Eq:Ftransform}
    \varphi(z)=\int\frac{\d q^z}{2\pi}\,\e^{{\i q^z z}}\widetilde{\varphi}(q^z)\,, \qquad \widetilde{\varphi}(q^z)=\int \d z\,\e^{-{\i q^z z}}\varphi(z)\,,
\end{align}
with $\widetilde\varphi(-q^z)=\widetilde\varphi(q^z)^*$ due to the reality of $\varphi(z)$,
Eq.\,\eqref{eq:P_vertex} becomes
\begin{align}
\label{eq:Pvertex-simple}
    {\P_{\rm vertex}=-\int \frac{\d q^z}{2\pi}\, (\i q^z)\,|\widetilde{\varphi}(q^z)|^2\,\widetilde{\pi}^{\rm R}_\varphi(-q^z) =-\int \frac{\d q^z}{2\pi}\, q^z\, |\widetilde{\varphi}(q^z)|^2\,{\rm Im}\widetilde{\pi}^{\rm R}_\varphi(q^z)\,.}
\end{align}
Above, we have used the relations\,\cite{Bellac:2011kqa}
\begin{align}
\label{eq:PiR-relations}
    {\rm Im}\widetilde{\pi}^{\rm R}_\varphi(q^z)= -{\rm Im}\widetilde{\pi}^{\rm R}_\varphi(-q^z)\,,\qquad {\rm Re}\widetilde{\pi}^{\rm R}_\varphi(q^z)= {\rm Re}\widetilde{\pi}^{\rm R}_\varphi(-q^z)\,,
\end{align}
and that since we are integrating over a symmetric interval, only the symmetric part of the integrand survives.

It is well known that the imaginary part of self-energies have an interpretation in terms of particle processes via the cutting rules~\cite{Cutkosky:1960sp,Weldon:1983jn,Kobes:1985kc,Kobes:1986za,Landshoff:1996ta,Gelis:1997zv,Bedaque:1996af}. To see this, we note that\,\cite{Bellac:2011kqa} 
\begin{align}
\label{eq:ImPi-varphi}
    {\rm Im} \, \widetilde{\pi}_\varphi^{\rm R}= -\frac{\i}{2} \left(\widetilde{\pi}_\varphi^{>}-\widetilde{\pi}_\varphi^{<}\right)\,,
\end{align}
which takes the form of a collision term. Note that the imaginary unit in the above equation is cancelled by the one contained in the self-energies, as can be seen in Eq.\,\eqref{Eq:Pi}, so that $\P_{\rm vertex}$ is explicitly real.

Focusing on the contribution to the self-energy in Eq.\,\eqref{eq:self-energy-phichichi}, Eq.\,\eqref{eq:ImPi-varphi} becomes
\begin{align}
    {\rm Im}\widetilde{\pi}^{\rm R}_\varphi (q^z) \supset &-\frac{g^2}{4} \int\frac{\d^4 p}{(2\pi)^4}\int\frac{\d^4 k_1}{(2\pi)^4}\int\frac{\d^4 k_2}{(2\pi)^4}\, (2\pi)^4 \delta^{(4)}(\hat{q}-p-k_1-k_2)\notag\\
    &\times\left[\widetilde{\Delta}_\phi^>(p)\widetilde{\Delta}^>_\chi (k_1) \widetilde{\Delta}^>_\chi (k_2)-\widetilde{\Delta}_\phi^<(p)\widetilde{\Delta}^<_\chi (k_1) \widetilde{\Delta}^<_\chi (k_2)\right] \notag \\
    = \: & -\frac{g^2}{4} \int\frac{\d^4 p}{(2\pi)^4}\int\frac{\d^4 k_1}{(2\pi)^4}\int\frac{\d^4 k_2}{(2\pi)^4}\, (2\pi)^4 \delta^{(4)}(\hat{q}-p+k_1+k_2)\notag\\
    &\times\left[\widetilde{\Delta}_\phi^>(p)\widetilde{\Delta}^<_\chi (k_1) \widetilde{\Delta}^<_\chi (k_2)-\widetilde{\Delta}_\phi^<(p)\widetilde{\Delta}^>_\chi (k_1) \widetilde{\Delta}^>_\chi (k_2)\right] \,,\label{eqn: imaginary pi r in terms of wightman propagators}
\end{align}
where we recall that $\hat q$ has been defined below Eq.\,\eqref{eq:relation_tildepiR_tildePiR}.
In the second step, we have changed the sign of~$k_1$ and~$k_2$ to draw closer to our intuition of an incoming $\phi$ particle with momentum~$p$ decaying into a pair of $\chi$ particles with momenta~$k_1$ and~$k_2$. This is reflected in the relative signs of the momenta inside the $\delta$ function; in the final expression of Eq.\,\eqref{eq:ImPi-varphi} one has the constraint $p-\hat q=k_1+k_2$ which has the physical interpretation of a $\phi$ particle with momentum $p$ which loses or gains q fraction of this momentum, $\hat q$,  to the wall, before decaying into a pair of $\chi$ particles with momenta $k_1,k_2$.
We have also used~${\widetilde{\Delta}^>(-k) = \:\widetilde{\Delta}^<(k)}$, which follows from taking Fourier transforms of \eqref{eq:symprop} with $a,b=+,-$. Equation~\eqref{eqn: imaginary pi r in terms of wightman propagators} implies the following notation for kinematics,
\begin{subequations}
\label{eq:kinematics}
\begin{align}
    &\phi:\quad\ \  p=(E^{(\phi)}_\vecp,\vecp_\perp,p^z)\,,\\
    &\chi_1:\quad k_1=(E^{(\chi)}_{\veck_1},\veck_{1,\perp},k_1^z)\,,\\
    &\chi_2:\quad k_2=(E^{(\chi)}_{\veck_2},\veck_{2,\perp},k_2^z)\,.
\end{align}
\end{subequations}

We now can use the quasi-particle limit of the Wightman functions in Wigner space given in Eqs.\,\eqref{eq:on-shell-limit}. Note that the Wigner transform coincides with the Fourier transform when there is full spacetime translation symmetry. 
It is convenient to use
\begin{align}
\label{eq:relation1}
    \delta(k^2-m^2_{\phi/\chi})\,\theta(\pm k^0)=\frac{\delta(k^0 \mp E_\veck^{(\phi/\chi)})}{2 E_\veck^{(\phi/\chi)}}\,,
\end{align}
where $E_\veck^{(\phi/\chi)}=\sqrt{\veck^2+m_{\phi/\chi}^2}$,
so that we can write the Wightman functions in a more compact form
\begin{subequations}
\begin{align}
     \Delta^<_{\phi/\chi}(p) = \: 2\pi \frac{1}{2E^{(\phi/\chi)}_{\vec p}} \sum_{\alpha=\pm} \delta\left(p^0-\alpha E^{(\phi/\chi)}_{\vec p}\right) \left( \frac{1-\alpha}{2} + f_{\phi/\chi} (\alpha \vec p) \right) \,,\\
    \Delta^>_{\phi/\chi}(p) = \: 2\pi \frac{1}{2E^{(\phi/\chi)}_{\vec p}} \sum_{\alpha=\pm} \delta\left(p^0-\alpha E^{(\phi/\chi)}_{\vec p}\right) \left( \frac{1+\alpha}{2} + f_{\phi/\chi} (\alpha \vec p) \right) \,.
\end{align}
\end{subequations}
Plugging this into Eq.\,\eqref{eqn: imaginary pi r in terms of wightman propagators}, we find
\begin{align}
\label{eq:Im-Pi-phi-to-2chi}
    {\rm Im}\widetilde{\pi}^{\rm R}_\varphi(q^z) \supset \: &-\frac{g^2}{4} \sum_{\alpha,\beta_1,\beta_2=\pm} \int_{\vec p,\phi} \int_{\vec k_1,\chi} \int_{\vec k_2,\chi} (2\pi)^3 \delta^{(3)}(\hat{\vecq} - \vec p +\vec k_1 + \vec k_2) \, 2\pi \,\delta(\alpha E^{(\phi)}_\vecp -\beta_1 E^{(\chi)}_{\veck_1}-\beta_2 E^{(\chi)}_{\veck_2})\notag\\ &\times\Big\{\left[\frac{1+\alpha}{2}+f_\phi(\alpha \vec p)\right] \left[\frac{1-\beta_1}{2}+f_\chi(\beta_1 \vec k_1)\right] \left[\frac{1-\beta_2}{2}+f_\chi(\beta_2 \vec k_2)\right] \notag\\
&\quad\quad - \left[\frac{1-\alpha}{2}+f_\phi(\alpha \vec p)\right] \left[\frac{1+\beta_1}{2}+f_\chi(\beta_1 \vec k_1)\right] \left[\frac{1+\beta_2}{2}+f_\chi(\beta_2 \vec k_2)\right]\Big\} \,,
\end{align}
where $\hat{\vecq}=(0,0,q^z)$.
The above expression describes all possible processes and the corresponding inverse processes involving one $\phi$ particle and two~$\chi$ particles, corresponding to the second diagram in Eq.\,\eqref{eq:diagram-2PI-2loop}.
In the following, we will avoid specifying the particle species in the integrals, since it is clear that momentum~$\vec p$ refers to the~$\phi$ particle and momenta~$\vec k_1$ and~$\vec k_2$ to the~$\chi$ particles.

To simplify our expression, we notice that the combinations~${\{\alpha,\beta_1,\beta_2\}=\{+,-,-\}}$ and ${\{\alpha,\beta_1,\beta_2\}=\{-,+,+\}}$ are excluded, since for these the energy conserving~$\delta$-function vanishes.
Let us write the remaining combinations, ignoring the integral and the three-momentum conserving~$\delta$-function for the time being
\begin{align}\label{eq:condensate-induced-processes}
    & \delta\left( E^{(\phi)}_{\vec p} - E^{(\chi)}_{\vec k_1} - E^{(\chi)}_{\vec k_2}\right) \Big\{ \underbrace{[1+f_\phi(\vec p)] f_\chi(\vec k_1) f_\chi(\vec k_2)}_{\chi-\text{pair annihilation}} - \underbrace{f_\phi(\vec p) [1+ f_\chi(\vec k_1)] [1+ f_\chi(\vec k_2)]}_{\chi-\text{pair production}} \Big\} \notag \\[1.5ex]
    & + \delta \left( E^{(\phi)}_{\vec p} + E^{(\chi)}_{\vec k_1} - E^{(\chi)}_{\vec k_2}\right) \Big\{ \underbrace{(1+f_\phi(\vec p))(1+f_\chi(-\vec k_1))f_\chi(\vec k_2)}_{\phi-\text{emission}} - \underbrace{f_\phi(\vec p) f_\chi(-\vec k_1) (1+ f_\chi(\vec k_2))}_{\phi-\text{absorption}} \Big\} \notag \\[1.5ex]
    & + \delta \left( E^{(\phi)}_{\vec p} + E^{(\chi)}_{\vec k_2} - E^{(\chi)}_{\vec k_1}\right) \Big\{ \underbrace{(1+f_\phi(\vec p))(1+f_\chi(-\vec k_2))f_\chi(\vec k_1)}_{\phi-\text{emission}} - \underbrace{f_\phi(\vec p) f_\chi(-\vec k_2) (1+ f_\chi(\vec k_1))}_{\phi-\text{absorption}} \Big\} \notag \\[1.5ex]
    & - \left( \vec p, \vec k_1, \vec k_2 \longleftrightarrow - \vec p, - \vec k_1, -\vec k_2 \right)\,.
\end{align}
Note that the expression is anti-symmetric under~${\{ \vec p, \vec k_1 , \vec k_2\}\to\{ -\vec p, -\vec k_1 , -\vec k_2\}}$, making the entire contribution anti-symmetric under~${\hat{\vecq}\to -\hat{\vecq}}$, as expected. In Eq.\,\eqref{eq:condensate-induced-processes} we have labelled the different contributions with their physical interpretation, which can be read off from the $\delta$ function constraint on the energies, as well as the fact that initial states appear weighted with the number distribution functions $f_X$, while final states are weighted with Bose-enhancement factors $1+f_X$. 
Note that here the Bose-enhancement factors are not inserted ad hoc,  but appear automatically in the formalism. Furthermore, it is easy to check that the above physical interpretation matches with the 3 momentum constraint $\delta^{(3)}(\hat{\vecq}-\vec p+\vec k_1+\vec k_2)$ appearing in\,\eqref{eq:Im-Pi-phi-to-2chi}. For example consider the contribution labelled as ``$\chi$-pair production'', which is proportional to $f_\phi({\bf p})(1+f_\chi({\bf k}_1))(1+f_\chi({\bf k}_2))$. Reading off the three-momenta appearing in the distribution functions, this has the natural interpretation of an initial $\phi$ particle with three-momentum ${\bf p}$ emitting two $\chi$ particles with momenta $\bf k_1$ and $\bf k_2$. This is compatible with the $\delta^{(3)}$ constraint enforcing ${\bf p}-\hat{\vecq} =\vec k_1+\vec k_2$, where $-\hat{\vecq}$ can be interpreted as momentum taken away from the wall. The rest of the contributions in Eq.\,\eqref{eq:condensate-induced-processes} can also be interpreted in a similar way.

Observe that if the~$\chi$-particles are very heavy, namely~$m_\chi^2\gg m_\phi^2$, the particle density~$f_\chi(\vec k)$ is strongly suppressed for momenta of the scale of the phase transition temperature~$\beta^{-1}\sim m_\phi$. 
Then, all terms containing~$f_\chi$ are exponentially suppressed, and the only process that plays a role is the decay of~$\phi$ particles, namely the~$\chi$ pair production.

\begin{figure}
    \centering
    \includegraphics[width=0.5\linewidth]{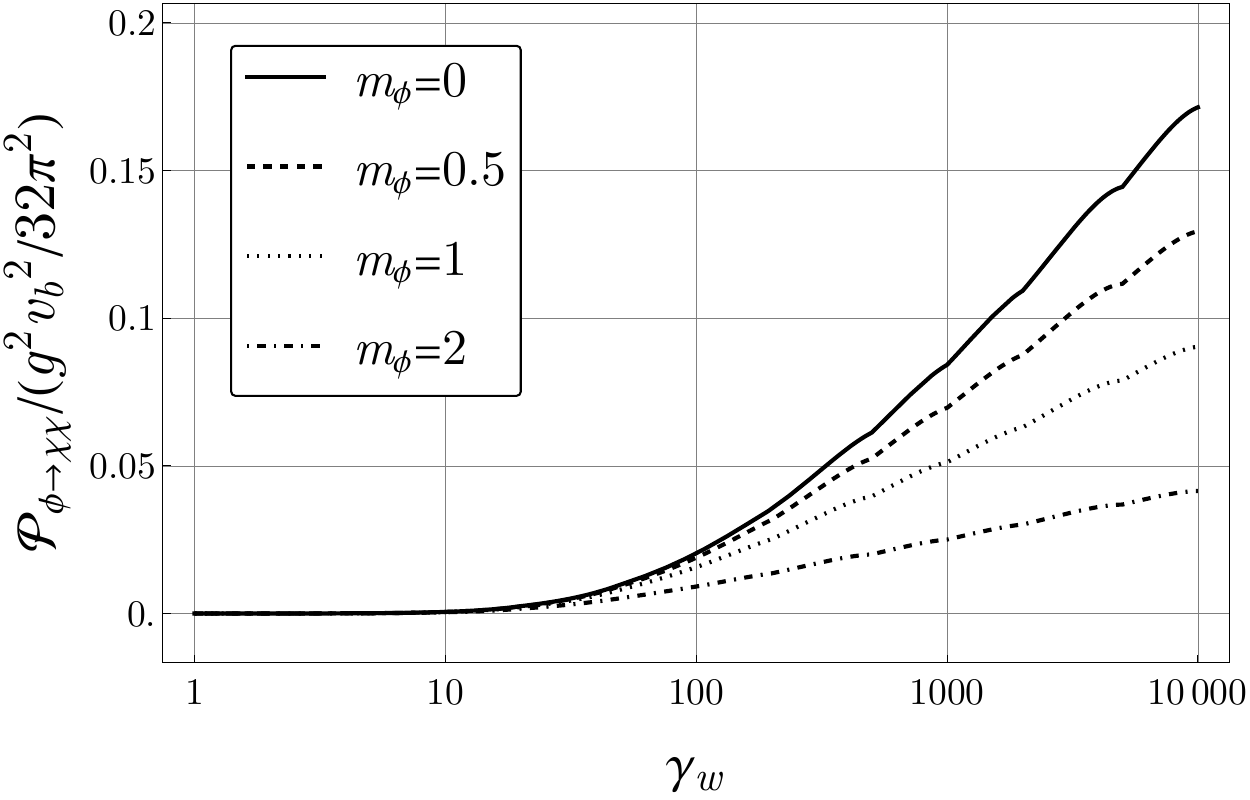}
    \caption{Pressure induced by pair production obtained via the numerical evaluation of Eq.\,\eqref{eqn: p phi to chichi full in text} varying the mass of the incoming particle~$\phi$. Reference values~$T=1$,~$L_w=1$ and~$m_\chi=5$.}
    \label{fig: pressure at different masses}
\end{figure}

Then, under the assumption that $m^2_\chi\gg m_\phi^2\sim T^2$, we can drop all terms except the one describing the pair production of~$\chi$ particles, and our expression simplifies enormously
\begin{align}
    \left[{\rm Im}\widetilde{\pi}^{\rm R}_\varphi(q^z)\right]_{\phi\to\chi\chi} \approx & 
    \frac{g^2}{4} \int_{\vec p ,\vec k_1, \vec k_2} (2\pi)^4 \delta^{(3)}(\hat{\vecq} - \vec p +\vec k_1 + \vec k_2)  \, \delta( E^{(\phi)}_\vecp - E^{(\chi)}_{\veck_1}- E^{(\chi)}_{\veck_2}) \Big[ f_\phi(\vec p) - f_\phi (-\vec p) \Big] \,. \label{eqn: im pi phi to chichi}
\end{align}
Substituting Eqs.\,\eqref{eqn: im pi phi to chichi} into Eq.\,\eqref{eq:Pvertex-simple}, and using the~$\delta$ function over the momentum along the~$z$ direction to localise the~$q$ integral, we get
\begin{align}
\label{eq:P-phi-to-2chi}
    \P_{\rm vertex}\supset \P_{\phi\rightarrow\chi\chi} \equiv \: &  \frac{g^2}{2} \int_{\vec p ,\vec k_1, \vec k_2} (2\pi)^3 \delta^{(2)}(\vecp_\perp-\vec{k}_{1,\perp}-\vec{k}_{2,\perp}) \delta(E^{(\phi)}_\vecp - E^{(\chi)}_{\veck_1}- E^{(\chi)}_{\veck_2}) \, f_\phi(\vecp) \, \Delta p^z \, |\widetilde{\varphi}(\Delta p^z)|^2\,,
\end{align}
where
\begin{align}\label{eq:Deltapz}
 \Delta p^z=  -(p^z-k_1^z-k_2^z)\,.
\end{align}
Equation~\eqref{eq:P-phi-to-2chi} agrees precisely with the expression derived in Refs.\,\cite{Azatov:2020ufh,Ai:2023suz}. Note that above we have defined $\Delta p^z$ with an additional factor of $-1$. 
The sign is different because we define the wall to expand in the positive $z$ direction, while in Refs.\,\cite{Azatov:2020ufh, Ai:2023suz}, it is defined to expand in the negative $z$ direction. The quantity~$\Delta p^z$ in~\eqref{eq:Deltapz} corresponds to $-q^z$ inside ${\rm Im} \tilde\pi^{\rm R}_\varphi$ (see Eq.\,\eqref{eqn: im pi phi to chichi}) which in the case of the $\chi$ pair production we identified with the momentum taken away from the wall. A momentum loss, as required to compensate the driving force and reach a stationary state, corresponds to $-q^z=\Delta p^z>0$ giving a positive $\P_{\rm vertex}$, fitting our conventions in which the balance of forces is written as $\Delta V= \P_{\rm mass}+\P_{\rm vertex}>0$. 
In the expression\,\eqref{eq:P-phi-to-2chi} for the force, the quantity  $g^2 |\tilde\varphi(\Delta p^z)|^2$ can be interpreted as squared transition amplitude $|\M_{\phi\rightarrow \chi\chi}|^2$ for an incoming $\phi$ particle to extract momentum $\Delta p^z$ from the wall and split into two particles. To give the total force, this probability has to be weighted with the density of incoming particles, multiplied by the momentum lost by the wall and integrated over the phase space of the final $\chi$ particles, subject to the appropriate energy and momentum conservation constraints.
In Appendix\,\ref{app:simplify} we further simplify the expression for the pressure due to pair production by using the remaining~$\delta$ functions. We arrive at the following general formula
\begin{equation}
    \label{eqn: p phi to chichi full in text}
    \P_{\phi\rightarrow\chi\chi} = \: \frac{g^2}{32\pi^2} \int_{\vec p} f_\phi(\vecp) \times \left[ \vartheta(-p^z-\Delta) - \vartheta(p^z-\Delta) \right]\int_{q_-}^{q_+} \d q^z \,q^z |\widetilde{\varphi}(q^z)|^2 \sqrt{1- \frac{4 m_\chi^2}{m_\phi^2 + 2q^z|p^z| - (q^z)^2}}  \,,
\end{equation}
where the bounds of the~$q$ integral are~$q_\pm = \: |p^z| \pm \sqrt{(p^z)^2 - \Delta^2}$, and we have defined~${\Delta^2=4m_\chi^2-m_\phi^2}$.
Equation~\eqref{eqn: p phi to chichi full in text} lends itself to numerical evaluation. 
In Fig.\,\ref{fig: pressure at different masses} we show the contribution to the pressure due to the pair production of heavy particles as a function of the relativistic boost factor~$\gamma_w$, for different values of the mass of the incoming particle~$m_\phi$. We have taken here a tanh wall profile, as defined in Eq.\,\eqref{eq:wall-profile}. The temperature~$T$ has been chosen as the reference scale, and we show the result for reference values of~$L_w$ and~$m_\chi$.
First, we observe that for all values of~$m_\phi$, the pressure asymptotically grows logarithmically with~$\gamma_w$, and different slopes imply different pre-factors of the logarithm.
The logarithmic growth was observed numerically in Ref.\,\cite{Ai:2023suz} for~$m_\phi=0$, and we demonstrate it analytically at the end of this section, by studying the ultrarelativistic limit.
As the value of~$m_\phi$ grows, the pressure decreases because of the suppression due to the statistical distribution.

Accordingly, the self-energy in Eq.\,\eqref{Eq:Pi} and therefore the friction from Eq.\,\eqref{eq:P_vertex} also receives contributions from processes involving three~$\phi$-particles.
An analogous derivation to the one just done leads to
\begin{align}
\label{eq:P-phi-phi-phi}
    \P_{\rm vertex}\supset \P_{\phi\phi\phi} \equiv \: & \frac{\lambda_\phi^2}{2} \int_{\vec p, \vec k_1, \vec k_2} (2\pi)^3 \delta^{(2)}(\vecp_\perp-\vec{k}_{1,\perp}-\vec{k}_{2,\perp}) \delta(E^{(\phi)}_\vecp - E^{(\phi)}_{\veck_1}- E^{(\phi)}_{\veck_2}) \, \Delta p^z\, |\widetilde{\varphi}(\Delta p^z)|^2 \notag \\
    & \times \Big[ f_\phi(\vec p)(1+f_\phi(\vec k_1))(1+f_\phi(\vec k_2)) - (1+f_\phi(\vec p))f_\phi(\vec k_1)f_\phi(\vec k_2)\Big]\,,
\end{align}
where no term has been dropped, and no approximation has been made. The above expression for the force captures the momentum lost by the wall from the splittings of incoming~$\phi$ particles or the annihilation of incoming pairs of particles. 
In Eq.\,\eqref{eq:P-phi-phi-phi}, the three-momentum integrals all refer to the particle species~$\phi$.

\paragraph{The ultrarelativistic regime.}
As~$\gamma_w$ grows larger, we can obtain an analytical asymptotic expression for Eq.\,\eqref{eqn: p phi to chichi full in text}. The computation is presented step by step in Appendix\,\ref{app:simplify}, leading to the final result
\begin{align}
    \P^{\gamma_w \to \infty}_{\phi\rightarrow\chi\chi} 
   & \approx \frac{g^2 v_b^2 T^2}{24 \times 32\pi^2} \log\left( \frac{ \gamma_w T}{2\pi L_w m_\chi^2} \right)\vartheta \bigg(\frac{\gamma_w T}{2\pi L_w m_\chi^2}-1\bigg) 
   \notag
    \\
   & \approx \frac{g^2 v_b^2\mathcal{P}^{\gamma_w\rightarrow\infty  }_{\phi\to \phi}}{ 32 \pi^2 \Delta m_\phi^2  } \log\left( \frac{\gamma_w T}{2\pi L_w m_\chi^2} \right)\vartheta \bigg(\frac{\gamma_w T}{2\pi L_w m_\chi^2 }-1\bigg)\,,
   \label{eqn: pair pressure ultrarel}
\end{align}
where in the last line, we rephrased the pressure from pair production in terms of the pressure from the mass gain of \,$\phi$, namely\,${\mathcal{P}^{\gamma_w\rightarrow\infty}_{\phi\to \phi}=\Delta m_\phi^2 T^2/24}$. 
The~$\vartheta$-function has been introduced to make it explicit that this expression is only asymptotically valid and breaks down when~${\pi L m_\chi^2/(\gamma_w T)\to 1}$. 
Equation\,\eqref{eqn: pair pressure ultrarel} demonstrates the logarithmic growth with~$\gamma_wT$ observed already in Fig.\,\ref{fig: pressure at different masses}. {We emphasise again that the friction we are discussing here is not due to mass change but due to the direct momentum exchange between the particles and the wall. That is why in the first line above, the friction depends on neither $\Delta m_\phi^2$ nor $\Delta m^2_\chi$.}

\paragraph{Phenomenology of friction from pair production}
During a dark sector FOPT, the pressure~$\P_{\phi\rightarrow \chi\chi}$ can dominate over~$\mathcal{P}_{\phi\to \phi}$ for~${g^2v_b^2\gtrsim 32\pi^2 \Delta m^2_\phi}$. This condition can be satisfied if the change of the mass of~$\phi$ particles is loop-suppressed. For example, the quartic coupling of the~$\Phi$ field could be dominated by the one-loop contribution of the heavy~$\chi$ field,~$(1/4!)\lambda_{\phi,\rm 1-loop}\sim (1/4) g^2/16\pi^2 \gg \lambda_\phi/4!$. 
This would lead to~$\Delta m_\phi^2\sim 3 g^2 v_b^2/16\pi^2$.
In this case,~$\P_{\phi\rightarrow \chi\chi}$ dominates over~$\P_{\phi\rightarrow \phi}$ for large~$\gamma_w$.
This simple case is a natural realisation where the pressure from heavy particles becomes relevant.

During a first-order electroweak phase transition, one identifies the $\phi$ particle with the Higgs $h$. There are two unavoidable particle production processes: 1) the pair production of Higgs from one incoming Higgs particle, $ h\to h h$, $hh \to h$ and 2) the pair production of gauge bosons from an incoming Higgs $h \to W^+ W^-, ZZ$ and their permutations, e.g., $Z\rightarrow h Z$. The main difference from the computation above is that both the Higgs and the gauge bosons are light particles. Consequently, all these particles are not Boltzmann suppressed in the plasma. A more precise evaluation would require numerical methods, starting with the expression in Eq.\,\eqref{eq:P-phi-phi-phi}.

\section{Mixing and transition radiation frictions from the VIA}
\label{sec:small_field}

In the previous sections, we have studied the friction induced by the elementary type (1) $1\rightarrow 1$ process and type (4) $1\rightarrow 2$ particle production. In this section, we study the contribution from the type (2) $1\rightarrow 1$ mixing and type (3) $1\rightarrow 2$ transition radiation. As we mentioned earlier, these contributions are in principle captured by \,$\P_{\rm mass}$, once we insert the full propagators~$\Delta$ found as solutions of the EoMs for the two-point functions or, as a suitable approximation, of the Boltzmann equations derived in the quasi-particle limit and at the leading order of the gradient expansion.
In practice, this is only feasible numerically.
To reveal how mixing and transition radiation are captured by \,$\P_{\rm mass}$ in the formalism we employ, we perform an expansion of the two-point functions in the background field \,$\varphi$, and show how these contributions are encoded inside higher-loop self-energy terms.
This expansion amounts to treating the\,$\varphi$-dependent mass terms as perturbation vertices, an approach known as VEV insertion approximation or VIA. Of course, keeping only a few terms in the expansion does not give us the full result that one would obtain by solving the Boltzmann equations, but it is sufficient for our purposes.

\subsection{Pressure from scalar mixing: process (2)}
\label{sec:pres_scalar_mix}

The model considered so far assumes a diagonal mass matrix. In the presence of mass mixing, it was shown in \cite{Azatov:2020ufh} that the transition from a light state to a heavy state can also induce friction. In this section, we show how this friction from the $1\rightarrow 1$ mixing process is captured from our general framework. In another context, mass mixing was shown to be a new source of baryogenesis for electroweak baryogenesis at slow walls\,\cite{Liu:2011jh,Cline:2021dkf} and fast walls\,\cite{Azatov:2021irb}.

As an illustrative example, we consider the theory of two real scalar fields interacting via the bubble background~$\varphi$. The potential is then written as
\begin{align} 
-\mathcal{L}_{\rm mix} = \kappa\, \varphi \chi s + \frac{1}{2} m_s^2 s^2+\frac{1}{2}m_\chi^2\chi^2 \, ,
\end{align} 
where $\kappa$ is a coupling of mass dimension one, and the kinetic terms are implied. The extension of the 2PI as written in Eq.\,\eqref{eq:2PI-general-form} is straightforward. We assume that $\chi$ is heavy ($m_\chi\gg T$) so that $\chi$ particles are Boltzmann suppressed outside of the bubble. The  $s$ particles are light ($m_s < T$) so they are abundant in the plasma. 
During the expansion of the bubble, a light $s$ particle passing through the wall can transit to a heavy $\chi$ particle due to the mixing term\,$\kappa\varphi \chi s$, inducing a contribution to the friction on the wall. Below, we derive this contribution.

\subsubsection{Light-heavy case.}
 We shall show that the friction from mixing can be understood from the following term in the effective action, 
\begin{align}
\label{eq:Dvarphi}
    D[\varphi]\equiv \frac{\i}{2} {\rm Tr} \left[G^{-1}(\varphi)\Delta\right]=\frac{\i}{2} \sum_{ij}\sum_{a, \, b} \int\d^4 x\, \left.\left[G_{ij}^{ab,-1}(\varphi(x)) \Delta_{ji}^{ba}(x,y)\right]\right|_{y \, = \, x}
\end{align}
where $i, j = \chi, s$ are flavour indices, and we recall that $a,b =\pm$ are Keldysh polarisation indices.
In the flavour basis, $\Delta^{ab}$ with fixed $a, b$ is a matrix of the form 
\begin{align}
    \Delta_{\rm flavor}^{ab} = \begin{pmatrix}
        \Delta^{ab}_{\chi \chi} &\Delta^{ab}_{ \chi s}
        \\ 
        \Delta^{ab}_{s \chi} &\Delta^{ab}_{ss} 
    \end{pmatrix}
\end{align} 
and the quadratic fluctuation operator reads
\begin{align}
\label{eq:G-1In}
    G_{\rm flavor}^{ab,-1} = \i c^{ab} \begin{pmatrix}
       \Box + m_\chi^2 &\kappa \varphi^a
       \\ \kappa  \varphi^a & \Box + m_s^2
    \end{pmatrix}\,,
\end{align}
with~$c = \mathrm{diag}(+1,-1)$.

We are interested in the terms explicitly depending on $\varphi$ which contribute to the condensate EoM. These are
\begin{align}
    D[\varphi] \supset  &-\frac{1}{2}\int\d^4 x\,\kappa\varphi^+(x) \left(\Delta^{++}_{\chi s}(x,x)+\Delta^{++}_{s\chi}(x,x)\right) \notag\\
    &+\frac{1}{2}\int\d^4x\, \kappa\varphi^-(x) \left(\Delta_{\chi s}^{--}(x,x)+\Delta_{s\chi}^{--}(x,x)\right) \, ,
\end{align}
In the condensate EoM \eqref{eq:eom-condensate-CTP}, the above contributes to a term
\begin{align}
    -\left.\frac{\delta D[\varphi]}{\delta\varphi^+(x)}\right|_{\varphi^+=\varphi^-=\varphi} \supset\: \frac{1}{2}\kappa \left.\left(\Delta^{++}_{\chi s}(x,x;\varphi)+\Delta^{++}_{s\chi}(x,x;\varphi)\right)\right|_{\varphi^+=\varphi^-=\varphi}\,.
\end{align}
Note that we have indicated here an implicit dependence of the two-point functions $\Delta$ on the background field, i.e. the one-point function, $\varphi$. The effective action is a functional of the two-point and one-point functions, which are independent. However, when extremizing the effective action, we obtain the EoMs that are coupled integro-differential equations for the one-point and the two-point functions. Here we use an approximation where the dependence of the one-point function on the two-point functions is neglected---above, on the RHS we have assumed that the two-point functions have been obtained from their EoMs and thus obtain an implicit dependence on the one-point functions $\varphi^a$.

Now one can expand $\Delta^{++}_{\chi s}(x_1,x_2;\varphi)$ and $\Delta^{++}_{s\chi}(x_1,x_2;\varphi)$ in $\varphi^a$, and we have 
\begin{align}
\label{eq:expansion_exact}
\Delta^{++}_{\chi s}(x_1,x_2;\varphi) = 
- \i\sum_a\int \d^4 x'\, a\,\kappa\,\varphi^a(x') \Delta^{+a}_{\chi\chi}(x_1, x';0) \Delta^{a+}_{ss}(x',x_2;0) + \mathcal{O}\bigg(\frac{\kappa\varphi }{m_
\chi^2}\bigg)^2\,,
\end{align}
where $\Delta_{\chi\chi}(x_1,x_2;0)$ and $\Delta_{ss}(x_1,x_2;0)$ are the two-point functions obtained by solving the Schwinger--Dyson equations with $\varphi =0$. We confirm explicitly in Appendix \ref{app:expansion_Boltz} that the expansion in Eq.\,\eqref{eq:expansion_exact} is the solution at leading order of the exact kinetic equation for the two-point functions. A similar expansion applies for $\Delta^{++}_{s\chi}(x_1,x_2;\varphi)$.

After this expansion, we obtain a self-energy-like term,
\begin{align}
\label{eq:Dvarphi-expansion}
   - \left.\frac{\delta D[\varphi]}{\delta\varphi^+(x)}\right|_{\varphi^+=\varphi^-=\varphi} \supset & - \i \,\kappa \int\d^4 x'\, \kappa\varphi(x') \left[\Delta_{\chi\chi}^{++}(x,x';0)\Delta_{ss}^{++}(x',x;0)-\Delta_{\chi\chi}^{+-}(x,x';0)\Delta_{ss}^{-+}(x',x;0)\right]\notag\\
   \equiv & \int\d^4 x'\, \Pi^{\rm R}_{\varphi;\rm mix}(x,x')\varphi(x')
    \,.
\end{align}
where in the last line, we have defined the {\it induced} self-energy
\begin{align}
    \Pi^{\rm R}_{\varphi;\rm mix}(x,x')=- \i  \kappa^2 \left[\Delta^{++}_{\chi\chi}(x,x';0)\Delta^{++}_{ss}(x,x';0)- \Delta_{\chi\chi}^{+-}(x,x';0)\Delta_{ss}^{+-}(x,x';0) \right]\,.
\end{align}
The above induced self-energy can be seen to arise from contributions to the effective action of the following form,
\begin{align}
\begin{tikzpicture}[baseline={-0.025cm*height("$=$")}]
\draw[draw=black,thick] (-0.5,0) circle (0.15) ;
\draw[draw=black,thick] (-0.5-0.11,0+0.11) -- (-0.5+0.11,0-0.11);
\draw[draw=black,thick] (-0.5-0.11,0-0.11) -- (-0.5+0.11,0+0.11);
\draw[draw=black,thick] (-0.35,0) -- (0.25,0);
\filldraw (0.25,0) circle (1.5pt) node {} ;
\filldraw (1.25,0) circle (1.5pt) node {} ;
\draw[draw=black,dashed] (0.25,0) arc (180: 0: 0.5);
\draw[draw=black] (0.25,0) arc (-180: 0: 0.5);
\draw[draw=black,thick] (1.25,0) -- (1.85,0);
\draw[draw=black,thick] (2,0) circle (0.15) ;
\draw[draw=black,thick] (2-0.11,0+0.11) -- (2+0.11,0-0.11);
\draw[draw=black,thick] (2-0.11,0-0.11) -- (2+0.11,0+0.11);
\end{tikzpicture}
\end{align}
where a dashed/solid line represents $\Delta_{\chi\chi}(0)/\Delta_{ss}(0)$. The dots represent the ``vertex'' given by the mixing term $\kappa\varphi \chi s$ and can be of either $+$ or $-$ type. Note that this diagram is two-particle-{\it reducible}. However, it does not come from $\Gamma_2$ in the 2PI effective action but only emerges from the one-loop term~$D[\varphi]$ after the field expansion.

Similar to $\P_{\rm vertex}$ in Eq.\,\eqref{eq:P_vertex}, the new self-energy term in Eq.\,\eqref{eq:Dvarphi-expansion} induces a pressure 
\begin{align}
\label{eq:pres_mix}
    \P_{\rm mix}=-\int_{-\delta}^{\delta}\d z\, \frac{\d \varphi}{\d z}\int \d z'  \pi^{\rm R}_{\varphi;\rm mix}(z,z')\varphi(z')\,,
\end{align}
where $\pi_{\varphi;\rm mix}^{\rm R}$ is obtained from Eq.\,\eqref{eq:piR} with $\Pi^{\rm R}_\varphi$ replaced by $\Pi^{\rm R}_{\varphi;\rm  mix}$.

Ignoring the plasma inhomogeneities in the $z$-direction or assuming large $\gamma_w$, $\Delta_{\chi\chi}(x,x';0)$ and $\Delta_{ss}(x,x';0)$ have spacetime-translation invariance, noting that we do not have additional $\varphi$-dependent mass terms since we are working within the VIA. Then, in analogy with the discussion in Section\,\ref{subsec: friction from pair production} and Eq.\eqref{eq:Pvertex-simple}, Eq.\,\eqref{eq:pres_mix} can be written as 
\begin{align}
    \P_{\rm mix}=-\int\frac{\d q^z}{2\pi}\, q^z\, |\widetilde{\varphi}(q^z)|^2\, {\rm Im}\widetilde{\pi}^{\rm R}_{\varphi;\rm mix}(q^z)\,.
\end{align}
Following the derivation of Eq.\,\eqref{eq:P-phi-to-2chi}, we can obtain
\begin{align}
    \P_{\rm mix} = & - \kappa^2 \int \frac{\d q^z}{2\pi} q^z |\widetilde{\varphi}(q^z)|^2 \int_{\vec p, \vec k} (2\pi)^3 \delta^{(3)}(\vec q - \vec p + \vec k) \, 2\pi\delta(E^{(s)}_\vecp - E^{(\chi)}_{\veck}) \left[ f_s(\vec p) - f_s (- \vec p)\right] \notag \\
    = \: & - \kappa^2 \int\frac{\d q^z}{2\pi} q^z |\widetilde{\varphi}(q^z)|^2 \int_{\vec p} \frac{1}{2E_{\vec p}^{(s)}} 2\pi \delta(E_{\vec p}^{(s)} - E_{\vec p -\vec q}^{(\chi)} )\left[ f_s(\vec p) - f_s (- \vec p)\right] \,,
\end{align}
where we have assumed that the~$\chi$-particles are very heavy and thus absent in the plasma.
We can derive an analytical estimate of the pressure due to mixing in the ultrarelativistic limit following the same steps as explained in Appendix\,\ref{app:simplify} for the pressure due to pair production. 
First, we use the energy conserving~$\delta$ function to fully localise the~$q^z$ integral, and in doing that imposes a constraint on~$p^z$
\begin{align}
     \P_{\rm mix} = \: & - \frac{\kappa^2}{2} \sum_{\alpha=\pm} \int \d p^z \vartheta((p^z)^2-m_\chi^2+m_s^2)\frac{q_\alpha^z}{q^z_\alpha-p^z} |\widetilde{\varphi}(q_\alpha^z)|^2 F(p^z) \notag \\
     = \: & - \kappa^2 \int_\Delta^\infty \d p^z G_{\rm mix}(p^z) F(p^z)\,.
     \label{eqn: pressure mixing full}
\end{align}
In the second step, we have used symmetry of the argument under~$p^z\to-p^z$ to restrict the integration domain. We have defined~$\Delta^2 = m_\chi^2-m_s^2$ and the momentum transfer
\begin{equation}
    q^z_\pm = \: p^z \pm \sqrt{(p^z)^2 - m_\chi^2 + m_s^2} \,.
\end{equation}
Function~$F$ follows the definition of Eq.\,\eqref{eqn: definition F}, and function~$G_{\rm mix}$ is defined as follows
\begin{equation}
    \label{eqn: definition G mix}
    G_{\rm mix}(p^z) = \: \sum_{\alpha=\pm} \frac{q_\alpha^z}{q^z_\alpha-p^z} |\widetilde{\varphi}(q_\alpha^z)|^2 \,.
\end{equation}
Equation\,\eqref{eqn: pressure mixing full} is the full pressure due to mixing, valid also away from the ultrarelativistic limit.
Using equilibrium distributions in the wall frame, the function~$F$ can be computed analytically as shown in Eq.\,\eqref{eqn: analytical F}, upon making the exchange~${m_\phi\to m_s}$.
In the ultrarelativistic limit, the mixing pressure can be evaluated in terms of a cumulant expansion, as explained in Appendix~\ref{app:simplify} for the pressure due to pair production in Eq.\,\eqref{eqn: pressure pair cumulant expansion}.
Following the same procedure for the pressure due to mixing we find
\begin{equation}
    \P_{\rm mix}^{\gamma_w\to\infty} = \: - \kappa^2 \mathcal{N}_F G_{\rm mix}(\langle p^z\rangle)\,.
\end{equation}
The normalisation~$\mathcal{N}_F$ and the average~$z$ momentum~$\langle p^z\rangle$ are computed in Appendix~\ref{app:simplify}
\begin{equation}
    \mathcal{N}_F = \: \frac{T^2}{24} \left( 1 + \mathcal{O}(\gamma_w^{-1}) \right) \,, \qquad \langle p^z \rangle = \: \sigma \gamma_w T \left( 1 + \mathcal{O}(\gamma_w^{-1}) \right) \,,
\end{equation}
where the proportionality constant is~$\sigma = 12\zeta_3/\pi^2$.
Since the wall Fourier transform~$\widetilde{\varphi}$ is strongly suppressed for large arguments, when computing~${G_{\rm mix}(\sigma \gamma_wT)}$ only the one with~$\alpha=-$ in Eq.\,\eqref{eqn: definition G mix} survives.
Using a tanh wall and expanding at leading order in~$\gamma_w$ we find
\begin{equation}
  L_w m_\chi^2/T \lesssim  \gamma_w  \qquad \Rightarrow \qquad \P_{\rm mix}^{\gamma_w\to\infty} = \: \frac{2 \kappa^2 v_b^2}{m_\chi^2} \frac{T^2}{24} \left( 1 + \mathcal{O}(\gamma_w^{-1}) \right)\,,
    \label{eqn: ultrarel mixing pressure}
\end{equation}
where we have also taken the limit~$m_\chi\gg m_s$.
Note that this result applies to general wall profiles. In fact, the only properties of the Fourier transformed wall profile~$\widetilde{\varphi}(q)$ we have used are that it is exponentially suppressed for large~$q$ and that it falls as~$1/q$ near the origin. The first is satisfied if the field approaches the minima~$\varphi=0$ and~$\varphi=v_b$ exponentially fast away from the wall. This is always valid for problems with a characteristic scale, here~$T$. 
The second property is equivalent to asking that for large length scales the field profile looks like a step function, which is generally true for a background field interpolating between the two minima.

Equation~\eqref{eqn: ultrarel mixing pressure} matches the result obtained in Eq.\,(81) of Ref.\,\cite{Azatov:2020ufh} up to a factor of two, where our~$\kappa$ equals to~$2B$ of Ref.\,\cite{Azatov:2020ufh}. The same analysis can be applied to the case of fermion mixing, presented in Appendix\,\ref{app:pres_fermion_mix}. In Fig.\,\ref{fig: mixing pressure} we show the mixing pressure computed via numerical evaluation of Eq.\,\eqref{eqn: pressure mixing full} compared to the ultrarelativistic asymptotic expression of Eq.\,\eqref{eqn: ultrarel mixing pressure}. 
We find good agreement between the two when~$\gamma_w$ is very large. As the wall becomes less relativistic, the pressure is suppressed and the asymptotic approximation breaks down for values~$L_wm_\chi^2/T \gtrsim  \gamma_w$.

\begin{figure}
    \centering
    \includegraphics[width=0.5\linewidth]{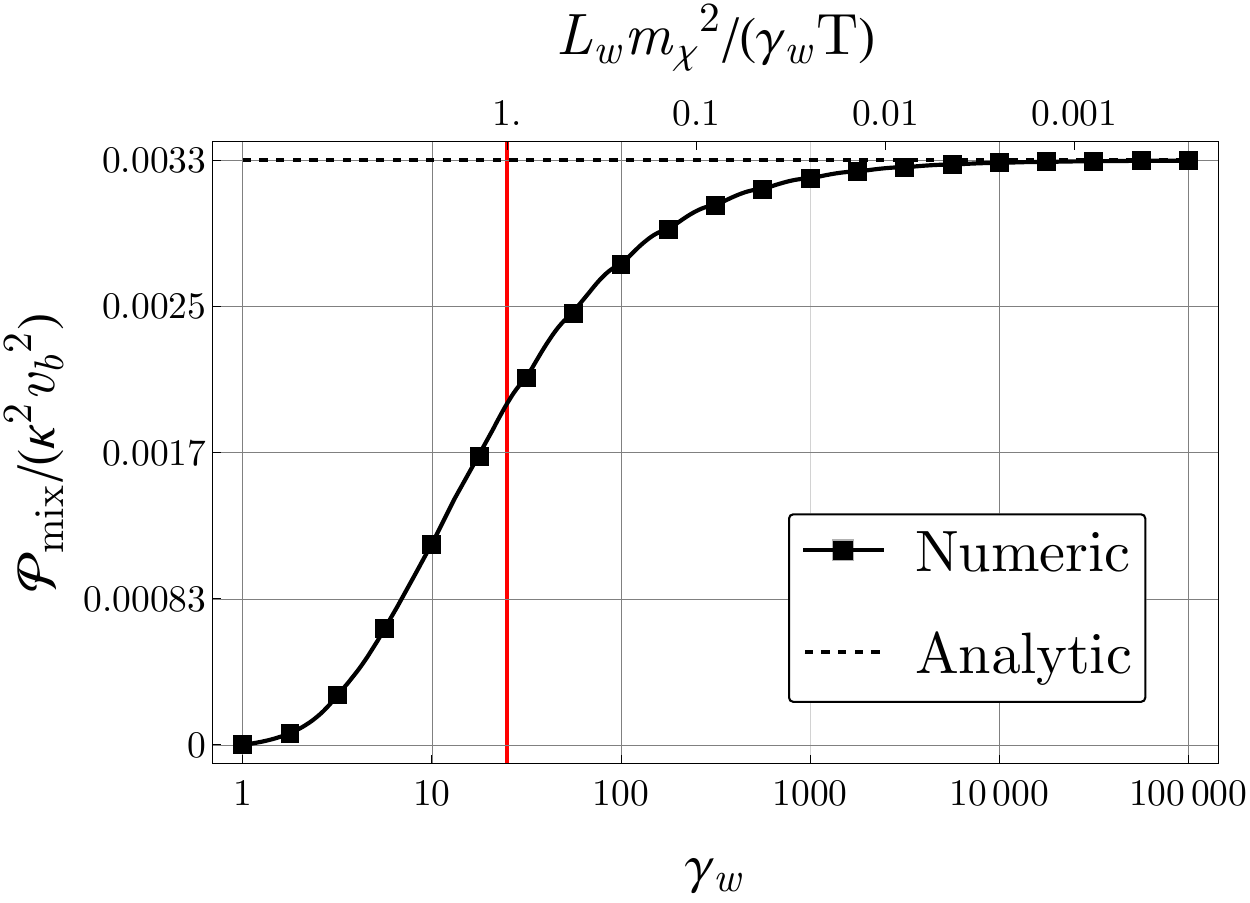}
    \caption{Pressure due to mixing as defined in Eq.\,\eqref{eqn: pressure mixing full}, compared to the asymptotic result valid for ultrarelativistic walls obtained in Eq.\,\eqref{eqn: ultrarel mixing pressure}. The reference values are~$T=1$,~$L_w=1$ and~$m_\chi=5$. Black squares represent the values used for the interpolation.}
    \label{fig: mixing pressure}
\end{figure}

If there is a $\varphi$-dependent mass term for $s$, which gives a gain in its squared mass $\Delta m^2_s$ upon entering the bubble, then there is also the B\"{o}deker-Moore pressure\,\cite{Bodeker:2009qy} (cf.~Eq.\,\eqref{eq:pres_BM}) for the elementary $1\rightarrow 1$ process of $s$-particles, $\P_{\rm BM,s}\approx \Delta m_s^2 T^2/24$. However, the contribution to the pressure from the mixing can dominant over $\P_{\rm BM,s}$ if $\Delta m^2_s/v^2_b \ll \kappa^2/m^2_\chi$.

\subsubsection{Light-light case.}

We now turn to the case when both scalar fields have a small mass ${m_s  \lsim m_\chi \lesssim T}$, which can thus be neglected in the computation. In this case, we directly work in the mass eigenspace of states. 
The first step is to diagonalise the inverse propagator in Eq.\,\eqref{eq:G-1In} and then go to the mass eigenbasis with a unitary transformation $U$: $\chi, s \to \chi', s'$. Doing so, we obtain the two eigenstates 
\begin{align}
   M^2_{1,2}= \frac{m_s^2 + m_\chi^2}{2} \pm \frac{(m_\chi^2- m_s^2)}{2}\sqrt{1 +4 (\kappa\varphi)^2/(m_\chi^2- m_s^2)^2}
\end{align}
and the inverse propagator can be written in the form 
\begin{align}
     G^{-1}_{\rm mass} =  \i c^{ab}\begin{pmatrix}
       \Box + M_1^2 & 0
       \\ 0 & \Box + M_2^2
    \end{pmatrix} \, . 
\end{align}
The one-loop term in the EoM for the condensate can then be written in the mass eigenbasis as
\begin{align}
    D[\varphi]\equiv \frac{\i}{2}{\rm Tr} \left[G_{{\rm mass}}^{-1}(\varphi)\Delta_{\rm mass}\right]=\frac{\i}{2} \sum\limits_{i=1}^2 \sum_{a, \, b}\int\d^4 x\, \left.\left[G^{ab,-1}_i (\varphi(x)) \Delta_i^{ba}(x,y)\right]\right|_{y \, = \, x}\,,
\end{align}
where now $i=1,2$ corresponds to the two mass eigenstates.  The $\varphi$-dependent part is
\begin{align}
    D[\varphi]\supset & -\frac{1}{2}\frac{(m_\chi^2- m_s^2)}{2}\sqrt{1 +4 (\kappa\varphi^+)^2/(m_\chi^2- m_s^2)^2} (\Delta^{++}_1 - \Delta^{++}_2)\notag\\
    &+ \frac{1}{2}\frac{(m_\chi^2- m_s^2)}{2}\sqrt{1 +4 (\kappa\varphi^-)^2/(m_\chi^2- m_s^2)^2} (\Delta^{--}_1 - \Delta^{--}_2)  \, .  
\end{align}
The two terms in the sum contribute to the EoM of the condensate $\varphi^\pm$, reflecting that we can consistently have $\varphi^+=\varphi^-$. If the two particles have $m_\chi \approx m_s \lesssim T$, then they have roughly identical thermal distributions, leading to~${\Delta^{++}_1 - \Delta^{++}_2 \to 0}$, and consequently the pressure from such mixing is suppressed. Therefore, we see that to have a sizeable mixing pressure, one of the states should be Boltzmann suppressed. 
As a cross-check, it can be seen that when carrying out the computation with the VIA approach but dropping the assumption of a heavy~$\chi$ particle, the resulting pressure goes to zero in the limit $m_\chi\rightarrow m_s$, because the kinematics enforces a zero value of the momentum loss of the wall,~$q^z$.

\subsection{Pressure from transition radiation:  process (3)}

To extract the contribution to the pressure due to the transition radiation, we now study the dynamics of a fermion~$\psi$ and a gauge boson~$A_\mu$, the latter being coupled to the condensate. The relevant part in the Lagrangian is 
\begin{align} 
\mathcal{L} \supset -g_1\bar \psi \gamma_\mu A^\mu \psi + \frac{g_2^2 \Phi^2}{4} A_\mu A^\mu \, . 
\end{align}
We take the~$\psi$-particles to be light ($m_\psi\sim T$) so that they are abundant in the plasma during the transition, and also assume that they couple weakly to the condensate so that they do not change mass appreciably during the phase transition. A particle having these properties is, for example, the electron during the electroweak phase transition. 

\subsubsection{Transition radiation and its relation to the CTP formalism}

When the fermion passes through the wall, it can emit a gauge boson due to the $\varphi$-dependent mass of the latter,\footnote{Transition radiation can happen if either the fermion or the gauge boson has a $\varphi$-dependent mass. We focus on this second scenario, as it is known that only this case brings a $\gamma_w$ increasing pressure\,\cite{Bodeker:2017cim, Gouttenoire:2021kjv}.} giving $\psi\rightarrow \psi A$, in a way analogous to a charged particle crossing a boundary separating two media with different refractive index and thus emitting light. 
As the bubble expands, particles enter it and gain momentum, which then results in transition radiation, thus inducing additional friction on the wall.

Perhaps unexpectedly, the contribution of the transition radiation to the friction becomes particularly relevant when the wall becomes highly relativistic and was found, within the ballistic approximation, to scale like $\gamma_w$\,\cite{Bodeker:2017cim}.  From a phenomenological point of view, this contribution is crucial as it almost excludes the possibility of runaway solutions for gauged phase transitions. This has very important consequences on the shape of the gravitational wave spectrum, the yield of baryogenesis and the production of dark matter during bubble expansion. 

Similarly to the mixing pressure, the pressure from the transition radiation process is, in principle, captured by\,$\P_{\rm mass}$. 
However, to see this without solving the Boltzmann equations, we again perform an expansion of the two-point functions of the gauge boson in insertions of the background~$\varphi$. The diagram responsible for the emission of the gauge bosons is 
\begin{align}
\label{eq:DA++;Feynman_bis}
&\Delta^{\mu\nu, ++}_A(x, x;\varphi)  \supset  
\begin{tikzpicture} [baseline={-0.025cm*height("$=$")}]
\draw[decorate,decoration={snake,amplitude=.5mm,segment length=1.5mm}] (-0.5,0) arc (180:120:0.5);
\draw[decorate,decoration={snake,amplitude=.5mm,segment length=1.5mm}] (0.5,0) arc (0:60:0.5);
\draw[decorate,decoration={snake,amplitude=.5mm,segment length=1.5mm}] (-0.5,0) arc (-180:0:0.5);
\fill (-0.5,0) circle (0.1)  node[left] {$\scriptstyle (+,x)$};
\draw[thick] (0,0.5) circle (0.25);
\draw (0-0.15,0.75+0.05) -- (0,0.75);
\draw (0-0.12,0.75-0.12) -- (0,0.75);
\fill (0.5,0) circle (0.1);
\draw (0.5,0) -- (1,0.5);
\draw (0.5,0) -- (1,-0.5);
\draw[thick,black] (0.9,0.4) circle (0.15);
\draw (0.75+0.05,0.55-0.05) -- (1.05-0.05,0.25+0.05);
\draw[thick,black] (0.9,-0.4) circle (0.15);
\draw (0.75+0.05,-0.55+0.05) -- (1.05-0.05,-0.25-0.05);
\end{tikzpicture}\,.
\end{align}
This generates new diagrams in the effective action that are not necessarily two-particle-irreducible. We will see that the friction caused by the transition radiation can be understood analogously to~$\P_{\rm vertex}$ from the induced self-energy terms after the field expansion.

Although VIA is useful for revealing transition radiation without solving the Boltzmann equations, it has some limitations. When treating $g_2^2\varphi^2 A_\mu A^\mu$ as a perturbation vertex term, the gauge boson has a $\varphi$-independent dispersion relation. And this means that for the elementary $1\to 1$ process in the VIA, the gauge boson can either enter the bubble without momentum exchange with the wall, or be reflected. This is, of course, an artefact of the VIA. The mass-gain process cannot be captured by any finite-order truncation in the VIA, and must be computed with the original term $\P_{\rm mass}$.

\subsubsection{Computation of the transition radiation in the CTP formalism}
We now present the complete derivation of the transition radiation in the CTP+2PI formalism. We start by writing down the two-point functions of the fermions and gauge bosons. The fermion two-point functions on the CTP contour are defined as 
\begin{subequations}
 \begin{align}
     &S^<_\psi(x,y)\equiv S_\psi^{+-}(x,y)= \langle T_\C \psi^+(x)\overbar{\psi}^-(y)\rangle=-\langle \overbar{\psi}(y)\psi(x)\rangle\,,\\
     & S^>_\psi(x,y) \equiv S_\psi^{-+}(x,y) = \langle T_\C \psi^-(x)\overbar{\psi}^+(y)\rangle=\langle \psi(x)\overbar{\psi}(y)\rangle\,,\\
     &S^T_\psi(x,y) \equiv S_\psi^{++}(x,y) = \langle T_\C \psi^+(x)\overbar{\psi}^+(y)\rangle=\langle T \psi(x)\overbar{\psi}(y)\rangle\,,\\
     &S^{\overbar{T}}_\psi(x,y)\equiv S_\psi^{--}(x,y) = \langle T_\C \psi^-(x)\overbar{\psi}^-(y)\rangle=\langle \overbar{T} \psi(x)\overbar{\psi}(y)\rangle\,.
 \end{align}   
\end{subequations}
In the on-shell limit and at the leading order in the gradient expansion, these read in Wigner space $\{k,x\}$\,\cite{Garbrecht:2018mrp}\footnote{Again, note the difference in the notation; Our $S_\psi$ is the $\i S_\psi$ in Ref.\,\cite{Garbrecht:2018mrp}. }
\begin{subequations}
\label{prop:N:expl}
\begin{align}
\overbar{S}_{\psi}^{<}(k,x)
&=-2\pi\delta(k^2-m_\psi^2)(\slashed{k}+m_\psi)\left[
\vartheta(k^0)f_\psi(\mathbf{k},x)
-\vartheta(-k^0)(1-\bar f_\psi(-\mathbf{k},x))
\right]\,,\\
\overbar{S}_\psi^{>}(k,x)
&=-2\pi\delta(k^2-m_\psi^2)(\slashed{k}+m_\psi)\left[
-\vartheta(k^0)(1-f_\psi(\mathbf{k},x))
+\vartheta(-k^0)\bar f_\psi(-\mathbf{k},x)
\right]\,,\\
\label{S:T}
\overbar{S}_\psi^{T}(k,x)
&=
\frac{{\rm i}(\slashed{k}+m_\psi)}{k^2-m_\psi^2+{\rm i}\varepsilon}
-2\pi\delta(k^2-m_\psi^2)(\slashed{k}+m_\psi)\left[
\vartheta(k^0)f_\psi(\mathbf{k},x)
+\vartheta(-k^0)\bar f_\psi(-\mathbf{k},x)
\right]\,,\\
\label{S:Tbar}
\overbar{S}_\psi^{\overbar T}(k,x)
&=
\frac{-{\rm i}(\slashed{k}+m_\psi)}{k^2-m_\psi^2-{\rm i}\varepsilon}
-2\pi\delta(k^2-m_\psi^2)(\slashed{k}+m_\psi)\left[
\vartheta(k^0)f_\psi(\mathbf{k},x)
+\vartheta(-k^0)\bar f_\psi(-\mathbf{k},x)
\right]
\,,
\end{align}
\end{subequations}
where $f_\psi$ is the distribution function for fermions and $\bar{f}_\psi$ for anti-fermions. 
We can define analogously the gauge-boson two-point functions on the CTP contour\,${\langle T_\C A^\mu A^\nu\rangle}$, which in Wigner space read
\begin{subequations}
\label{eq:on-shell-limit-gauge}
\begin{align}
 \overbar{\Delta}_A^{\mu\nu, <}(k,x)&=
2\pi P^{\mu\nu} \delta(k^2-m_A^2)\left[
\vartheta(k^0) f_A(\mathbf k,x)
+\vartheta(-k^0) (1+ f_A(-\mathbf k,x))\right]
\,,
\\
\overbar\Delta_A^{\mu\nu,>}(k,x)&=
2\pi P^{\mu\nu} \delta(k^2-m_A^2)\left[
\vartheta(k^0) (1+f_A(\mathbf k,x))
+\vartheta(-k^0)  f_A(-\mathbf k,x)\right]
\,,
\\
\overbar{\Delta}_A^{\mu\nu,T}(k,x)&=
\frac{\i P^{\mu\nu} 
}{k^2-m_A^2+{\rm i}\varepsilon}+
2\pi P^{\mu\nu} \delta(k^2-m_A^2)\left[
\vartheta(k^0) f_A(\mathbf k,x)
+\vartheta(-k^0)  f_A(-\mathbf k,x)\right]
\,, \label{eqn:barDelta++gauge}
\\
\overbar{\Delta}_A^{\mu\nu,\bar T}(k,x)&=
-\frac{\i P^{\mu\nu}
}{k^2-m_A^2-{\rm i}\varepsilon}+
2\pi P^{\mu\nu} \delta(k^2-m_A^2)\left[
\vartheta(k^0) f_A(\mathbf k,x)
+\vartheta(-k^0)  f_A(-\mathbf k,x)\right]
\,,
\end{align}
\end{subequations}
where
\begin{align}
    P^{\mu\nu}=\sum_{\lambda} \epsilon^{\mu}_\lambda (k) \epsilon^{\nu,*}_{\lambda}(k)\,, 
\end{align}
and $\epsilon_\lambda$ are the two transversal polarisation vectors, labelled by $\lambda$, of the gauge bosons emitted on shell.

Again, we start by looking at the one-loop term in the 2PI effective action, 
\begin{align}
    \label{eqn: one-loop term gauge}
    D_A[\varphi]\equiv \frac{\i}{2} {\rm Tr} \left[G_A^{-1}(\varphi)\Delta^{}_A\right]=\frac{\i}{2} \sum_{\mu,\nu} \sum_{a, \, b} \int\d^4 x\, \left.\left[G_{A,\mu\nu}^{ab,-1}(\varphi(x)) \Delta_A^{\nu\mu,ba}(x,y)\right]\right|_{y \, = \, x} \,,
\end{align}
where $G_{A,\mu\nu}^{ab,-1}(\varphi) =- \i\left[\Box+ g_2^2 (\varphi^{a})^2/2 + {\rm gauge\ dependent}\right] c^{ab}\eta_{\mu\nu}$ is the kinetic operator for the gauge field.
In what follows the sum over repeated spacetime indices is implied.
Taking the functional derivative of Eq.\,\eqref{eqn: one-loop term gauge} with respect to the condensate gives
\begin{align}
\label{eq:funct_deriv}
    -\left.\frac{\delta D_A[\varphi]}{\delta \varphi^+(x)}\right|_{\varphi^+=\varphi^-=\varphi}=-\frac{g_2^2}{2}\eta_{\mu\nu} \varphi(x) \left.\Delta_A^{\mu\nu,++}(x,x;\varphi)\right|_{\varphi^+=\varphi^-=\varphi}\,,
\end{align}
In the following, all two-point functions are to be understood as physical ones.

Now we expand the two-point function in insertions of the background field. 
Care is needed since the zero-temperature mass of the gauge boson is zero when~$\varphi=0$, we must understand what the perturbative parameter one is expanding in is.
A solution is provided by the non-zero thermal mass of the gauge boson~$m_A= m_{A;\rm therm}$, which suggests the natural expansion parameter to be~$g_2\varphi/m_{A}$. We then have  
\begin{align}
\label{eq:DeltaA-expansion}
   \Delta^{\mu\nu,ab}_A(x, y;\varphi) &= \Delta_{A}^{\mu\nu,ab} (x,y;0)\notag\\
   & + \frac{\i}{2 }\,g_2^2\sum_{c} c \,\eta_{\alpha\beta}\int \d^4 x'\, \Delta_A^{\mu\alpha,ac}(x, x';0) 
   (\varphi^c(x'))^2 \Delta^{\beta\nu,cb}_A(x', y;0) +\O((g_2\varphi)^4) \,. 
\end{align}
At this point, the two-point function $\Delta_A(x, y;0)$ is the resummed propagator in the absence of the background field. This expansion is verified in Appendix\,\ref{app:expansion_Boltz}, and we relate it to the EoM of the two-point function. After expanding the CTP propagators in the mass of the gauge boson ($g_2 \ll 1 $), we can still expand it in terms of free thermal propagators ($g_1 \ll 1$). We have 
\begin{align}
\label{Eq:free_expansion}
    &\Delta^{\mu\nu, ab}_A(x,y;0) =  \Delta_{A;\rm free}^{\mu\nu, ab}(x,y)\notag\\
    &+ g_1^2{\rm tr_s} \sum_{cd}(cd)\int\d^4 x_1  \d^4 x_2\, \Delta_{A;\rm free}^{\mu\alpha,ac}(x, x_1)  S^{cd}_{\psi;\rm free}(x_1, x_2)\gamma_{\beta} S^{dc}_{\psi;\rm free}(x_2, x_1)\gamma_{\alpha} \Delta^{\beta\nu,db}_{A;\rm free}(x_2, y)+\O(g^4_1)\,,
\end{align}
where ${\rm tr}_s$ is the trace in spinor space and $\Delta_{A,\rm free}^{\mu\nu, ab}(x,y)$ is the gauge boson two-point function in absence of the fermionic loop corrections. Above, the minus sign from $(-\i g_1)\times (-\i g_1)$ has been cancelled by the minus sign coming from the fermion loop: $\langle \bar{\psi}(x_1)\psi(x_1)\cdots \bar{\psi}(x_2)\psi(x_2)\rangle= -{\rm tr} \left(S_\psi(x_1,x_2)\cdots  \ S_\psi(x_2,x_1))\right) $.
This expansion is also verified in Appendix \ref{app:expansion_Boltz}. Substituting the expansion above into Eq.\,\eqref{eq:DeltaA-expansion} generates a series of terms, which can be represented as Feynman diagrams, as follows\footnote{This is only to show schematically the expansion and we have ignored any numerical factors.}
\begin{align}
\label{eq:DA++;Feynman}
&\Delta^{\mu\nu, ++}_A(x, x;\varphi)  =  
\begin{tikzpicture}[baseline={-0.025cm*height("$=$")}]
\draw[double,decorate,decoration={snake,amplitude=.5mm,segment length=1.5mm}] (0,0) circle (0.5);
\fill (-0.5,0) circle (0.1) node[left] {$\scriptstyle (+,x)$};
\end{tikzpicture}
\notag\\
&\qquad\qquad\ =  
\begin{tikzpicture}[baseline={-0.025cm*height("$=$")}]
\draw[decorate,decoration={snake,amplitude=.5mm,segment length=1.5mm}] (0,0) circle (0.5);
\fill (-0.5,0) circle (0.1) node[left] {$\scriptstyle (+,x)$};
\end{tikzpicture}
\ +\
\begin{tikzpicture}[baseline={-0.025cm*height("$=$")}]
\draw[decorate,decoration={snake,amplitude=.5mm,segment length=1.5mm}] (0,0) circle (0.5);
\fill (-0.5,0) circle (0.1) node[left] {$\scriptstyle (+,x)$};
\fill (0.5,0) circle (0.1);
\draw (0.5,0) -- (1,0.5);
\draw (0.5,0) -- (1,-0.5);
\draw[thick,black] (0.9,0.4) circle (0.15);
\draw (0.75+0.05,0.55-0.05) -- (1.05-0.05,0.25+0.05);
\draw[thick,black] (0.9,-0.4) circle (0.15);
\draw (0.75+0.05,-0.55+0.05) -- (1.05-0.05,-0.25-0.05);
\end{tikzpicture}
    \ + \     
\begin{tikzpicture} [baseline={-0.025cm*height("$=$")}]
\draw[decorate,decoration={snake,amplitude=.5mm,segment length=1.5mm}] (-0.5,0) arc (180:120:0.5);
\draw[decorate,decoration={snake,amplitude=.5mm,segment length=1.5mm}] (-0.5,0) arc (-180:60:0.5);
\fill (-0.5,0) circle (0.1)  node[left] {$\scriptstyle (+,x)$};
\draw[thick] (0,0.5) circle (0.25);
\draw (0-0.15,0.75+0.05) -- (0,0.75);
\draw (0-0.12,0.75-0.12) -- (0,0.75);
\end{tikzpicture}
\ + \ 
\begin{tikzpicture} [baseline={-0.025cm*height("$=$")}]
\draw[decorate,decoration={snake,amplitude=.5mm,segment length=1.5mm}] (-0.5,0) arc (180:120:0.5);
\draw[decorate,decoration={snake,amplitude=.5mm,segment length=1.5mm}] (0.5,0) arc (0:60:0.5);
\draw[decorate,decoration={snake,amplitude=.5mm,segment length=1.5mm}] (-0.5,0) arc (-180:0:0.5);
\fill (-0.5,0) circle (0.1)  node[left] {$\scriptstyle (+,x)$};
\draw[thick] (0,0.5) circle (0.25);
\draw (0-0.15,0.75+0.05) -- (0,0.75);
\draw (0-0.12,0.75-0.12) -- (0,0.75);
\fill (0.5,0) circle (0.1);
\draw (0.5,0) -- (1,0.5);
\draw (0.5,0) -- (1,-0.5);
\draw[thick,black] (0.9,0.4) circle (0.15);
\draw (0.75+0.05,0.55-0.05) -- (1.05-0.05,0.25+0.05);
\draw[thick,black] (0.9,-0.4) circle (0.15);
\draw (0.75+0.05,-0.55+0.05) -- (1.05-0.05,-0.25-0.05);
\end{tikzpicture}
+ \O (g_1^4,g_2^4)
\end{align}
The first diagram in the second line is the free propagator in the absence of the background field, while all other diagrams are from higher-order terms in the expansion in $g^2_1$, $g^2_2$. All these diagrams are included when solving the Boltzmann equation for~$f_A$. 
The transition radiation contribution originates from the last diagram on the second line, containing one insertion of the condensate and one fermionic loop. At face value, its contribution is suppressed by additional powers of the couplings, however, we will see that its impact on the pressure is proportional to $\gamma_w$ and thus can be dominant in the ultrarelativistic regime\,\cite{Bodeker:2017cim}. Note that there is another diagram in which the fermion loop is inserted into the gauge propagator in the lower half of the diagram, which contributes to a factor of two.
We write the last diagram in terms of free two-point functions 
\begin{align}
   &\Delta^{\mu\nu, ++}_A(x, x;\varphi) \supset  \frac{\i}{2}  g_1^2 g_2^2 \sum_{cde} (cde) {\rm tr}_s\int \d^4 x'\d^4 x_1\d^4 x_2\notag\\
   &\notag\times\,\Delta_{A;\rm free}^{\mu\alpha, +c}(x, x') \eta_{\alpha\beta}  (\varphi^c(x'))^2 \Delta_{A;\rm free}^{\beta\delta, cd}(x',x_1) S^{de}_{\psi;\rm free}(x_1, x_2)\gamma_\sigma S^{ed}_{\psi;\rm free}(x_2, x_1)\gamma_{\delta} \Delta^{\sigma\nu, e+}_{A;\rm free}(x_2, x)  \\
   &+(\mu\leftrightarrow\nu) \,  .
\end{align}
This translates, using Eq.\,\eqref{eq:funct_deriv}, to 
\begin{align}
\label{eq:DA-expansion}
    -\left.\frac{\delta D_A[\varphi]}{\delta \varphi^+(x)}\right|_{\varphi^+=\varphi^-  = \varphi}\supset   \varphi(x)\int\d^4 x'\,  \Pi_{\varphi;\supset\rm TR}^{\rm R}(x,x') \varphi^2(x')  \, ,
\end{align}
where 
\begin{align}
\label{eq:PiR_contain_TR}
    &\Pi^{\rm R}_{\varphi;\supset \rm TR}(x,x')  = -\frac{\i}{2} \eta_{\mu\nu}g_1^2 g_2^4 \sum_{cde} (cde) \Delta_{A;\rm free}^{\mu\alpha,+c}(x, x')\eta_{\alpha\beta} {\rm tr}_s \int\d^4 x_1\d^4 x_2\notag\\
    &\qquad\qquad\qquad\times \Delta_{A;\rm free}^{\beta\delta,cd}(x',x_1) S^{de}_{\psi;\rm free}(x_1, x_2)\gamma_{\sigma} S^{ed}_{\psi;\rm free}(x_2, x_1)\gamma_{\delta}\Delta^{\sigma\nu, e+}_{A;\rm free}(x_2, x)  \, .
\end{align}
Here, the subscript ``$\supset$TR'' indicates that this induced self-energy contains the transition radiation contribution.\footnote{The usual definition of retarded self-energy is $\Pi^{\rm R}(x,x')=\Pi^{++}(x,x')-\Pi^{+-}(x,x')=\sum_c (c) \Pi^{+c}(x,x')$, where $c$ is the Keldysh index on the vertex at $x'$. Equation~\eqref{eq:PiR_contain_TR} also takes this form.}

If one wishes to obtain Eq.\,\eqref{eq:DA-expansion} from the action principle, it corresponds to a contribution to the effective action~$\Gamma$ of the form
\begin{equation}
    {\Gamma\supset}-\frac{1}{4}\sum_{ab} (ab) \int\d^4 x_1 \d^4 x_2\, [\varphi^a(x_1)]^2 \Pi^{ab}_{\varphi;\supset \rm TR}(x_1,x_2) [\varphi^b(x_2)]^2 \sim
\begin{tikzpicture} [baseline={-0.025cm*height("$=$")}]
\draw[decorate,decoration={snake,amplitude=.5mm,segment length=1.5mm}] (-0.5,0) arc (180:120:0.5);
\draw[decorate,decoration={snake,amplitude=.5mm,segment length=1.5mm}] (0.5,0) arc (0:60:0.5);
\draw[decorate,decoration={snake,amplitude=.5mm,segment length=1.5mm}] (-0.5,0) arc (-180:0:0.5);
\fill (-0.5,0) circle (0.1)  node[left] {$\scriptstyle (a)$};
\draw[thick] (0,0.5) circle (0.25);
\draw (0-0.15,0.75+0.05) -- (0,0.75);
\draw (0-0.12,0.75-0.12) -- (0,0.75);
\fill (0.5,0) circle (0.1)  node[right] {$\scriptstyle (b)$};
\draw (0.5,0) -- (1,0.5);
\draw (0.5,0) -- (1,-0.5);
\draw[thick,black] (0.9,0.4) circle (0.15);
\draw (0.75+0.05,0.55-0.05) -- (1.05-0.05,0.25+0.05);
\draw[thick,black] (0.9,-0.4) circle (0.15);
\draw (0.75+0.05,-0.55+0.05) -- (1.05-0.05,-0.25-0.05);
\draw (-0.5,0) -- (-1,0.5);
\draw (-0.5,0) -- (-1,-0.5);
\draw[thick,black] (-0.9,0.4) circle (0.15);
\draw (-0.75-0.05,0.55-0.05) -- (-1.05+0.05,0.25+0.05);
\draw[thick,black] (-0.9,-0.4) circle (0.15);
\draw (-0.75-0.05,-0.55+0.05) -- (-1.05+0.05,-0.25-0.05);
\end{tikzpicture}     \notag 
    \,.
\end{equation}
Then applying $\delta \Gamma[\varphi^a]/\delta\varphi^+|_{\varphi^+=\varphi^-=\varphi}=0$ would give us a term  as  Eq.\,\eqref{eq:DA-expansion}. Here we note that the form of $\Pi^{\rm R}_{\varphi;\supset \rm TR}(x,y)$ is different from that of the self-energy defined through Eq.\,\eqref{eq:Gamma2-self-energy}.

Following the procedure outlined in Section\,\ref{subsec: friction from pair production}, we can obtain the contribution to the friction as
\begin{align}
    \P_{\supset \rm TR}&=-\frac{1}{ 2}\int_{-\delta}^\delta \d z\, \frac{\d\varphi^2(z)}{\d z}\int\d z'\, \pi_{\varphi;\supset \rm TR}^{\rm R} (z,z') \varphi^2(z')\notag\\
    &=
    -\frac{1}{ 2 } \int \frac{\d q^z}{2\pi} q^z |\widetilde{\varphi^2}(q^z)|^2\, {\rm Im} \widetilde{\pi}_{\varphi;\supset \rm TR}^{\rm R} (q^z) \label{eq:pres_TR}\,,
\end{align}
where 
\begin{align}
    \pi_{\varphi;\supset \rm TR}^{\rm R} (z,z')= \int \d t' \d \vecx'_\perp \Pi_{\varphi;\supset \rm TR}^{\rm R} (t-t',\vecx_\perp-\vecx'_\perp; z,z')\,.
\end{align}
Note that the free two-point functions both for the fermions and for the bosons are completely independent of the background field~$\varphi$, 
and in particular, enjoy full translational invariance, such that~${\pi^{\rm R}_{\varphi;\supset \rm TR}(z,z')=\pi^{\rm R}_{\varphi;\supset \rm TR}(z-z')}$. If we take $g\varphi-g v_b$ as a perturbation, then the gauge boson has a nonvanishing but $z$-independent mass. We observe that, differently from the pressure induced by pair production or by mixing, which are quadratic in the background field, the pressure due to transition radiation is quartic in~$\varphi$. In particular, the Fourier transform of the square of the background field~$\widetilde{\varphi^2}$ appears.
As we did for the general ${\rm Im}\widetilde{\pi}_{\varphi}^{\rm R}$, to compute ${\rm Im}\widetilde{\pi}^{\rm R}_{\varphi;\supset \rm TR}$ it is simpler to use the relation
\begin{align}
    {\rm Im }\widetilde{\pi}^{\rm R}_{\varphi;\supset \rm TR}= -\frac{\i}{2}\left(\widetilde{\pi}^>_{\varphi;\supset \rm TR}- \widetilde{\pi}^{<}_{\varphi;\supset\rm TR}\right)\,,
\end{align}
where $\widetilde{\pi}^{\gtrless}_{\varphi;\rm TR}$ correspond to 
\begin{subequations}
\begin{align}
&\Pi^{>}_{\varphi;\supset \rm TR}(x,x')  =-\frac{\i}{2} \eta_{\mu\nu}g_1^2 g_2^4 \sum_{de} (de) \Delta_{A;\rm free}^{\mu\alpha, -+}(x, x')\eta_{\alpha\beta
}  {\rm tr}_s \int\d^4 x_1\d^4 x_2\notag\\
&\qquad\qquad\qquad\quad\times \Delta_{A;\rm free}^{\beta\delta,+d} (x',x_1) S^{de}_{\psi;\rm free}(x_1, x_2) \gamma_{\sigma} S^{ed}_{\psi;\rm free}(x_2, x_1) \gamma_{\delta} \Delta^{\sigma\nu, e -}_{A;\rm free}(x_2, x)  \, ,\\
&\Pi^{<}_{\varphi;\supset \rm TR}(x,x')  =-\frac{\i}{2} \eta_{\mu\nu} g_1^2 g_2^4 \sum_{de} (de) \Delta_{A;\rm free}^{\mu\alpha,  +-}(x, x')\eta_{\alpha\beta} {\rm tr}_s  \int\d^4 x_1\d^4 x_2\notag\\
&\qquad\qquad\qquad\quad\times  \Delta_{A;\rm free}^{\beta\delta,  -d}(x',x_1) S^{de}_{\psi;\rm free}(x_1, x_2)\gamma_{\sigma} S^{ed}_{\psi;\rm free}(x_2, x_1) \gamma_{\delta} \Delta^{\sigma\nu,  e+}_{A;\rm free}(x_2, x)\,.
\end{align}
\end{subequations}
In terms of Feynman diagrams, these can be represented as 
\begin{subequations}
\begin{align}
 &\Pi^>_{\varphi;\supset \rm TR}(x,x')\quad\sim \quad \sum_{d,e}
\begin{tikzpicture} [baseline={-0.025cm*height("$=$")}]
\draw[decorate,decoration={snake,amplitude=.5mm,segment length=1.5mm}] (-0.5,0) arc (180:120:0.5);
\draw[decorate,decoration={snake,amplitude=.5mm,segment length=1.5mm}] (0.5,0) arc (0:60:0.5);
\draw[decorate,decoration={snake,amplitude=.5mm,segment length=1.5mm}] (-0.5,0) arc (-180:0:0.5);
\fill (-0.5,0) circle (0.1)  node[left] {$\scriptstyle (-,x)$};
\draw[thick] (0,0.5) circle (0.25);
\fill (0-0.25,0.5) circle (0.05) node[left,xshift=.05cm] {$\scriptstyle e$} ;
\draw (0-0.15,0.75+0.05) -- (0,0.75);
\draw (0-0.12,0.75-0.12) -- (0,0.75);
\fill (0+0.25,0.5) circle (0.05) node[right,xshift=-.05cm,yshift=.05cm] {$\scriptstyle d$} ;
\fill (0.5,0) circle (0.1)  node[right] {$\scriptstyle (+,x')$};
\end{tikzpicture} \,,\\
&\Pi^<_{\varphi;\supset\rm TR}(x,x')\quad \sim\quad  \sum_{d,e}
\begin{tikzpicture} [baseline={-0.025cm*height("$=$")}]
\draw[decorate,decoration={snake,amplitude=.5mm,segment length=1.5mm}] (-0.5,0) arc (180:120:0.5);
\draw[decorate,decoration={snake,amplitude=.5mm,segment length=1.5mm}] (0.5,0) arc (0:60:0.5);
\draw[decorate,decoration={snake,amplitude=.5mm,segment length=1.5mm}] (-0.5,0) arc (-180:0:0.5);
\fill (-0.5,0) circle (0.1)  node[left] {$\scriptstyle (+,x)$};
\draw[thick] (0,0.5) circle (0.25);
\fill (0-0.25,0.5) circle (0.05) node[left,xshift=.05cm] {$\scriptstyle e$} ;
\draw (0-0.15,0.75+0.05) -- (0,0.75);
\draw (0-0.12,0.75-0.12) -- (0,0.75);
\fill (0+0.25,0.5) circle (0.05) node[right,xshift=-.05cm,yshift=.05cm] {$\scriptstyle d$} ;
\fill (0.5,0) circle (0.1)  node[right] {$\scriptstyle (-,x')$};
\end{tikzpicture}\,,
\end{align}    
\end{subequations}
where we have also indicated the Keldysh indices. Each function is a sum of four terms. Different choices of $\{d,e\}$ give different cuttings of the diagram and thus different particle processes. Since in the transition radiation process, both fermions are on-shell external states, and since it is the fermion Wightman functions~$\overbar{S}_\psi^<$ and~$\overbar{S}_\psi^>$ that enforce on-shell conditions by being proportional to~${\delta(k^2-m^2)}$ (see Eqs.\,\eqref{prop:N:expl}), and furthermore carry the required statistical factors of~$f_\psi$ and~$1-f_\psi$, the relevant term we are tracing should have~$d$ and~$e$ taking different signs. 

Our formula for the transition radiation pressure amounts to a leading-order term in the expansion in $g_2^2\varphi^2 A_\mu A^\mu/4$ of the following diagram
\begin{align}
\label{eq:transition_radiation_full_mass}
    \begin{tikzpicture}
[baseline={-0.025cm*height("$=$")}]
    \draw[thick] (-2,0) -- (2,0);
    \fill (0,0) circle (0.1);
    \draw[line width=0.5mm, decorate,decoration={snake,amplitude=.5mm,segment length=1.5mm}] (0,0) -- (2,1);
\end{tikzpicture}\,,  
\end{align}
where the thick wavy line represents the free gauge boson with the $\varphi$-dependent mass term taken into account. More explicitly, we have 
\begin{align}
    \begin{tikzpicture}
     [baseline={-0.025cm*height("$=$")}]   \draw[line width=0.5mm, decorate,decoration={snake,amplitude=.5mm,segment length=1.5mm}] (-1,0) -- (1,0); 
    \end{tikzpicture}
   =
  \begin{tikzpicture}
      \draw[decorate,decoration={snake,amplitude=.5mm,segment length=1.5mm}] (-1,0) -- (1,0); 
  \end{tikzpicture}
  +
   \begin{tikzpicture}   \draw[decorate,decoration={snake,amplitude=.5mm,segment length=1.5mm}] (-1,0) -- (1,0); 
   \fill (0,0) circle (0.1);
   \draw[thick] (0,0) -- (-.6,0.6);
   \draw[thick] (0,0) -- (0.6,0.6);
   \draw[thick] (-.5,0.5) circle (0.13);
   \draw[thick] (0.5, 0.5) circle (0.13);
   \draw[thick] (-0.5-0.1, 0.5-0.1) -- (-0.5+0.1, 0.5+0.1);
   \draw[thick] (0.5-0.1, 0.5+0.1) -- (0.5+0.1, 0.5-0.1) ;
  \end{tikzpicture}
  +
  \begin{tikzpicture}   \draw[decorate,decoration={snake,amplitude=.5mm,segment length=1.5mm}] (-1,0) -- (2.5,0); 
   \fill (0,0) circle (0.1);
   \draw[thick] (0,0) -- (-.6,0.6);
   \draw[thick] (0,0) -- (0.6,0.6);
   \draw[thick] (-.5,0.5) circle (0.13);
   \draw[thick] (0.5, 0.5) circle (0.13);
   \draw[thick] (-0.5-0.1, 0.5-0.1) -- (-0.5+0.1, 0.5+0.1);
   \draw[thick] (0.5-0.1, 0.5+0.1) -- (0.5+0.1, 0.5-0.1) ;
   \fill (1.5,0) circle (0.1);
   \draw[thick] (1.5,0) -- (-.6+1.5,0.6);
   \draw[thick] (1.5,0) -- (0.6+1.5,0.6);
   \draw[thick] (-0.5+1.5,0.5) circle (0.13);
   \draw[thick] (0.5+1.5, 0.5) circle (0.13);
   \draw[thick] (-0.5-0.1+1.5, 0.5-0.1) -- (-0.5+0.1+1.5, 0.5+0.1);
   \draw[thick] (0.5-0.1+1.5, 0.5+0.1) -- (0.5+0.1+1.5, 0.5-0.1) ;
  \end{tikzpicture}
  +...\,.
\end{align}
To be more specific, the transition radiation process is represented by 
\begin{align}
\label{eq:transition_radiation_Feyn}
\begin{tikzpicture}
[baseline={-0.025cm*height("$=$")}]
    \draw[thick] (-2,0) -- (2,0);
    \fill (0,0) circle (0.1);\draw[decorate,decoration={snake,amplitude=.5mm,segment length=1.5mm}] (0,0) -- (2,1);
    \fill (1.,0.5) circle (0.1) ;
    \draw[thick] (1, 0.5) -- (0.3, 0.85);
    \draw[thick] (0.3, 0.85) -- (0.3, 0.85);
    \draw[thick] (1, 0.5) -- (1.03,  1.2);
    \draw[thick] (0.4, 0.8) circle (0.13);
    \draw[thick] (0.4-0.06, 0.8-0.1) -- (0.4+0.07,  0.8+0.1) ;
    \draw[thick] (1.02, 1.1) circle (0.13);
    \draw[thick] (1.02 -0.14 , 1.1) -- (1.02 + 0.14, 1.1);
\end{tikzpicture}  
\end{align}
where we see an internal, off-shell gauge propagator.
This means that the terms in $\Pi^>_{\varphi;\supset \rm TR}$ and $\Pi^<_{\varphi;\supset\rm TR}$ that give rise to the transition radiation process should have $d$ and $e$ taking the same sign as their nearest neighbour gauge mass vertex ($g_2^2\varphi^2A_\mu A^\mu$) so that we would have the internal gauge propagator. Therefore we choose the $\{d=+,e=-\}$ term from $\Pi^{>}_{\varphi;\supset\rm TR}$, and the $\{d=-,e=+\}$ term from $\Pi^{<}_{\varphi;\supset\rm TR}$. In conclusion, we have identified the relevant terms that contribute to the transition radiation process:
\begin{subequations}
\begin{align}
&\Pi^{>}_{\varphi;\rm TR}(x,x') 
 \quad \sim \quad
\begin{tikzpicture} [baseline={-0.025cm*height("$=$")}]
\draw[decorate,decoration={snake,amplitude=.5mm,segment length=1.5mm}] (-0.5,0) arc (180:120:0.5);
\draw[decorate,decoration={snake,amplitude=.5mm,segment length=1.5mm}] (0.5,0) arc (0:60:0.5);
\draw[decorate,decoration={snake,amplitude=.5mm,segment length=1.5mm}] (-0.5,0) arc (-180:0:0.5);
\fill (-0.5,0) circle (0.1)  node[left] {$\scriptstyle (-,x)$};
\draw[thick] (0,0.5) circle (0.25);
\fill (0-0.25,0.5) circle (0.05) node[left,xshift=.05cm] {$\scriptstyle -$} ;
\draw (0-0.15,0.75+0.05) -- (0,0.75);
\draw (0-0.12,0.75-0.12) -- (0,0.75);
\fill (0+0.25,0.5) circle (0.05) node[right,xshift=-.05cm] {$\scriptstyle +$} ;
\fill (0.5,0) circle (0.1)  node[right] {$\scriptstyle (+,x')$};
\end{tikzpicture}\notag\\
&\ =   {\frac{\i}{2}}\eta_{\mu\nu}g_1^2 g_2^4\Delta_{A;\rm free}^{\mu\alpha,>}(x, x')\eta_{\alpha\beta}\notag\\
&\qquad\qquad\quad\times {\rm tr}_s \int\d^4 x_1\d^4 x_2 \Delta_{A;\rm free}^{\beta\delta,++}(x',x_1) S^{<}_{\psi;\rm free}(x_1, x_2)\gamma_{\sigma} S^{>}_{\psi;\rm free}(x_2, x_1)\gamma_{\delta} \Delta^{\sigma\nu,--}_{A;\rm free}(x_2, x)  \,,\\
&\Pi^{<}_{\varphi;\rm TR}(x,x') \quad\sim \quad
\begin{tikzpicture} [baseline={-0.025cm*height("$=$")}]
\draw[decorate,decoration={snake,amplitude=.5mm,segment length=1.5mm}] (-0.5,0) arc (180:120:0.5);
\draw[decorate,decoration={snake,amplitude=.5mm,segment length=1.5mm}] (0.5,0) arc (0:60:0.5);
\draw[decorate,decoration={snake,amplitude=.5mm,segment length=1.5mm}] (-0.5,0) arc (-180:0:0.5);
\fill (-0.5,0) circle (0.1)  node[left] {$\scriptstyle (+,x)$};
\draw[thick] (0,0.5) circle (0.25);
\fill (0-0.25,0.5) circle (0.05) node[left,xshift=.05cm] {$\scriptstyle +$} ;
\draw (0-0.15,0.75+0.05) -- (0,0.75);
\draw (0-0.12,0.75-0.12) -- (0,0.75);
\fill (0+0.25,0.5) circle (0.05) node[right,xshift=-.05cm] {$\scriptstyle -$} ;
\fill (0.5,0) circle (0.1)  node[right] {$\scriptstyle (-,x')$};
\end{tikzpicture}\notag\\
&\ =\frac{\i}{2}\eta_{\mu\nu}g_1^2 g_2^4 \Delta_{A;\rm free}^{\mu \alpha,<}(x, x') \eta_{\alpha\beta}\notag\\
&\qquad\qquad\quad\times {\rm tr}_s \int\d^4 x_1\d^4 x_2 \Delta_{A;\rm free}^{\beta\delta, --}(x',x_1) S^{>}_{\psi;\rm free}(x_1, x_2) \gamma_{\sigma} S^{<}_{\psi;\rm free}(x_2, x_1)\gamma_{\delta} \Delta^{\sigma\nu, ++}_{A;\rm free}(x_2, x)\,.
\end{align}
\end{subequations}
Other choices of $\{d,e\}$  are not relevant for the present discussion.
We are ready to compute the pressure from the transition radiation~$\P_{\rm TR}$ using the previous expressions but with the subscript ``$\supset${\rm TR}'' replaced by ``${\rm TR}$''. Taking the Fourier transform, we get
\begin{align}
    {\rm Im} \widetilde{\pi}^{\rm R}_{\varphi;\rm TR} (q) & \subset \frac{g_1^2 g_2^4}{4 } \eta_{\mu\nu}\eta_{\alpha\beta}\int\frac{\d^4 k_1}{(2\pi)^4}\int\frac{\d^4 k_2}{(2\pi)^4} \int\frac{\d^4 k_3}{(2\pi)^4} (2\pi)^4 \delta^{(4)}(\hat{q}+k_1-k_2-k_3)
    \notag\\
    &\times {\rm tr}_s \left[\widetilde{\Delta}_{A;\rm free}^{\mu\alpha,>}(k_3)  \widetilde{\Delta}_{A;\rm free}^{\beta\delta,++}(k_1-k_2)   \widetilde{S}^<_{\psi;\rm free} (k_1) \gamma_{\sigma} \widetilde{S}^>_{\psi;\rm free} (k_2) \gamma_{\delta} \widetilde{\Delta}_{A;\rm free}^{\sigma\nu,--} ({k_1-k_2})\right.\notag\\
    &\qquad\left. -\widetilde{\Delta}_{A;\rm free}^{\mu\alpha,<}(k_3)  \widetilde{\Delta}_{A;\rm free}^{\beta\delta,--}(k_1-k_2)   \widetilde{S}^>_{\psi;\rm free} (k_1) \gamma_{\sigma} \widetilde{S}^<_{\psi;\rm free} (k_2) \gamma_{\delta} \widetilde{\Delta}_{A;\rm free}^{\sigma\nu,++} ({k_1-k_2}) \right]\,.
\end{align}
For our purpose, we will take $q=\hat{q}\equiv (0,0,0,q^z)$, which is the external four-momentum provided by the bubble wall. 
One can already compare this form with Eq.\,\eqref{eqn: imaginary pi r in terms of wightman propagators}. We have used ``$\subset$'' instead of ``$=$'' above because the RHS contains all processes allowed by the vertex $g_1\overbar{\psi} \gamma_\mu A^\mu \psi$ and kinematics. For example, one can have $\psi+A \rightarrow \psi$ or $A\rightarrow \psi+\bar{\psi}$ if kinematics allows. We wish to extract the transition radiation process $\psi\rightarrow \psi + A$ and $\overbar{\psi}\rightarrow \overbar{\psi}+A$. 
To this end, we rewrite the Fourier transform of the free fermion Wightman functions as
\begin{subequations}
\begin{align}
\widetilde{S}_{\psi}^{<}(k)
&=-\frac{2\pi(\slashed{k}+m_\psi)}{2 E_\veck^{(\psi)}}\left[
\underbrace{\delta(k^0-E_\veck^{(\psi)})f_\psi(\mathbf{k})}_{\rm incoming\ fermion}
-\underbrace{\delta(k^0+E_\veck) (1-\bar f_\psi(-\mathbf{k}))}_{\rm outgoing\ antifermion}
\right]\,,\\
\widetilde{S}_\psi^{>}(k)
&=-\frac{2\pi(\slashed{k}+m_\psi)}{2E_\veck^{(\psi)}}\left[
-\underbrace{\delta(k^0-E^{(\psi)}_\veck) (1-f_\psi(\mathbf{k}))}_{\rm outgoing\ fermion}
+\underbrace{\delta(k^0+E_\veck^{(\psi)})\bar f_\psi(-\mathbf{k})}_{\rm incoming\  antifermion}
\right]\,,  
\end{align}    
\end{subequations}
where we have indicated the interpretation of each term. For the fermion terms, we need to recognise the factor $\slashed{k}+m_\psi$ as the sum over spins for fermion external states
\begin{align}
    \sum_{\rm spins} u^s (k) \bar{u}^s(k) = \slashed{k}+m_\psi\,.
\end{align}
While for the antifermion terms, we need to make a change of variable $\veck\rightarrow -\veck$, and recognise 
\begin{align}
    \sum_{\rm spins} v^s (k) \bar{v}^s(k) = -[-\slashed{k}+m_\psi]\,.
\end{align}
For simplicity, we will focus on the fermion terms giving the process~${\psi\rightarrow \psi+ A}$. The process~${\bar\psi \to \bar\psi + A}$ simply contributes to a factor of two for the friction. To identify the external gauge boson, we can write 
\begin{subequations}
\begin{align}
 \widetilde{\Delta}_A^{\mu\nu, <}(k)&=
\frac{2\pi P^{\mu\nu}}{2 E_\veck^{(A)}}\left[
\underbrace{\delta(k^0-E_\veck^{(A)}) f_A(\mathbf k)}_{\rm gauge\ boson\ absorption}
+\underbrace{\delta(k^0+E_\veck^{(A)}) (1+ f_A(-\mathbf k))}_{\rm gauge\ boson\ emission}\right] \,,
\\
\widetilde{\Delta}_A^{\mu\nu,>}(k)&=
\frac{2\pi P^{\mu\nu}}{2 E_\veck^{(A)}}\left[
\underbrace{\delta(k^0- E_\veck^{(A)}) (1+f_A(\mathbf k))}_{\rm gauge\ boson\ emission}
+\underbrace{\delta(k^0+E_\veck^{(A)})  f_A(-\mathbf k)}_{\rm gauge\ boson\ absorption }\right]\,.
\end{align}  
\end{subequations}
With these at hand, we can immediately recognise the terms relevant for the transition radiation
\begin{align}
\label{Eq:Tr_rad}
     {\rm Im} \widetilde{\pi}^{\rm R}_{\varphi;\rm TR} (\hat{q}) = \: &-  \frac{g_2^4}{ 4 }\int_{\vec k_1, \psi} \int_{\vec k_2, \psi} \int_{\vec k_3, A} (2\pi)^4 \delta^{(4)}(\hat{q}+k_1-k_2-k_3)\, |\M_{\psi\rightarrow \psi+A}|^2 
    \notag\\
    &\times \Big\{f_\psi(\veck_1) [1-f_\psi (\veck_2)] [1+f_A(\veck_3)]- [1-f_\psi(\veck_1)] f_\psi (\veck_2) f_A(\veck_3)\Big\}\notag\\
    &+ (\bar{\psi}\rightarrow \bar{\psi}+A)\,,
\end{align}
where $|\M_{\psi(k_1)\rightarrow \psi(k_2)+A(k_3)}|^2$ is the squared transition amplitude of the diagram in Eq.\,\eqref{eq:transition_radiation_Feyn}, with polarisations summed over and $g_2^4\varphi^4/4$ factored out. From the conventional Feynman rules, one has for the emission of an off-shell photon $A^\mu$ from a fermion
\begin{align}
    \M_{\psi(k_1)\rightarrow \psi(k_2)+A(k_3)}=  g_1\bar{u}^s(k_2) \gamma_\delta \Delta^{\delta\beta,++}_{A;\rm free}(k_1-k_2) \eta_{\beta\alpha}\epsilon^{\alpha*}_\lambda (k_3) u^{s'}(k_1)\,,
\end{align}
where $s$, $s'$ are spin polarisation indices, and $\lambda$ the gauge boson polarisation index.

To perform our computation, we shall ignore the finite-temperature part in the gauge boson propagators $\Delta_{A;\rm free}^{++}$ and $\Delta_{A;\rm free}^{--}$ as these parts will make the virtual gauge boson on-shell which is forbidden by kinematics. After squaring the invariant transition amplitude and summing over the polarisations,  
we obtain in the 't Hooft-Feynman gauge ($\xi=1$),
\begin{align}
&|\M_{\psi(k_1)\rightarrow \psi(k_2)+A(k_3)}|^2= 
\notag
\\
&- 4g_1^2\sum_\lambda \frac{  \big[(m_\psi^2-k_1 \cdot k_2) \varepsilon_\lambda(k_3)\cdot \varepsilon_\lambda^*(k_3)+ (k_2 \cdot \varepsilon_\lambda^*(k_3))(k_1\cdot \varepsilon_\lambda(k_3))+(k_1 \cdot \varepsilon_\lambda^*(k_3))(k_2\cdot \varepsilon_\lambda(k_3)) \bigg]}{[(k_1-k_2)^2-m_A^2]^2}\,.
\end{align}
For simplicity and to show that the transition radiation does not depend on the emitter (here the fermion) being massive, from here onwards we take $m_\psi=0$.
We now choose the following kinematics for the two fermions ($k_1$ and $k_2$) and the gauge boson ($k_3$)
\begin{subequations}
\begin{align}
k_1&=(p^0, 0, 0, - p^0)\,,\\
k_2&=((1-x) p^0, k_\perp, 0, - \sqrt{(1-x)^2 (p^0)^2- k_\perp^2 })\,,\\
k_3&=(x p^0, -k_\perp, 0, -\sqrt{x^2 (p^0)^2- k_\perp^2- m_{A}^2})\,.
\end{align} 
\end{subequations}
In the soft limit $x\ll 1$, 
$k_1$ becomes almost aligned with $k_2$. 
For the transverse polarisations, one has\,\cite{Bodeker:2017cim,Azatov:2020ufh,Gouttenoire:2021kjv}
\begin{align}
\label{eq:M2-TR}
|\M_{\psi(k_1)\rightarrow \psi(k_2)+A_T(k_3)}|^2\approx 4 g_1^2 C_2[R] \frac{k_\perp^2}{x^2}\frac{1}{[k_\perp^2/(1-x)+m_A^2]^2}\approx  4 g_1^2 C_2[R] \frac{k_\perp^2}{x^2}\frac{1}{[k_\perp^2+m_A^2]^2}\,,
\end{align} 
where $C_2[R]$ is the second Casimir of the representation of the gauge group of the incoming particle. Using $\d k_2^z/k_2^0=\d k^0_2/k_2^z$ and integrating over $\veck_{2,\perp}$ and $k_2^0$, we get
\begin{align}
     {\rm Im} \widetilde{\pi}^{\rm R}_{\varphi;\rm TR} (q^z) \approx & - \frac{g_2^4}{ 4}\int\frac{\d^3 \veck_1}{(2\pi)^3 4 (p^0)^2 } \int\frac{\d^3 \veck_3}{(2\pi)^3  2 k^0 }(2\pi)\delta(q^z+k^z_1-\bar{k}^z_2-k^z_3)
    \notag\\
    &\times |\M_{\psi(k_1)\rightarrow \psi(\bar{k}_2)+A_T(k_3)}|^2 f_\psi(\veck_1) \,,
\end{align}
where we have used $k_2^z\approx k_2^0\approx p^0$ and $\bar{k}_2=(k_1^0-k_3^0,\veck_{1,\perp}-\veck_{3,\perp},\sqrt{(k_1^0-k_3^0)^2-(\veck_{1,\perp}-\veck_{3,\perp})^2})$. We can directly substitute this equation into $\P_{\rm TR}$. Integrating over $q^z$ removes the remaining Dirac delta function and we obtain
\begin{align}
    \P_{\rm TR} \approx & \frac{g_2^4}{8}\int\frac{\d^3 \veck_1}{(2\pi)^3 4 (p^0)^2 } \int\frac{\d^3 \veck_3}{(2\pi)^3  2 k^0 }\times \Delta p^z\times |\widetilde{\varphi^2}(\Delta p^z)|^2\times
     |\M_{\psi(k_1)\rightarrow \psi(\bar{k}_2)+A_T(k_3)}|^2 f_\psi(\veck_1)\,,
\end{align}
where $\Delta p^z=-(k_1^z-\bar{k}_2^z -k_3^z)$. Compared to the B\"{o}deker-Moore method, one may identify (see Eq.\,(16) of\,\cite{Bodeker:2017cim}) 
\begin{align}
\label{eq:compare-with-BM}
    |\M|^2_{\rm (BM)} \sim  \frac{g_2^4}{ 8} |\widetilde{\varphi^2}(\Delta p^z)|^2\times
     |\M_{\psi(k_1)\rightarrow \psi(\bar{k}_2)+A_T(k_3)}|^2\,.
\end{align}
Again $|\widetilde{\varphi^2}(\Delta p^z)|^2$ is exponentially suppressed for large $\Delta p^z\gg 1/L_w$. As a rough estimate, here we replace it with its asymptotic expression at $\Delta p^z L_w\rightarrow 0$.
From the profile \eqref{eq:wall-profile}, we have 
\begin{align}
\label{eq:Fourier-varphi2}
    |\widetilde{\varphi^2}(\Delta p^z)|^2 \xrightarrow{\Delta p^z L_w\rightarrow 0} \frac{v_b^4}{(\Delta p^z)^2}\,.
\end{align}
Further, we have 
\begin{align}
\label{eq:Deltapz gauge}
    \Delta p^z \approx \frac{k_\perp^2+m_A^2}{2 x p^0}\,.
\end{align}
Substituting Eqs.\,\eqref{eq:M2-TR},\,\eqref{eq:Fourier-varphi2},\,\eqref{eq:Deltapz gauge} into the RHS of Eq.\,\eqref{eq:compare-with-BM}, we obtain
\begin{align}
   \frac{g_2^4}{8} |\widetilde{\varphi^2}(\Delta p^z)|^2\times
     |\M_{\psi(k_1)\rightarrow \psi(k_2)+A_T(k_3)}|^2 \approx  \frac{1}{2} \times 4 g_1^2 C_2[R] \frac{k_\perp^2}{x^2}\times \frac{4(p^0)^2 x^2}{(k_\perp^2+m^2_A)^2} \times   \frac{m^4_{A,h}}{(k_\perp^2+m_A^2)^2}\,,
\end{align}
where $m_{A,h}=g_2 v_b/\sqrt{2}$ is the Higgsed mass of the gauge boson. The above expression is close to Eq.\,(24) of Ref.\,\cite{Bodeker:2017cim} with a few differences: (1) It has an additional factor $1/2$; (2) Since we have used the VIA, the gauge boson does not have a Higgsed mass in our case, and $m_A$ is a screening mass due to thermal corrections; (3) The $1/k_\perp^4$ in Ref.\,\cite{Bodeker:2017cim} is replaced with $1/(k_\perp^2+m^2_A)^2$. Nevertheless, the similarity between these two expressions immediately allows us to conclude, following B\"{o}deker-Moore's analysis, that the transition radiation pressure is of the type
\begin{equation}
\P_{\rm TR} \propto \gamma_w T^3 v_b
\end{equation} 
in the ultrarelativistic regime.  The most salient feature of this result is the dramatic increase of the pressure with an increasing boost factor $\gamma_w$, thus preventing runaway in most of the parameter space of gauged FOPTs. A rigorous computation of the pressure requires a precise understanding of the IR cut-off and of possible saturation of the gauge boson phase space. This is left for future studies.   

In this section, we focused for simplicity on the case of a fermion emitting a gauge boson. However, the same exercise can be performed for a scalar emitter, like the Higgs boson, or for a gauge boson emitter, like the $W$-boson.

\section{Localised bubble equation of motion}
\label{sec:localisation}

In Section\,\ref{sec:2PI}, we have identified a self-energy term which descends from~$\Gamma_2$ in the 2PI effective action, namely the proper self-energy in Eq.\,\eqref{eq:Piphiab}. 
The proper self-energy term contributes to the equation of motion for the bubble, and that has been overlooked in the previous literature on bubble wall dynamics. 
In the VIA, one can have additional self-energy-like terms generated from the field expansion, as we have shown in the discussions on mixing and transition radiation in Section\,\ref{sec:small_field}.
These self-energy terms are non-local, i.e., they contain an integral over the entire past lightcone. Furthermore, the condensate EoM couples to the EoM for the two-point functions, which describe the evolution of particles and are also non-local, making it extremely difficult to fully solve the dynamics.
Our discussions on the non-local self-energy terms so far have relied on certain approximations that made it possible to write them in a closed form. 
For the proper self-energy term in the condensate EoM, we have assumed an ultrarelativistic bubble wall, while for the self-energy terms generated in the VIA, we have assumed the plasma to be homogeneous across the wall. 
All these assumptions serve to restore the $z$-translational invariance in the wall frame so that the self-energy terms trivially localise (see the discussion below Eq.\,\eqref{eq:self-energy-phichichi}). 
However, these assumptions are not generic. For example, in many models, the bubble wall velocity may not be relativistic or only mildly relativistic, may never reach ultrarelativistic velocities, and typically the plasma across the wall is not homogeneous in the $z$-direction.

Upon applying a localisation procedure, the EoMs for the two-point functions can be reduced to the familiar Boltzmann equations\,\cite{Chou:1984es,Calzetta:1986cq} (for a pedagogical introduction, see\,\cite{Ai:2023qnr}). It would then also be useful to have a localised condensate EoM that can be more easily adopted in numerical computations of the bubble wall velocity, e.g., in the open source software {\tt WallGo}\,\cite{Ekstedt:2024fyq}.  

We perform the gradient expansion for the proper self-energy term in the condensate EoM. This is viable if the bubble wall width is larger than the typical wavelength perpendicular to the wall of the particles, when compared in the same frame, 
\begin{align}
\label{eq:gradient_expansion_cond}
  \partial_z < p^z \qquad \text{(gradient expansion)} \quad \Rightarrow\quad  L_w \gtrsim \lambda^{\rm (wall)}_z = \frac{1}{p^z} \sim \frac{1}{\gamma_w T}\,. 
\end{align}
In realistic models, the wall thickness is typically $L_w \sim \O(10)/T$\,\cite{Moore:1995si}. Since $\gamma_w\geq 1$, the above condition can be satisfied in many models even far away from the ultrarelativistic limit.
Doing the expansion
\begin{subequations}
\begin{align}
    &\varphi(y)=\varphi(x)+(y^\mu-x^\mu)\partial_\mu\varphi(x)+\O(\partial^2)\,,\label{eq:varphi-expansion}  \\
    &\overbar{\Pi}_\varphi^{\rm R}\left(q,\frac{x+y}{2}\right)=\overbar{\Pi}^{\rm R}_\varphi(q,x)+\frac{1}{2}(y^\mu-x^\mu) \partial_\mu \overbar{\Pi}^{\rm R}_\varphi(q,x)+\O(\partial^2)\,,
\end{align}    
\end{subequations}
and performing the Wigner transform, we obtain 
\begin{align}
    &\int \d^4 y\, \Pi^{\rm R}_\varphi(x,y) \varphi(y)\notag\\
    &=\varphi(x) \overbar{\Pi}^{\rm R}_\varphi(q=0,x)+\frac{\i}{2} \varphi(x)\partial_{\mu}\partial^{\mu}_{(q)} \overbar{\Pi}^{\rm R}_\varphi(q,x)\big|_{q^\mu=0}+\i (\partial_\mu\varphi)\partial^{\mu}_{(q)} \overbar{\Pi}^{\rm R}_\varphi(q,x)\big|_{q^\mu=0} +\O(\partial^2)\,,
\end{align}
where we assign for the derivative with respect to $q$ a subscript $(q)$. We observe that the last term has the form of a dissipative damping term. 
Using the relations\,\eqref{eq:PiR-relations}, valid when generalized to the Wigner space, we obtain (see also \cite{Chadha-Day:2022inf} for a similar result)
\begin{align}
\label{eq:grad-expansion-self-energy-term}
    &\int \d^4 y\, \Pi^{\rm R}_\varphi(x,y)\varphi(y)\notag\\
    &=\varphi(x) \overbar{\Pi}^{\rm R}_\varphi(0,x)-\frac{\varphi(x)}{2}\lim_{q^\mu\rightarrow 0}\frac{\partial_{\mu} {\rm Im}\overbar{\Pi}^{\rm R}_\varphi(q,x)}{q_\mu} -\underbrace{ (\partial_\mu\varphi)\lim_{q^\mu\rightarrow 0}\frac{{\rm Im}\overbar{\Pi}^{\rm R}_\varphi(q,x)}{q_\mu}}_{\rm standard\ form\ for\ a\ friction\ term} +\O(\partial^2)\,.
\end{align}
Since ${\rm Im}\bar{\Pi}_\varphi^{\rm R}$ takes the form of a collision term, $\lim_{q^\mu\rightarrow 0}{\rm Im} \Bar{\Pi}^{\rm R}_\varphi(q,x)$ describes particle processes via the VEV-dependent vertices in the limit of vanishing four-momentum exchange between the wall and the particles.
Recalling that the tree-level mass of $\varphi$ enters the equation of motion through the linear term $m^2\varphi\supset V'_{\rm eff}(\phi)$, the first and second terms in Eq.~\eqref{eq:grad-expansion-self-energy-term} can be seen to contribute as a higher-order correction, although being spacetime-dependent in general, to the mass (or more generally the effective potential $V_{\rm eff}(\varphi)$). 
We denote this mass correction as
\begin{align}
    \Delta m^2_{\Pi_\varphi}(x)\equiv \overbar{\Pi}^{\rm R}_\varphi(0,x)-\frac{1}{2}\lim_{q^\mu\rightarrow 0}\frac{\partial_{\mu} {\rm Im}\overbar{\Pi}^{\rm R}_\varphi(q,x)}{q_\mu}\,.
\end{align}
Then at the next-to-leading order ($\O(\partial)$) in the gradient expansion, we obtain the EoM of the condensate
\begin{align}
\label{eq:eom-final form2}
    \boxed{\Box\varphi + V' (\varphi)+\sum_{i=
    \phi,
    \chi}\frac{\d m^2_i(\varphi)}{\d\varphi}\int_{\vec k, i}\,{f_i(\vec k,x)} + \Delta m^2_{\Pi_\varphi}(x)\varphi- (\partial_\mu\varphi) \lim_{q^\mu\rightarrow 0}\frac{{\rm Im}\overbar{\Pi}^{\rm R}_\varphi(q,x)}{q_\mu} =0\,.}
\end{align}
This equation can be immediately applied for numerical computation of the bubble wall velocity. 
To have an expression of the corresponding frictional pressure, we look at the last term in the wall frame. In this case, both~$\varphi$ and~$\overbar{\Pi}^{\rm R}_\varphi$ do not depend on $t$ and $\vecx_\perp$. 
Using the same notation as in the previous sections, we write explicitly
\begin{align}
    (\partial_\mu\varphi)\lim_{q^\mu\rightarrow 0}\frac{{\rm Im}\overbar{\Pi}^{\rm R}_\varphi(q,z)}{q_\mu}= (\partial_z\varphi)\lim_{q^z\rightarrow 0}\frac{{\rm Im}\overbar{\Pi}^{\rm R}_\varphi(q^z,\vecq_\perp=0,z)}{q_z}\equiv (\partial_z\varphi)\lim_{q^z\rightarrow 0}\frac{{\rm Im}\overbar{\pi}^{\rm R}_\varphi(q^z,z)}{q_z}\,,
\end{align}
where we have defined $\bar{\pi}^{\rm R}_\varphi$ in the last equality.
We then get the new friction term at~$\O(\partial)$ in the gradient expansion,
\begin{align}
    \P_{\rm vertex;\, GE}= \int_{-\delta}^{\delta}\d z\, \left[\left(\frac{\d \varphi}{\d z}\right)^2 \lim_{q^z\rightarrow 0}\frac{{\rm Im}\overbar{\pi}^{\rm R}_\varphi(q^z,z)}{q_z}\right]\,,
\end{align}
where the ``GE'' subscript stands for the leading term in the \emph{gradient expansion}.

\section{Discussions and conclusion}
\label{sec:Conc}

FOPTs are a promising source of primordial gravitational waves and can catalyse baryogenesis, magnetogenesis, dark matter production and PBH production and have attracted great attention in recent years. A precise understanding of all these phenomena requires, as a first step, the precise determination of the relevant thermodynamic quantities and the subsequent calculation of the bubble nucleation rate from a given underlying particle physics model, see e.g.\,\cite{Croon:2020cgk,Chala:2024xll,Kierkla:2023von, Kierkla:2025qyz,Bernardo:2025vkz}. Since the phase transition happens near thermal equilibrium, the nucleation rate can be computed using equilibrium techniques, although out-of-equilibrium effects are a matter of ongoing investigation\,\cite{Gould:2024chm,Hirvonen:2024rfg}.
After the bubble nucleates, it expands and perturbs the plasma near its boundary, namely the bubble wall. 
The dynamics of the bubble is a genuinely out-of-equilibrium phenomenon and must be investigated using appropriate tools.
Of particular relevance is the terminal bubble wall velocity, which not only affects the gravitational wave signal produced during the FOPT but also significantly impacts FOPT-related phenomenology.

In this paper, we have conducted a first-principles study of the bubble wall dynamics based on nonequilibrium quantum field theory. We have derived the EoM of the bubble wall\,\eqref{eq:eom-final form}, using the 2PI effective action and the CTP formalism. Compared with the conventional condensate EoM\,\eqref{eq:condensate-eom}, we identify a new phenomenologically relevant term---the last two terms in \,\eqref{eq:eom-final form}---that describes the direct interaction between the bubble wall and the particles via condensate-particle vertices, e.g. $\varphi\phi\chi^2$ where $\varphi$ is the VEV describing the bubble wall. 

With our complete condensate EoM, we have shown how all the particle processes studied in the kick approach and listed in the Introduction of this work can be captured by the kinetic approach as well, thus filling a standing conceptual gap between the two approaches. Processes of type (1) are captured by the one-loop term in the condensate EoM, as shown in Ref.\,\cite{Ai:2024shx}.
In the ultrarelativistic wall regime, we have shown that the new term in the condensate EoM is responsible for the heavy dark matter production studied in Refs.\,\cite{Azatov:2021ifm,Azatov:2024crd,Ai:2024ikj} and friction on the bubble wall via the type $(4)$ processes. 
We have argued that processes of type $(2)-(3)$ can be captured by the conventional condensate EoM and the Boltzmann equations, provided one takes into account the $\varphi$-dependent masses in solving the Boltzmann equations. By adopting the VEV insertion approximation, we have shown explicitly how the expansion of the propagators in insertions of the background~$\varphi$ can generate new self-energy terms from the last term in Eq.\,\eqref{eq:condensate-eom}, which can then describe the 1-to-1 mixing and the 1-to-2 transition radiation.   

More generically, for slower bubble walls, the self-energy term is non-local, making the condensate EoM challenging to solve. We have performed a localisation procedure to bring it to a local form in Eq.\,\eqref{eq:eom-final form2}, that can readily be used in future numerical computations and incorporated into, e.g., {\tt WallGo}\,\cite{Ekstedt:2024fyq}.  
Although not the focus of this work, we believe that the $\varphi$-dependent vertices could also be important in the Boltzmann equations and should be considered in collision terms. This will be the focus of future investigations.

\section*{Acknowledgments}

It is a pleasure to thank Marieke Postma and Alberto Mariotti for useful discussions. We also thank Alberto Mariotti for comments on the draft. The work of WYA is supported by the European Union (ERC, NLO-DM, 101044443).
MV is supported by the ``Excellence of Science - EOS'' - be.h project n.30820817, and by the Strategic Research Program High-Energy Physics of the Vrije Universiteit Brussel. C.T. acknowledges support from the Cluster of Excellence ``Precision Physics, Fundamental Interactions, and
Structure of Matter'' (PRISMA+ EXC 2118/1) funded by the Deutsche Forschungsgemeinschaft (DFG, German Research Foundation) within the German Excellence Strategy
(Project No. 390831469).

\newpage

\begin{appendix}
\renewcommand{\theequation}{\Alph{section}\arabic{equation}}

\section{Derivation of the pressure induced by pair production}
\label{app:simplify}

In this Appendix we present the derivation of Eq.\,\eqref{eqn: p phi to chichi full in text} and of its asymptotic form in the ultrarelativistic limit, given in Eq.\,\eqref{eqn: pair pressure ultrarel}.

\subsection{Phase-space integral}
The integral in Eq.\,\eqref{eq:P-phi-to-2chi} is not a simple decay integral, as the conservation of momentum along the $z$ direction has been relaxed, and its evaluation must be done with care. 
First of all, we go back to Eq.\,\eqref{eqn: im pi phi to chichi}, plug it into Eq.\,\eqref{eq:Pvertex-simple} and use that the integrand is symmetric under~${q^z\to -q^z}$ as well as the sign flip of the incoming and outgoing momenta, to write
\begin{align}
    \P_{\phi\rightarrow\chi\chi} = - \frac{g^2}{2} \int_{-\infty}^\infty \frac{\d q^z}{2\pi} \,q^z |\widetilde{\varphi}(q^z)|^2 \int_{\vec p, \vec k_1,\vec k_2} (2\pi)^3 \, \delta^{(3)}(\vec q - \vecp+\vec{k}_{1}+\vec{k}_{2})\, (2\pi)\,\delta(E^{(\phi)}_\vecp - E^{(\chi)}_{\veck_1}- E^{(\chi)}_{\veck_2}) f_\phi(\vecp) \,,
\end{align}
where $\vecq= (0,0,q^z)$.
Using the three-momentum conserving $\delta$-function to get rid of the $\vec k_2$ integration and relabelling $\veck_1$ as $\veck$,  we arrive at
\begin{equation}
    \label{eqn: phi-chi-chi pressure in terms of I}
    \P_{\phi\rightarrow\chi\chi} = \: - \frac{g^2}{2} \int_{-\infty}^\infty \frac{\d q^z}{2\pi} \,q^z |\widetilde{\varphi}(q^z)|^2 \int_{\vec p} f_\phi(\vecp) I(\vec q, \vec p) \,,
\end{equation}
where we have defined
\begin{equation}
\label{eqn: k1 integral}
    I(\vec q, \vec p) = \: \int_{\vec k} \frac{1}{2E^{(\chi)}_{\vec q - \vec p + \vec k}}\,(2\pi)\,\delta\left(E^{(\phi)}_\vecp - E^{(\chi)}_{\veck}- E^{(\chi)}_{\vec q - \vec p +\vec k}\right)\,.
\end{equation}
To compute Eq.\,\eqref{eqn: k1 integral}, we introduce spherical coordinates for the momentum\,$\vec k$
\begin{equation}
    \int_{\vec k} = \: \frac{1}{(2\pi)^3}\int_0^\infty \d k \,  \frac{k^2}{2E^{(\chi)}_k}\int_0^{2\pi} \d\phi \int_{-1}^1 \d\cos\theta \,,
\end{equation}
and choose the angle\,$\theta$ such that
\begin{equation}
    (\vec q -\vec p)\cdot \vec k= \: |\vec q-\vec p| k \cos\theta \,.
\end{equation}
Then, we can use the energy conserving~$\delta$-function to localise the~$\cos\theta$ integral.
In fact, we can write
\begin{equation}
    \delta\left(E^{(\phi)}_\vecp - E^{(\chi)}_{\veck}- E^{(\chi)}_{\vec q - \vec p + \vec k}\right) = \: \frac{1}{|f'(\cos\theta^\star)|} \delta(\cos\theta-\cos\theta^\star)\,,
\end{equation}
where\,$\cos\theta^\star$ solves
\begin{equation}
    E^{(\phi)}_\vecp - E^{(\chi)}_{\veck}- E^{(\chi)}_{\vec q - \vec p + \vec k} = \: E^{(\phi)}_\vecp - E^{(\chi)}_{\veck} - \sqrt{|\vec q-\vec p|^2 + k^2 - 2 |\vec q-\vec p |k\cos\theta^\star+m_\chi^2} = \: 0\,,
\end{equation}
and
\begin{equation}
    f(\cos\theta^\star) = \: \sqrt{|\vec q-\vec p|^2 + k^2 - 2 |\vec q-\vec p |k\cos\theta^\star+m_\chi^2}\,.
\end{equation}
We find
\begin{equation}
    \cos\theta^\star = \: \frac{\left(E_k^{(\chi)}\right)^2 + |\vec q-\vec p|^2 - \left(E^{(\phi)}_\vecp - E^{(\chi)}_{k}\right)^2}{2|\vec q - \vec p|k} \,,
\end{equation}
and
\begin{equation}
    \frac{1}{|f'(\cos\theta^\star)|} = \: \frac{E^{(\phi)}_\vecp - E^{(\chi)}_{k}}{|\vec q- \vec p|k} \,.
\end{equation}
Plugging this into Eq.\,\eqref{eqn: k1 integral}, we find
\begin{align}
    I(\vec q, \vec p)
    = \: &\frac{1}{8\pi|\vec q - \vec p|} \int_0^\infty \d k \frac{k}{E_k^{(\chi)}} H (\vec q, \vec p; k)\,,
\end{align}
where the function~$H$ implements the constraints
\begin{equation}
    -1\leq\cos\theta^\star\leq1 \longrightarrow -1\leq \frac{\left(E_k^{(\chi)}\right)^2 + |\vec q-\vec p|^2 - \left(E^{(\phi)}_\vecp - E^{(\chi)}_{k}\right)^2}{2|\vec q - \vec p|k} \leq 1 \,,
\end{equation}
and
\begin{equation}
   0<  E_k^{(\chi)} < E_{\vec p}^{(\phi)} \,.
\end{equation}
The last inequality follows from restricting the~$k$-integral to the region where the energy-conserving~$\delta$-function has support.
We change the integration variable from $k$ to $E = E_k^{(\chi)}$, using $E\d E = k \d k$, so that we are left with computing the integral
\begin{equation}
    I(\vec q, \vec p) = \: \frac{1}{8\pi|\vec q - \vec p|} \int_{m_\chi}^{E_{\vec p}^{(\phi)}} \d E \, H (\vec q, \vec p; E) \,.
\end{equation}
Note that we already implemented the constraint on the energy as an upper bound for the integration. We should also translate the constraint\,$H$, onto a constraint on the variable\,$E$.
We have a chain of inequalities
\begin{equation}
   - 2|\vec q - \vec p| \sqrt{E^2 - m_\chi^2} \leq |\vec q - \vec p|^2 - (E_{\vec p}^{(\phi)})^2 + 2E_{\vec p}^{(\phi)} E \leq 2|\vec q - \vec p| \sqrt{E^2 - m_\chi^2} \,.
\end{equation}
It is convenient to introduce
\begin{equation}
    E_\Delta \equiv \: \frac{\left(E_{\vec p}^{(\phi)}\right)^2 - |\vec q - \vec p|^2}{2E_{\vec p}^{(\phi)}} \, \longrightarrow \, |\vec q - \vec p|^2 = \: E_{\vec p}^{(\phi)} \left( E_{\vec p}^{(\phi)} - 2E_\Delta\right)\,,
\end{equation}
together with the dimensionless variables~$x$,~$y$ and~$z$ defined through the following relations
\begin{equation}
    E_\Delta\equiv m_\chi x\,, \qquad E_{\vec p}^{(\phi)} \equiv m_\chi y \,, \qquad E \equiv m_\chi z\,.
\end{equation}
We can then re-write the constraints as
\begin{equation}
    - \sqrt{y-2x} \, \sqrt{z^2 - 1} \leq \sqrt{y} \, (z - x) \leq \sqrt{y-2x} \, \sqrt{z^2 - 1} \,,
\end{equation}
or equivalently
\begin{equation}
    \sqrt{y} \, |z - x| \leq \sqrt{y-2x} \, \sqrt{z^2 - 1} \,.
\end{equation}
The inequality is always well defined since by definition of variables\,$x$,\,$y$ and\,$z$ we have\,$y>2x$ and\,$z>1$, as well as\,$y>2$ which is imposed by the kinematics.
Squaring the inequality, one obtains a condition on\,$x$ and\,$z$, namely
\begin{equation}\label{eqn: inequality in xyz}
    P(x, y, z) \equiv z^2-yz-\left(1-\frac{xy}{2} - \frac{y}{2x}\right) \,,
 \qquad 2x \, P(x, y, z) \leq 0 \,.
\end{equation}
The quadratic polynomial\,$P(x, y, z)$ in\,$z$ has two roots
\begin{equation}
    z_{\pm} = \: \frac{y}{2} \left( 1 \pm \sqrt{1-\frac{2x}{y}} \sqrt{1-\frac{2}{xy}} \right) \,,
\end{equation}
between which the polynomial is negative.
If~$0<x<2/y$, the polynomial is always positive, and Eq.\,\eqref{eqn: inequality in xyz} then has no solutions. We are left to analyse the two cases~$x<0$ and~$2/y<x<y/2$ for which the two roots\,$z_\pm$ exist.
For~$2/y<x<y/2$, Eq.\,\eqref{eqn: inequality in xyz} is solved for~$z\in [z_-,z_+]\subset[1,y]$.
On the other hand, if~$x<0$, the inequality\,\eqref{eqn: inequality in xyz} is satisfied only for~$z>z_+$ and~$z<z_-$. Yet, if~$x<0$ we see that~$z_+\geq y$ and~$z_-<1$, meaning the inequality\,\eqref{eqn: inequality in xyz} is never satisfied in the integration domain.
So we compute the integral in Eq.\,\eqref{eqn: k1 integral} to be
\begin{align}
\label{eq:final_I}
    I(\vec q, \vec p) = \: & \frac{m_\chi}{8\pi|\vec q - \vec p|} \int_1^y \d z \, H(z) = \: \frac{m_\chi}{8\pi|\vec q - \vec p|} \vartheta(xy-2) (z_+ - z_-)
    \\ \notag 
    = \: & \vartheta\left( \left(E_{\vec p}^{(\phi)}\right)^2 - |\vec q - \vec p|^2 - 4m_\chi^2\right) \frac{1}{8\pi} \sqrt{1- \frac{4 m_\chi^2}{\left(E_{\vec p}^{(\phi)}\right)^2-|\vec q - \vec p|^2}}
\end{align}
We find that the integral is only nonzero if~${\left(E_{\vec p}^{(\phi)}\right)^2 \geq |\vec q- \vec p|^2 + 4m_\chi^2}$, a constraint that arises purely from the kinematics.

We now analyse what constraint arises on the momentum transfer~$q^z$ from this requirement, in particular recalling~$\vec q = (0,0,q^z)$ we have
\begin{equation}
    (q^z)^2- 2 q^zp^z + \Delta^2 \leq 0 \,, \qquad \mathrm{with} \quad \Delta^2 = \: 4m_\chi^2 - m_\phi^2
\end{equation}
which is solved by
\begin{equation}
    q^z\in [\tilde q_-,\tilde q_+]\,,\qquad \mathrm{where} \quad \tilde{q}_\pm = \: p^z \pm \sqrt{(p^z)^2 -\Delta^2}\,.
\end{equation}
This only has a solution if~$(p^z)^2\geq  \Delta^2$. For the moment, let us assume~$m_\phi^2<4m_\chi^2$, i.e.~$\Delta^2>0$. Note that this is in line with the rest of our calculations: we assumed the~$\chi$-particles to be very massive and thus absent in the plasma.
Then, the constraint on~$p^z$ translates into either~${p^z<-\Delta<0}$ or~${p^z>\Delta>0}$.
In the first case, the incoming particle is \emph{entering the bubble}. The momentum transfer~$q^z\in[\tilde{q}_-,\tilde{q}_+]$ is strictly negative, implying the wall loses momentum to the particle, and friction is generated. 
In the second case, the incoming particle is \emph{exiting the bubble}. The momentum transfer is then strictly positive, implying the wall is extracting momentum from the particle, thus accelerating.
As a final comment, we observe that if~${\Delta^2=4m_\chi^2-m_\phi^2<0}$ the decay is allowed at zero temperature. This manifests itself in the constraint on~$p_z$ being trivially satisfied, implying the process can happen at zero momentum transfer~$q^z=0$. Although this is a good sanity check of our result, we should stress that if the~$\chi$-particles are light, they are present in the plasma and other processes become equally or more relevant than pair production. We shall not consider this case in what follows.

We can now plug in our result for the integral~$I(\vec q,\vec p)$ into Eq.\,\eqref{eqn: phi-chi-chi pressure in terms of I} and obtain
\begin{align}
    \P_{\phi\rightarrow\chi\chi} = \: & - \frac{g^2}{32\pi^2} \int_{\vec p} f_\phi(\vecp) \times \Bigg\{ \vartheta(p^z-\Delta)\int_{q_-}^{q_+} \d q^z \,q^z |\widetilde{\varphi}(q^z)|^2 \sqrt{1- \frac{4 m_\chi^2}{m_\phi^2 + 2q^zp^z - (q^z)^2}} \notag \\
    & \qquad\qquad + \vartheta(-p^z-\Delta) \int_{-q_+}^{-q_-} \d q^z \, q^z |\widetilde{\varphi}(q^z)|^2 \sqrt{1- \frac{4 m_\chi^2}{m_\phi^2 + 2q^zp^z - (q^z)^2}} \Bigg\} \,,
\end{align}
where we have defined
\begin{align}
   q_\pm = \: |p^z| \pm \sqrt{(p^z)^2 -\Delta^2}\,.
\end{align}
In particular, this implies that the constraint on the component of the momentum through the wall~$p^z$ is not trivial.
By changing variable~$q^z\to-q^z$ in the second term, we arrive at the final form for the pressure induced by pair production
\begin{align}
    \label{eqn: p phi to chichi full}
    \P_{\phi\rightarrow\chi\chi} = \: & \frac{g^2}{32\pi^2} \int_{\vec p} f_\phi(\vecp) \times \left[ \vartheta(-p^z-\Delta) - \vartheta(p^z-\Delta) \right]\int_{q_-}^{q_+} \d q^z \,q^z |\widetilde{\varphi}(q^z)|^2 \sqrt{1- \frac{4 m_\chi^2}{m_\phi^2 + 2q^z|p^z| - (q^z)^2}} \,.
\end{align}
We observe once again that the term with positive~$p^z$ momentum describes particles exiting the bubble that lose momentum to the wall and, in fact, induce a negative pressure.

\subsection{Ultrarelativistic limit}
Up to now, our calculation is general. Next, we focus on ultra-relativistic walls, to show that we recover the known results in the literature.
It is convenient to introduce the functions
\begin{subequations}
\begin{align}
    F(p^z) = \: & \int \frac{\d^2 \vec p_\perp}{(2\pi)^3 2E_{\vec p}^{(\phi)}} \left[f_\phi(\vec p_\perp,-p^z) - f_\phi(\vec p_\perp,p^z) \right] \,, \label{eqn: definition F} \\    
    G(p^z) = \: & \int_{q_-}^{q_+} \d q^z \,q^z |\widetilde{\varphi}(q^z)|^2 \sqrt{1- \frac{4 m_\chi^2}{m_\phi^2 + 2q^z |p^z| - (q^z)^2}} \,, \label{eqn: definition G}
\end{align}
\end{subequations}
so that the pressure takes the form
\begin{equation}
    \label{eqn: pressure pair in terms of F and G}
    \P_{\phi\rightarrow\chi\chi} = \: \frac{g^2}{32\pi^2} \int_{\Delta}^\infty \d p^z \, F(p^z) G(p^z) \,,
\end{equation}
where we recall~$\Delta=\sqrt{4m_\chi^2-m_\phi^2}>0$.
Note that in going from Eq.\,\eqref{eqn: p phi to chichi full} to Eq.\,\eqref{eqn: pressure pair in terms of F and G} we split the integration domain over positive and negative~$p^z$, and changed~$p^z\to-p^z$ in the latter.
The integral in Eq.\,\eqref{eqn: pressure pair in terms of F and G} cannot, in general, be computed analytically.
We can, however, expand the function~$G$ around a special point and obtain an asymptotic result for the integral valid at least in the limit~$\gamma_w\gg1$.
Let us expand around a generic point~$p_0^z$
\begin{equation}
    G(p^z) = \: G(p_0^z) + G'(p_0^z) (p^z - p_0^z) + \frac{1}{2} G''(p_0^z) (p^z-p_0^z)^2 + \ldots \,,
\end{equation}
and plugging this into Eq.\,\eqref{eqn: pressure pair in terms of F and G} we find
\begin{equation}
    \P_{\phi\rightarrow\chi\chi} = \: \frac{g^2}{32\pi^2} \left\{ G(p_0^z) \int_{\Delta}^\infty \d p^z \, F(p^z) + G'(p_0^z) \int_{\Delta}^\infty \d p^z \, (p^z - p_0^z) F(p^z) + \ldots \right\}\,.
\end{equation}
For the linear term in the expansion to vanish, we see immediately that~$p_0^z$ should be the average~$z$-momentum, namely
\begin{equation}
    p_0^z = \langle p^z \rangle \equiv \: \mathcal{N}_F^{-1} \int_{\Delta}^\infty \d p^z \, p^z F(p^z)\,,
\end{equation}
having defined the normalisation constant
\begin{equation}
    \mathcal{N}_F = \: \int_{\Delta}^\infty \d p^z \,F(p^z) \,.
\end{equation}
Then, choosing to expand the function~$G$ around the average~$z$ component of the momentum of the incoming particle, we have
\begin{equation}
    \label{eqn: pressure pair cumulant expansion}
    \P_{\phi\rightarrow\chi\chi} = \: \frac{g^2}{32\pi^2} \mathcal{N}_F \left\{ G(\langle p^z\rangle) + \frac{1}{2} G''(\langle p^z\rangle) \left(\langle (p^z)^2\rangle - \langle p^z\rangle^2 \right) + \ldots \right\}\,,
\end{equation}
where the expectation value of a polynomial~$Q(p^z)$ has been defined as
\begin{equation}
    \langle Q(p^z) \rangle = \: \mathcal{N}_F^{-1} \int_{\Delta}^\infty \d p^z \, Q(p^z) F(p^z)\,.
\end{equation}
We refer to the expansion in Eq.\,\eqref{eqn: pressure pair cumulant expansion} as a \emph{cumulant expansion}. Note that it is not guaranteed that this expansion is convergent or that it can be organised in terms suppressed with~$\gamma_w$, and in fact, it will generally not be. We will come back to this point when we compare our analytical formula to the numerical results.

So far we have not made use of working in the ultrarelativistic limit. Our next step is to find the normalisation~$\mathcal{N}_F$ and the average~$z$-momentum~$\langle p^z\rangle$ when~$\gamma_w\gg1$.
We use the equilibrium distribution functions in the wall frame, where the plasma is boosted by a velocity~$-v_w$
\begin{equation}
    f_\phi(p_\perp, p^z) = \: \left(\e^{\frac{\gamma_w}{T} \left(E_{\vec p}^{(\phi)} + v_w p^z\right)} - 1\right)^{-1} \,,
\end{equation}
where we recall that the energy is~$E_{\vec p}^{(\phi)} = \: \sqrt{p_\perp^2+(p^z)^2+m_\phi^2}$.
The function~$F$ as defined in Eq.\,\eqref{eqn: definition F} can be then found analytically
\begin{align}
    F(p^z) = \: & \int_0^\infty \frac{\d p_\perp \, p_\perp}{(2\pi)^2 2 E_{\vec p}^{(\phi)}} \left[f_\phi (p_\perp, - p^z) - f_\phi (p_\perp, p^z) \right] \notag \\
    = \: & - \frac{T}{8\pi^2\gamma_w} \left[  \log\left( 1- \e^{-\frac{\gamma_w}{T} \left( \sqrt{(p^z)^2 + m_\phi^2} -v_w p^z\right)} \right) - \log\left( 1- \e^{-\frac{\gamma_w}{T} \left( \sqrt{(p^z)^2 + m_\phi^2} + v_w p^z\right)} \right) \right] \,.
    \label{eqn: analytical F}
\end{align}
In the limit~$\gamma_w\gg1$, the second term is always exponentially suppressed and never contributes. For completeness, we perform all our calculations keeping it, and will only remove it after integration.
We recall now that we always assume~$\phi$ particles to be light, in particular~$m_\phi\ll m_\chi\ll\gamma_wT$. In the following, we drop~$m_\phi$ from the function~$F(p^z)$, allowing us to perform the next step analytically. 
Corrections can be computed perturbatively or included numerically, as we will demonstrate below.
By calling~${c_\pm = \gamma_w(1\pm v_w) = \sqrt{(1\pm v_w)/(1\mp v_w)}}$, we compute the normalisation factor analytically
\begin{align}
    \mathcal{N}_F = \: & \int_{\Delta}^\infty \d p^z \, F(p^z) = \: - \frac{T^2}{8\pi^2\gamma_w} \left[ c_+ \int_{c_-T\Delta}^\infty \d x \, \log(1-\e^{-x}) - c_- \int_{c_+T\Delta}^\infty \d x \, \log(1-\e^{-x}) \right] \notag \\
    = \: & \frac{T^2}{8\pi^2\gamma_w} \left[ c_+ \mathrm{Li}_2(\e^{-c_-T\Delta}) - c_- \mathrm{Li}_2(\e^{-c_+T\Delta}) \right] \,.
\end{align}
In the ultrarelativistic limit, namely~$\gamma_w\gg1$, we have~$c_-\simeq1/(2\gamma_w)\ll1$ and~$c_+\simeq2\gamma_w\gg1$. 
Since the dilog function for small argument is linear,~$\mathrm{Li}_2(x) = x + \mathcal{O}(x^2)$, the second term is of order~$\e^{-2\gamma_wT\Delta}$ and can be dropped.
Expanding the first term, we find the normalisation constant
\begin{align}
    \mathcal{N}_F = \: & \frac{T^2}{8\pi^2\gamma_w} \left[ \frac{\pi^2}{3}\gamma_w - \frac{\Delta}{T} \log \left( \frac{2\mathbb{e}\gamma_wT}{\Delta} \right) + \mathcal{O} (\gamma_w^{-1}) \right] = \: \frac{T^2}{24} + \mathcal{O}(\gamma_w^{-1} \log\gamma_w)\,.
    \label{eqn: normalisation constant F}
\end{align}
Next, we turn to the computation of the average~$z$ momentum
\begin{align}
    \langle p^z \rangle = \: & \mathcal{N}_F^{-1} \int_{\Delta}^\infty \d p^z \, p^z F(p^z) \notag \\
    = \: & \left(\frac{24}{T^2} + \mathcal{O}(\gamma_w^{-1})\right) \left( -\frac{T^3}{8\pi^2\gamma_w}\right) \left[ c_+^2 \int_{c_-T\Delta}^\infty \d x \, x \log(1- \e^{-x}) - c_-^2 \int_{c_+T\Delta}^\infty \d x \, x \log(1- \e^{-x}) \right] \notag \\
    = \: & \left(\frac{24}{T^2} + \mathcal{O}(\gamma_w^{-1})\right) \frac{T^3}{8\pi^2\gamma_w} \left[ c_+^2 \left( c_-T\Delta \mathrm{Li}_2 (\e^{-c_-T\Delta}) + \mathrm{Li}_3(\e^{-c_-T\Delta}) \right) - (+ \leftrightarrow -) \right] \notag \\
    = \: & \frac{12\zeta_3}{\pi^2} \gamma_w T + \mathcal{O}(\gamma_w^{-1} \log\gamma_w) \,.
    \label{eqn: average pz momentum}
\end{align}
In the following, we call the proportionality constant~${\sigma\equiv12\zeta_3/\pi^2=1.46\ldots}$.

Having found the normalisation~$\mathcal{N}_F$ and the average~$z$ momentum~$\langle p^z\rangle$, we are now ready to compute the leading order term in the cumulant expansion of Eq.\,\eqref{eqn: pressure pair cumulant expansion}. In particular, we must evaluate the function~$G$ as defined in Eq.\,\eqref{eqn: definition G} at the value~$p^z = \sigma \gamma_wT$ and expand for large~$\gamma_w$.
In the limit~$\gamma_w\gg 1$, the~$q$ integration limits become
\begin{equation}
    \label{eqn: ultrarelativistic bounds for q integral}
     [q_-,q_+] \to \left[\frac{\Delta^2}{2\sigma\gamma_w T},2\sigma\gamma_w T\right]\,.
\end{equation}
As for the square root factor,
\begin{equation}
     \sqrt{1- \frac{4m_\chi^2}{m_\phi^2+2 q^z\sigma\gamma_wT - (q^z)^2}} \longrightarrow \: 1 - \frac{1}{2}\frac{4m_\chi^2}{2q^z\sigma\gamma_wT} + \mathcal{O}(\gamma_w^{-2})\,,
\end{equation}
where in the last step we must use that the Fourier transform of the bubble~$|\widetilde{\varphi}(q^z)|^2$ is heavily suppressed for momenta~$q\simeq \gamma_wT$, which follows from the wall being thick in the wall frame.
Now using~$m_\phi\ll m_\chi$, we replace~$\Delta\to2m_\chi$, and the function~$G$ computed at~$\langle p^z\rangle$ becomes
\begin{align}
    G(\langle p^z \rangle) = \: & \int_{\frac{2m_\chi^2}{\sigma\gamma_wT}}^{2\sigma\gamma_wT} \d q^z \,q^z |\widetilde{\varphi}(q^z)|^2  \left(  1 - \frac{m_\chi^2}{q^z\sigma\gamma_wT} \right)  \,.
    \label{eqn: G ultrarel}
\end{align}
Plugging this back inside Eq.\,\eqref{eqn: pressure pair in terms of F and G}, we find the pressure in the ultrarelativistic limit to be
\begin{equation}
    \mathcal{P}_{\phi\to\chi\chi}^{\gamma_w\to\infty} = \: \frac{g^2}{32\pi^2} \frac{T^2}{24} \int_{\frac{2m_\chi^2}{\sigma\gamma_wT}}^{2\sigma\gamma_w T} \d q^z \,q^z |\widetilde{\varphi}(q^z)|^2\,.
    \label{eqn: pressure pair ultrarelativistic limit}
\end{equation}
Using a step wall approximation, we recover the result presented in Eq.~(93) of\,\cite{Azatov:2020ufh}, up to a log which was missed, see\,\cite{Ai:2023suz}.

A more realistic wall profile is the hyperbolic tangent
\begin{align}
\label{eq:wall-profile}
      \varphi(z)= \frac{v_b}{2}\left[1-\text{tanh}\left(\frac{z}{L_w}\right) \right] \, .
\end{align}
Its Fourier transform reads\footnote{Note that the above expression differs from that given in Ref.\,\cite{Ai:2023suz}. The mentioned reference overlooked a difference between the used Fourier transform $\widetilde{\varphi}(q^z)=\int \d z\,\e^{-\i q^z z} \varphi(z)$ and the default definition in Mathematica: $\widetilde{\varphi} (q^z) = \int \frac{\d z}{\sqrt{2\pi}}\,\e^{\i q^z z} \varphi(z)$. \label{fn:Fourier-transform} }
\begin{align}
  \label{eq:shape_of_wall}
      \widetilde \varphi(q^z) = \frac{\i v_b L_w  \pi  }{2} \text{csch} \bigg(\frac{\pi L_w q^z}{2}\bigg) + v_b \pi \delta(q^z)\,,
  \end{align}
where the term with $\delta(q^z)$ is irrelevant for the present purpose as it leads to no momentum transfer between the particles and the wall, thus not contributing to\,${\cal P}_{\rm vertex}$, see Eq.\,\eqref{eq:Pvertex-simple}. 
Inserting Eq.\,\eqref{eq:shape_of_wall} into Eq.\,\eqref{eqn: G ultrarel}, we obtain 
\begin{align}
    G(\langle p^z \rangle) = \: &
    v_b^2 \int_{\frac{2m_\chi^2}{\sigma\gamma_wT}}^{2\sigma\gamma_wT} \d q^z \,q^z \frac{ \pi^2L_w^2 }{4} \times \text{csch}^2 \bigg(\frac{\pi L_w q^z }{2}\bigg) \left(  1 - \frac{m_\chi^2}{q^z\sigma\gamma_wT} \right) \notag \\
    = \: & v_b^2 \int_{\xi_0}^{\xi_1} \d \xi \, \xi \, \mathrm{csch}^2 \xi \left( 1 - \frac{\xi_0}{2\xi} \right)
    \,,
\end{align}
where~$\xi_0 = \: \pi L_w m_\chi^2/(\sigma\gamma_wT)$ and~${\xi_1=\pi L_w \sigma\gamma_w T = (\pi L_w m_\chi)^2/\xi_0}$.
We can compute the integral analytically, and expand it for small~$\xi_0$ and large~$\xi_1$. We find
\begin{align}
    \int_{\xi_0}^{\xi_1} \d \xi \, \xi \, \mathrm{csch}^2 \xi  \left( 1 - \frac{\xi_0}{2\xi} \right)= \: & \xi_0 \coth \xi_0 - \xi_1 \coth\xi_1 + \log \frac{\sinh\xi_1}{\sinh\xi_0} -\frac{\xi_0}{2} \left(\coth\xi_0 - \coth\xi_1\right) \notag \\
    = \: & 1 - \log 2\xi_0 - \frac{1}{2} + \mathcal{O}\left(\xi_0,\e^{-2\xi_1}\right) \notag \\
    = \: & \log \frac{\sqrt{\e} \sigma\gamma_w T}{2\pi L m_\chi^2} + \mathcal{O}\left( \frac{\pi L m_\chi^2}{\gamma_wT}, \e^{-2\pi L \sigma\gamma_w T} \right) \,.
\end{align}
Putting everything together, we have an analytic expression for the contribution to the pressure from pair production in the ultra-relativistic limit, and it reads
\begin{equation}
    \label{eqn: ultrarelativistic p phi to chichi}
    \P^{\gamma_w \to\infty}_{\phi\rightarrow\chi\chi} = \, \frac{g^2 v_b^2}{32\pi^2} \frac{T^2}{24} \log\left( \frac{\sqrt{\e} \sigma\gamma_w T}{2\pi L_w m_\chi^2} \right) + \mathcal{O}(\gamma_w^{-1} \log\gamma_w) \,.
\end{equation}

\subsection{Comparing exact numerics with asymptotic analytics}

We now compare the analytic asymptotic formula we have obtained in Eq.\,\eqref{eqn: ultrarelativistic p phi to chichi} with the full expression for the pressure given in Eq.\,\eqref{eqn: p phi to chichi full}, which we compute numerically.
For the numerical evaluation, we use the formula for the pressure in Eq.\,\eqref{eqn: pressure pair in terms of F and G}, where the analytical expression for~$F$ is given in Eq.\,\eqref{eqn: analytical F}. Just as we did in our analytical expansion, we take~${m_\phi\to0}$ and consider a tanh wall profile. Note that different values of the mass~$m_\phi$ are taken into account in Fig.\,\ref{fig: pressure at different masses} in Section\,\ref{subsec: friction from pair production}.

\begin{figure}
    \centering
    \includegraphics[width=0.5\textwidth]{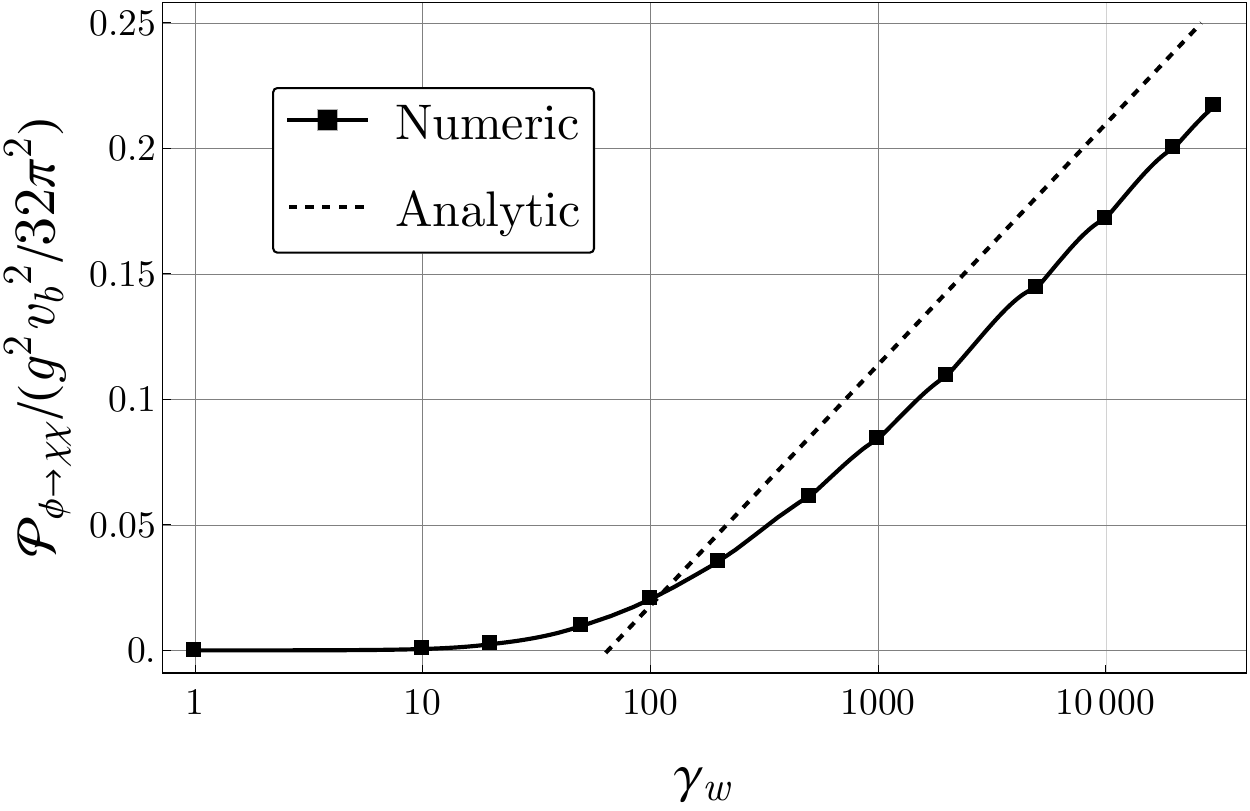}
    \caption{Pressure induced by pair production obtained via the numerical evaluation of Eq.\,\eqref{eqn: pressure pair in terms of F and G} (continuous) compared to the analytical approximation in Eq.\,\eqref{eqn: ultrarelativistic p phi to chichi}. Reference values~$T=1$,~$L_w=1$ and~$m_\chi=5$. Black squares represent the values used for the interpolation.}
    \label{fig: numerics vs analytics with offset}
\end{figure}

Figure~\ref{fig: numerics vs analytics with offset} shows the comparison between the numerical evaluation of the full result and the analytical approximation of the pressure for ultrarelativistic walls.
The temperature has been chosen as the reference scale by setting~$T=1$. The figure shows the results for some reference values for~$L_w$ and~$m_\chi$. Changing these values within the validity of our approximations does not affect the qualitative behaviour.
Alas, the asymptotic behaviour is the same, but we are off by a constant term!
And yet we made sure to keep our expansion under control, and we estimate the error to be of~$\mathcal{O}(\gamma_w^{-1}\log\gamma_w)$.

To explain the constant offset that we observe in Fig.\,\ref{fig: numerics vs analytics with offset}, we look back at the expansion in Eq.\,\eqref{eqn: pressure pair cumulant expansion} with greater suspicion and wonder whether this made sense at all.
The asymptotic behaviour of the second derivative of~$G$ for~$\gamma_w\gg1$ is readily evaluated
\begin{equation}
    G(\langle p^z\rangle) \overset{\langle p^z\rangle \gg1}{\sim} \log \langle p^z\rangle \sim \log \gamma_w  \longrightarrow G''(\langle p^z\rangle) \sim \frac{1}{\gamma_w^2} \,,
\end{equation}
which seems to suggest that the expansion is convergent in the ultrarelativistic limit.
However, let us compute the second cumulant of~$F$, which is multiplying~$G''(\langle p^z\rangle)$ in the cumulant expansion of Eq.\,\eqref{eqn: pressure pair cumulant expansion}
\begin{align}
    \langle (p^z)^2 \rangle = \: & \left( \frac{24}{T^2} + \mathcal{O}(\gamma_w^{-1}) \right) \left( -\frac{T^4}{8\pi^2\gamma_w}\right) \left[ c_+^3 \int_{c_-\Delta}^\infty \d x \, x^2 \log (1 - e^{-x} ) - ( + \leftrightarrow - ) \right] \notag \\
    = \: & \frac{8\pi^2}{15} \gamma_w^2T^2 + \mathcal{O}(\gamma_w)\,,
\end{align}
from which we compute the second cumulant to leading order in the Lorentz factor
\begin{equation}
    \langle (p^z)^2 \rangle - \langle p^z\rangle^2 = \: \left( \frac{8\pi^2}{15} - \left( \frac{12\zeta_3}{\pi^2}\right)^2  \right) \gamma_w^2T^2 = \: 3.13 \ldots\times \gamma_w^2T^2\,,
\end{equation}
which multiplied by the second derivative~$G''(\langle p^z\rangle)$ yields a factor of order one.
More dramatically, we can easily convince ourselves that each higher-order term will also be of the same order, namely independent of~$\gamma_wT$ in the ultrarelativistic limit. 
The series can still be expected to converge, thanks to the factorial suppression of higher-order terms, but it will converge to a constant term that cannot be dropped.
This means that the cumulant expansion of Eq.\,\eqref{eqn: pressure pair cumulant expansion} in the ultrarelativistic limit can be rewritten as
\begin{equation}
    \P_{\phi\rightarrow\chi\chi} = \: \frac{g^2}{32\pi^2} \mathcal{N}_F \left\{ G(\langle p^z\rangle) + \mathcal{O}(1) \right\}\,.
\end{equation}
We understand that the cumulant expansion in Eq.\,\eqref{eqn: pressure pair cumulant expansion} cannot be organised in terms that are less and less relevant as~$\gamma_w$ grows, as only the leading order term depends on the Lorentz factor in this limit. It follows, that the constant~$\mathcal{O}(1)$ term cannot possibly be fixed in this expansion, as it would require computing all higher-order terms.
Thus, it is no surprise that in Fig.\,\ref{fig: numerics vs analytics with offset}, we see an offset by a constant, and it would have been baffling otherwise.

\begin{figure}
    \centering
    \includegraphics[width=0.495\textwidth]{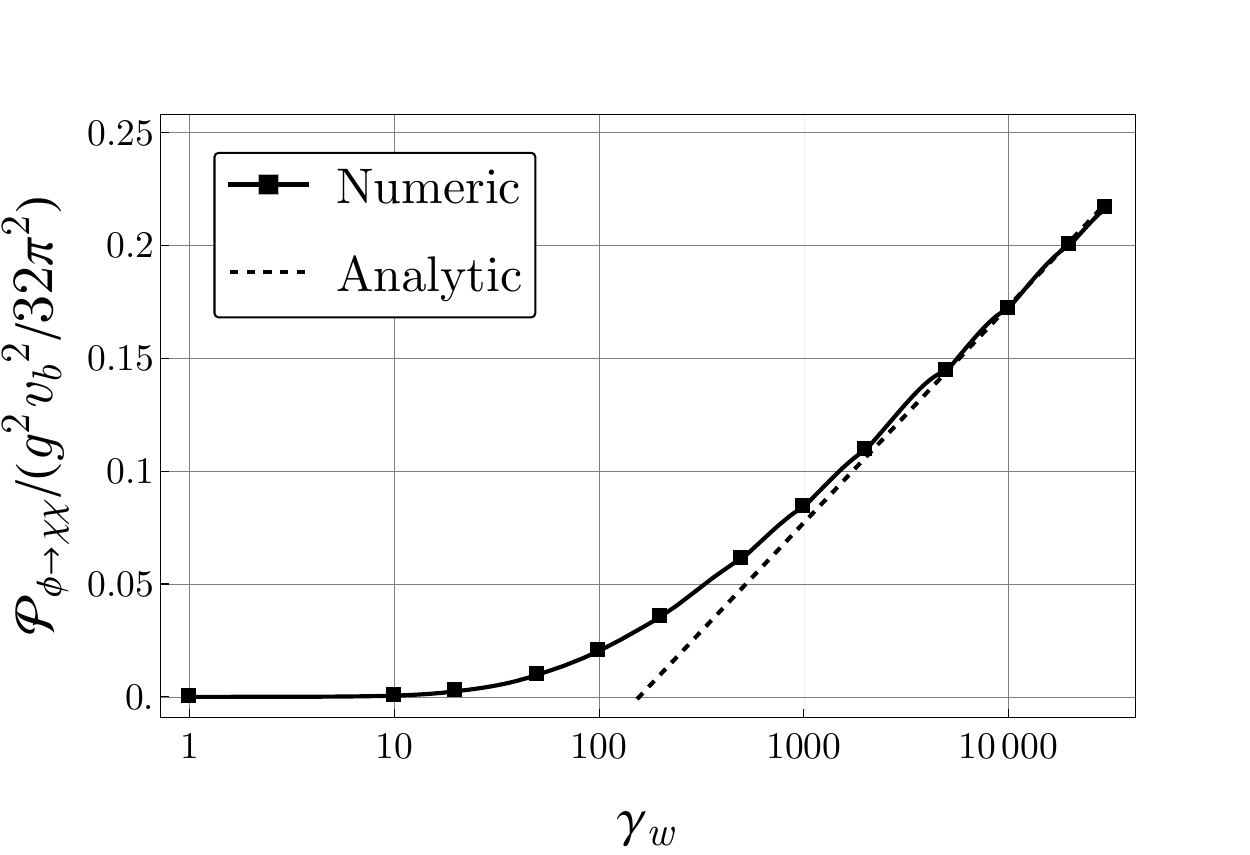}
    \includegraphics[width=0.495\textwidth]{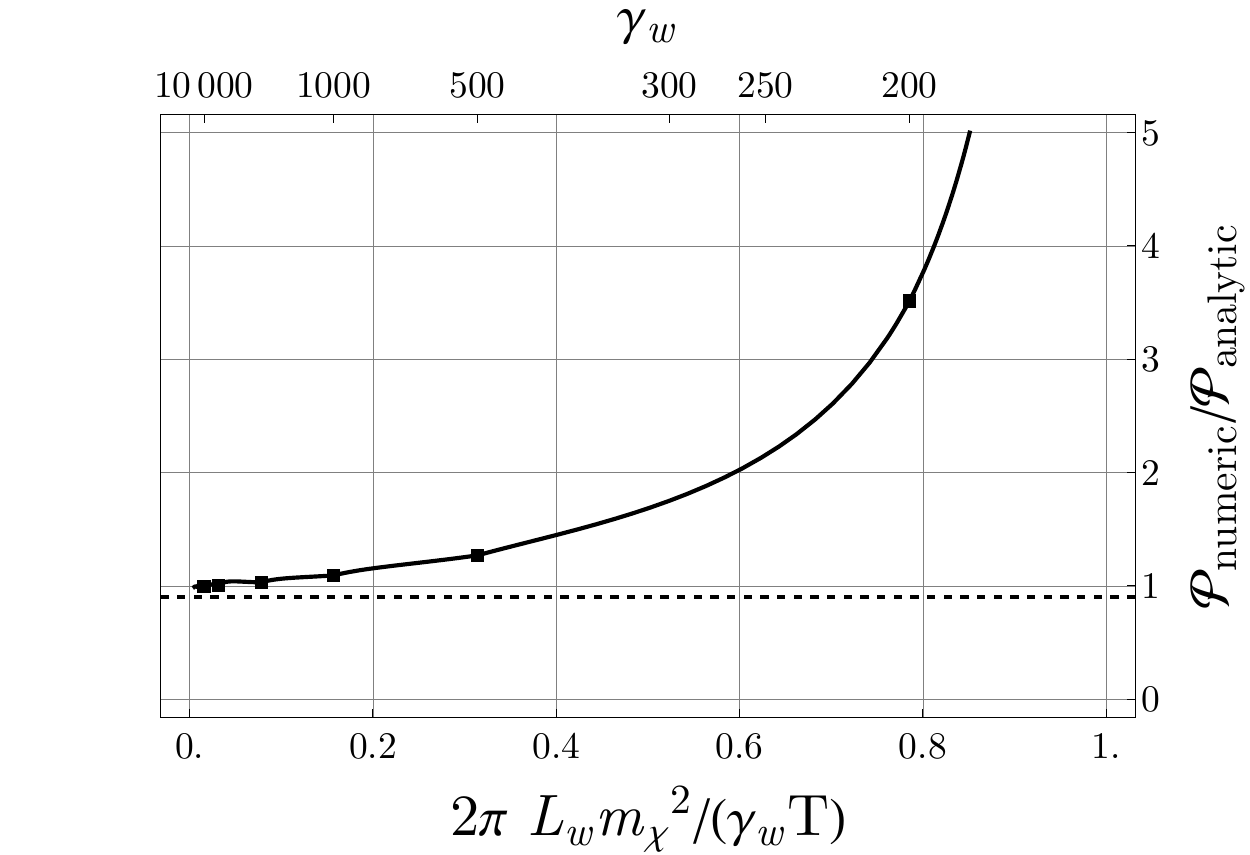}
    \caption{\textbf{Left:} Pressure induced by pair production obtained via the numerical evaluation of Eq.\,\eqref{eqn: pressure pair in terms of F and G} (continuous) compared to the analytical approximation in Eq.\,\eqref{eqn: ultrarelativistic p phi to chichi solving offset}. Reference values~$T=1$,~$L_w=1$ and~$m_\chi=5$. Black squares represent the values used for the interpolation.
    \textbf{Right:} Ratio of the pressure obtained numerically and analytically in the region where the latter expression is valid.}
    \label{fig: numerics vs analytics}
\end{figure}

To make the comparison between numerical and analytical more quantitative, we fix the constant by demanding that the numerical and the analytical estimates overlap in the asymptotic region~$\gamma_w\ggg1$. Interestingly, it is sufficient to eliminate the factor~$\sqrt{e}\sigma$ from the argument of the logarithm in Eq.\,\eqref{eqn: ultrarelativistic p phi to chichi}, namely, we use the pressure
\begin{equation}
    \label{eqn: ultrarelativistic p phi to chichi solving offset}
    \P^{\gamma_w \to \infty}_{\phi\rightarrow\chi\chi} \approx \, \frac{g^2 v_b^2}{32\pi^2} \frac{T^2}{24} \log\left( \frac{\gamma_w T}{2\pi L_w m_\chi^2} \right) \,.
\end{equation}
The comparison is then plotted in Fig.\,\ref{fig: numerics vs analytics}.
We see that even for large relativistic factors~$\gamma_w\sim250$ the analytic formula underestimates the contribution of pair production by a factor of two.

\section{Field expansion for the two-point functions}
\label{app:expansion_Boltz}

In this appendix, we show that the expansion of the CTP propagators that we have performed in the main text fulfills Dyson-Schwinger equations order by order in the coupling expansion and the VEV insertion approximation. 

\subsection{Scalar propagator}

We start from the tree-level EoM for the two-point functions
\begin{align} 
\label{eq:KB_off-diag}
-\sum_j(\Box_x\delta_{ij} + m_i^2\delta_{ij} + \delta M_{ij}(\varphi) ) \Delta^{ab}_{\, jk }(x,y; \varphi) = \i\,a \delta_{ik} \delta^{ab} \delta^{(4)}(x-y) \,, 
\end{align}
where~$\delta M_{ij}$ includes perturbatively small off-diagonal components that are responsible for the mixing.
We now show that Eq.\,\eqref{eq:KB_off-diag} is satisfied with the expansion presented in e.g., Eq.\,\eqref{eq:expansion_exact}, namely
\begin{align}
\label{eq:expansion_exact_bis}
\Delta^{ab}_{jk}(x_1,x_2;\varphi) = \Delta^{ab}_{jk}(x_1,x_2;0)
-\i  \sum_{c, \kappa, \ell} c \,\int \d^4 x'\,   \Delta^{ac}_{j\kappa}(x_1, x';0) \delta M_{\kappa\ell} \Delta^{cb}_{\ell k}(x',x_2;0) + \mathcal{O}\bigg(\frac{\delta M_{jk}}{m_i^2}\bigg)^2\,.
\end{align}

At zeroth order in the mixing, the two-point functions~$\Delta^{ab}_{ij}(x,y; 0)$ fulfill
\begin{align} 
-\sum_{j}\delta_{ij}( \Box_x + m_i^2 ) \Delta^{ab}_{\, jk}(x,y; 0) = \i\,a\delta_{ik} \delta^{ab} \delta^{(4)}(x-y) \, . 
\end{align}
By applying the operator~$\Box_x\delta_{ij} + m_i^2\delta_{ij} + \delta M_{ij}(\varphi)$ on Eq.\,\eqref{eq:expansion_exact_bis}, one obtains at leading order in the expanding parameter 
\begin{align}
&-\sum_j(\Box_{x}\delta_{ij} + m_i^2\delta_{ij} + \delta M_{ij}(\varphi) )\Delta^{ab}_{ jk }(x,y;\varphi) = \i\,a \delta_{ik}\delta^{ab} \delta^{(4)}(x-y) -\sum_j\delta M_{ij}(\varphi) \Delta^{ab}_{ \, jk }(x,y;0) \notag\\
&\qquad\qquad\qquad\qquad\quad +\i\,\sum_{c, j \kappa, \ell} c \,\int \d^4 z \, \delta_{ij}( \Box_{x} + m_i^2 ) \Delta^{ac}_{\, j\kappa }(x, z;0) \delta M_{\kappa\ell}(\varphi)  \Delta^{cb}_{\ell \, k}(z,y;0) \,.
\end{align}
The last two terms cancel, and Eq.\,\eqref{eq:KB_off-diag} is satisfied.

\subsection{Gauge boson propagator}

Let us first restate the Lagrangian of the model containing gauge bosons and fermions
\begin{align} 
\mathcal{L} = - \frac{1}{4} F_{\mu \nu}F^{\mu \nu} -g_1\bar \psi \gamma_\mu A^\mu \psi + \frac{g_2^2 \varphi^2}{4} A_\mu A^\mu \, . 
\end{align}

\paragraph{Free expansion.}
In the main text, we performed two controlled expansions of the two-point functions of the gauge bosons in Eq.\,\eqref{Eq:free_expansion} (around a free theory $g_1 \to 0$) and Eq.\,\eqref{eq:DeltaA-expansion} (around a massless gauge boson $g_2 \to 0$). The EoM governing the exact evolution of the two-point functions is given in the Feynman gauge $\xi=1$ by
\begin{align}
\label{eq:DeltaA-EoM}
&\Box_x\Delta_A^{\mu\nu, ab}(x, y)-\sum_c c \int \d^4z\, \Pi_A^{\mu\alpha, ac} (x, z)\Delta_A^{\alpha\nu,cb}(z, y)=\i a\delta^{ab} {\eta^{\mu\nu}}\delta^{(4)} (x-y)\,,
\end{align}
where we defined the {\it particle self-energies}\footnote{Note that the condensate self-energy $\Pi_\varphi$, defined in Eq.\,\eqref{eq:Piphiab}, and the particle self-energy $\Pi_\phi$ are distinguished from each other, as pointed out in Ref.\,\cite{Ai:2023qnr}. }
\begin{align}
\label{eq:Piphiab_bis}
    \Pi^{\alpha\beta, cd}_A(x,x')&\equiv - (cd) \frac{\delta\Gamma_2[\varphi, \Delta_A]}{\delta\Delta_A^{\alpha\beta, dc}(x',x)}\notag\\
    &=\i g_1^2{\rm tr_s} S^{cd}_{\psi;\rm free}(x_1, x_2)\gamma_{\beta} S^{dc}_{\psi;\rm free}(x_2, x_1)\gamma_{\alpha} +\O(g^4_1) \,.
\end{align}

We first consider the expansion in the fermion-gauge-boson interaction, setting $\varphi=0$. The free propagator satisfies 
\begin{align}
\Box_x\Delta^{\mu\nu, ab}_{A;\rm free}(x, y;0)= \i a\delta^{ab} \eta^{\mu\nu}\delta^{(4)} (x-y)\, .
\end{align}
We now show that the following expression, 
\begin{align}
\label{Eq:free_expansion2}
    &\Delta^{\mu\nu, ab}_A(x,y;0) =  \Delta_{A,\rm free}^{\mu\nu, ab}(x,y)\notag\\
    &+ g_1^2{\rm tr_s} \sum_{cd}(cd)\int\d^4 x_1  \d^4 x_2\, \Delta_{A;\rm free}^{\mu\alpha,ac}(x, x_1)  S^{cd}_{\psi;\rm free}(x_1, x_2)\gamma_{\beta} S^{dc}_{\psi;\rm free}(x_2, x_1)\gamma_{\alpha} \Delta^{\beta\nu,db}_{A;\rm free}(x_2, y)+\O(g^4_1)\,,
\end{align}
satisfies the EoM~\eqref{eq:DeltaA-EoM}.
Applying the operator $\Box_x$ on the above equation, one obtains
\begin{align}
&\Box_x\Delta^{\mu\nu, ab}_{A}(x,y) =  \i c^{ab}\eta^{\mu\nu}\delta^{(4)} (x-y)\notag\\
&+ g_1^2 \sum_{cd}(cd)\,{\rm tr_s} \int\d^4 x_1  \d^4 x_2\, \Box_x\Delta_{A;\rm free}^{\mu\alpha, ac}(x, x_1) S^{cd}_{\psi;\rm free}(x_1, x_2) \gamma_\beta S^{dc}_{\psi;\rm free}(x_2, x_1)\gamma_\alpha \Delta^{\beta\nu, db}_{A;\rm free}(x_2, y) \notag\\
&=\i a \delta^{ab}\eta^{\mu\nu} \delta^{(4)} (x-y) + \i g_1^2 \sum_{cd}(cd) a\delta^{ac} \delta^{\mu\alpha} {\rm tr_s}\int  \d^4 x_2\, S^{cd}_{\psi;\rm free}(x, x_2)\gamma_\beta S^{dc}_{\psi;\rm free}(x_2, x)\gamma_\alpha \Delta^{\beta\nu, db}_{A;\rm free}(x_2, y)
    \notag\\
&=\i c^{ab} \eta^{\mu\nu} \delta^{(4)} (x-y) + \sum_{d}d\int  \d^4 x_2\, {\Pi_A^{\mu\beta,ad}} (x, x_2) \Delta_{A}^{\beta\nu, db}(x_2, y)+\O(g_1^4)\,.
\end{align}

\paragraph{Field expansion.}
Next, let us turn to the field expansion. We will now show that the expansion in Eq.\,\eqref{eq:DeltaA-expansion}, 
\begin{align}
\label{eq:DeltaA-expansion_bis}
   \Delta^{ab}_A(x, y;\varphi) = \Delta_{A}^{ab} (x,y;0) + \frac{\i}{2}\sum_{c} c \int \d^4 x'\, \Delta_A^{ac}(x, x';0) g_2^2 (\varphi^c(x'))^2 \Delta^{cb}_A(x', y;0) +\O((g_2\varphi)^4) \, , 
\end{align}
is valid at leading order.
The EoMs for the two-point functions are
\begin{align}
  (\Box_x+ g_2^2 (\varphi^a)^2/2 )\Delta^{ab}_A(x, y;\varphi) =\i c^{ab} \delta^{(4)} (x-y)\, , \quad   \Box_x \Delta^{ab}_A(x, y;0) =\i c^{ab} \delta^{(4)} (x-y)\,.
\end{align}
Applying the operator $(\Box_x+ g_2^2 (\varphi^a)^2/2 )$ on the expansion in Eq.\,\eqref{eq:DeltaA-expansion_bis}, we obtain 
\begin{align}
  (\Box_x+ g_2^2 (\varphi^a)^2/2 )\Delta^{ab}_A(x, y;\varphi) = \i c^{ab} \delta^{(4)} (x-y)  +\O((g_2\varphi)^4) \, . 
\end{align}
where we used the fact that 
\begin{align}
  \sum_{c} c \int \d^4 x'\, \Box_x\Delta_A^{ac}(x, x';0) g_2^2 (\varphi^c(x'))^2 \Delta^{cb}_A(x', y;0)  =  \i g_2^2 (\varphi^a)^2 \Delta^{ab}_A(x, y;0) \, . 
\end{align}
This verifies the validity of our expansion.

In Appendix\,\ref{app:pres_fermion_mix}, we will compute the pressure from the up-scattering to a heavier fermion. This will require expanding the fermion CTP propagator. The proof of the validity of this expansion follows the same line as for the scalar and gauge propagators.

\section{Pressure from fermion mixing}
\label{app:pres_fermion_mix}
In the main text, we have studied the pressure coming from the up-scattering of scalars. In this appendix, we turn to the study of the more natural mixing among fermions. As an example, we consider the following Lagrangian 
\begin{align} 
-\mathcal{L}_{\rm mix} = Y \varphi \bar \chi N + Y^*\varphi \bar{N} \chi  + m_N \bar N N + m_\chi \bar \chi \chi  \, ,
\end{align} 
where $N, \chi$ are two Dirac fermions coupling to the condensate via the mixing term $Y \varphi \bar \chi N$. We consider that  $\chi$ is light so that it is abundant in the plasma, while $N$ is massive and Boltzmann suppressed in the plasma. The out-of-equilibrium production of $N$ from $\chi$, can induce a mixing pressure, $\P_{\chi \to N}+ \P_{\bar{\chi}\rightarrow \bar{N} }$.

To evaluate the pressure, we start with\,\cite{Berges:2004yj} 
\begin{align}
    D[\varphi]\equiv -\i\, {\rm Tr} \left[G^{-1}_{ik}(\varphi)S_{kj}\right]= -\i\, \text{tr}_s\bigg[\sum_{ik}\sum_{a, \, b} \int\d^4 x\, \left.\left[G_{ik}^{ab,-1}(\varphi(x)) S_{ki}^{ba}(x,y)\right]\right|_{y \, = \, x} \bigg] \,,
\end{align}
for $i, j, k = \chi, N$ and where the $\text{tr}_s$ denotes the trace over the spin space.
In the flavor basis, we have
\begin{align}
    S_{ij}^{ab} = \begin{pmatrix}
        S^{ab}_{\chi \chi} &S^{ab}_{ \chi N}
        \\ 
        S^{ab}_{N \chi} &S^{ab}_{NN} 
    \end{pmatrix} \, , 
    \qquad \qquad  G^{-1, ab}_{\rm flavor} = \i c^{ab} \begin{pmatrix}
       -\slashed{\partial}+ m_\chi & Y\varphi^a
       \\ Y^*  \varphi^a & -\slashed{\partial}+m_N
    \end{pmatrix} \, .
\end{align} 
We are interested in the term depending on $\varphi$,
\begin{align}
    D[\varphi] &\supset  {\rm tr_s}\int\d^4 x\,  \varphi^+(x) \left(Y S^{++}_{N\chi}(x,x)+ Y^* S^{++}_{\chi N}(x,x)\right)\notag\\
    &-{\rm tr_s} \int\d^4 x\, \varphi^-(x) \left(Y S^{--}_{N\chi}(x,x)+ Y^* S^{--}_{\chi N}(x,x)\right)\,.
\end{align}
From this, we have a contribution to the condensate EoM
\begin{align}
    -\left.\frac{\delta D[\varphi] }{\delta\varphi(x)}\right|_{\varphi^+=\varphi=\varphi} = {\rm tr_s} \left.\left(Y S^{++}_{N\chi}(x,x)+Y^* S^{++}_{\chi N}(x,x)\right)\right|_{\varphi^+=\varphi^-=\varphi}\,,
\end{align}
where again we have indicated the implicit dependence on $\varphi$ of the fermionic two-point functions.
We now can expand $S_{N\chi}^{++}$ as
\begin{align}
S^{++}_{N \chi}(x_1,x_2) =
- \i \sum_a\int \d^4 x'\, a\,Y^*\varphi^a(x') S^{+a}_{\chi\chi}(x_1, x') S^{a+}_{NN}(x',x_2) + \mathcal{O}\bigg(\frac{Y\varphi }{m_N}\bigg)^2 \, ,
\end{align}
and similarly for $S^{++}_{\chi N}$.

Then we obtain
\begin{align}
    -\left.\frac{\delta D[\varphi]}{\delta \varphi^+(x)}\right|_{\varphi^-=\varphi^+=\varphi}=\int \d^4 x'\, \Pi^{\rm R}_{\varphi;\rm mix} (x,x')\varphi(x')\,,
\end{align}
where
\begin{align}
    \Pi^{\rm R}_{\varphi;\rm mix}(x,x')\equiv -2\i\, |Y|^2 \text{tr}_s\bigg[\left[S^{++}_{\chi\chi}(x,x')S^{++}_{NN}(x',x)- S_{\chi\chi}^{+-}(x,x')S_{NN}^{-+}(x',x) \right] \bigg]\,.
\end{align}
Similarly, we would have 
\begin{align}
    \P_{\chi\rightarrow N}+ \P_{\bar{\chi}\rightarrow \bar{N}} &= -\int_{-\delta}^\delta \d z\, \frac{\d\varphi}{\d z}\pi^{\rm R}_{\varphi;\rm mix}(z,z')\varphi(z')\notag\\
    &=-\int\frac{\d q^z}{2\pi} \, q^z |\widetilde{\varphi}(q^z)|^2 {\rm Im }\pi^{\rm R}_{\varphi;\rm mix} (q^z) \,.
\end{align}
One can then proceed with estimating ${\rm Im}\pi^{\rm R}_{\varphi;\rm mix}(q^z)$ following the procedure outlined in the main text.

\end{appendix}

\bibliographystyle{utphys}
\bibliography{ref}{}

\providecommand{\href}[2]{#2}\begingroup\raggedright\begin{thebibliography}{100}

\bibitem{Witten:1984rs}
E.~Witten, ``{Cosmic Separation of Phases},''
\href{http://dx.doi.org/10.1103/PhysRevD.30.272}{{\em Phys. Rev.} {\bfseries
  D30} (1984) 272--285}.

\bibitem{Hogan_GW_1986}
C.~J. Hogan, ``{Gravitational radiation from cosmological phase transitions},''
  {\em Mon. Not. Roy. Astron. Soc.} {\bfseries 218} (1986) 629--636.

\bibitem{Kosowsky:1992vn}
A.~Kosowsky and M.~S. Turner, ``{Gravitational radiation from colliding vacuum
  bubbles: envelope approximation to many bubble collisions},''
  \href{http://dx.doi.org/10.1103/PhysRevD.47.4372}{{\em Phys. Rev.} {\bfseries
  D47} (1993) 4372--4391},
\href{http://arxiv.org/abs/astro-ph/9211004}{{\ttfamily arXiv:astro-ph/9211004
  [astro-ph]}}.

\bibitem{Kosowsky:1992rz}
A.~Kosowsky, M.~S. Turner, and R.~Watkins, ``{Gravitational waves from first
  order cosmological phase transitions},''
\href{http://dx.doi.org/10.1103/PhysRevLett.69.2026}{{\em Phys. Rev. Lett.}
  {\bfseries 69} (1992) 2026--2029}.

\bibitem{Kamionkowski:1993fg}
M.~Kamionkowski, A.~Kosowsky, and M.~S. Turner, ``{Gravitational radiation from
  first order phase transitions},''
  \href{http://dx.doi.org/10.1103/PhysRevD.49.2837}{{\em Phys. Rev.} {\bfseries
  D49} (1994) 2837--2851},
\href{http://arxiv.org/abs/astro-ph/9310044}{{\ttfamily arXiv:astro-ph/9310044
  [astro-ph]}}.

\bibitem{Kuzmin:1985mm}
V.~A. Kuzmin, V.~A. Rubakov, and M.~E. Shaposhnikov, ``{On the Anomalous
  Electroweak Baryon Number Nonconservation in the Early Universe},''
  \href{http://dx.doi.org/10.1016/0370-2693(85)91028-7}{{\em Phys. Lett. B}
  {\bfseries 155} (1985) 36}.

\bibitem{Shaposhnikov:1986jp}
M.~Shaposhnikov, ``{Possible Appearance of the Baryon Asymmetry of the Universe
  in an Electroweak Theory},'' {\em JETP Lett.} {\bfseries 44} (1986) 465--468.

\bibitem{Nelson:1991ab}
A.~E. Nelson, D.~B. Kaplan, and A.~G. Cohen, ``{Why there is something rather
  than nothing: Matter from weak interactions},''
  \href{http://dx.doi.org/10.1016/0550-3213(92)90440-M}{{\em Nucl. Phys. B}
  {\bfseries 373} (1992) 453--478}.

\bibitem{Carena:1996wj}
M.~Carena, M.~Quiros, and C.~E.~M. Wagner, ``{Opening the window for
  electroweak baryogenesis},''
  \href{http://dx.doi.org/10.1016/0370-2693(96)00475-3}{{\em Phys. Lett. B}
  {\bfseries 380} (1996) 81--91},
  \href{http://arxiv.org/abs/hep-ph/9603420}{{\ttfamily arXiv:hep-ph/9603420}}.

\bibitem{Long:2017rdo}
A.~J. Long, A.~Tesi, and L.-T. Wang, ``{Baryogenesis at a
  Lepton-Number-Breaking Phase Transition},''
  \href{http://dx.doi.org/10.1007/JHEP10(2017)095}{{\em JHEP} {\bfseries 10}
  (2017) 095}, \href{http://arxiv.org/abs/1703.04902}{{\ttfamily
  arXiv:1703.04902 [hep-ph]}}.

\bibitem{Bruggisser:2018mrt}
S.~Bruggisser, B.~Von~Harling, O.~Matsedonskyi, and G.~Servant, ``{Electroweak
  Phase Transition and Baryogenesis in Composite Higgs Models},''
  \href{http://dx.doi.org/10.1007/JHEP12(2018)099}{{\em JHEP} {\bfseries 12}
  (2018) 099}, \href{http://arxiv.org/abs/1804.07314}{{\ttfamily
  arXiv:1804.07314 [hep-ph]}}.

\bibitem{Bruggisser:2018mus}
S.~Bruggisser, B.~Von~Harling, O.~Matsedonskyi, and G.~Servant, ``{Baryon
  Asymmetry from a Composite Higgs Boson},''
  \href{http://dx.doi.org/10.1103/PhysRevLett.121.131801}{{\em Phys. Rev.
  Lett.} {\bfseries 121} no.~13, (2018) 131801},
  \href{http://arxiv.org/abs/1803.08546}{{\ttfamily arXiv:1803.08546
  [hep-ph]}}.

\bibitem{Morrissey:2012db}
D.~E. Morrissey and M.~J. Ramsey-Musolf, ``{Electroweak baryogenesis},''
  \href{http://dx.doi.org/10.1088/1367-2630/14/12/125003}{{\em New J. Phys.}
  {\bfseries 14} (2012) 125003},
  \href{http://arxiv.org/abs/1206.2942}{{\ttfamily arXiv:1206.2942 [hep-ph]}}.

\bibitem{Azatov:2021irb}
A.~Azatov, M.~Vanvlasselaer, and W.~Yin, ``{Baryogenesis via relativistic
  bubble walls},'' \href{http://dx.doi.org/10.1007/JHEP10(2021)043}{{\em JHEP}
  {\bfseries 10} (2021) 043}, \href{http://arxiv.org/abs/2106.14913}{{\ttfamily
  arXiv:2106.14913 [hep-ph]}}.

\bibitem{Baldes:2021vyz}
I.~Baldes, S.~Blasi, A.~Mariotti, A.~Sevrin, and K.~Turbang, ``{Baryogenesis
  via relativistic bubble expansion},''
  \href{http://dx.doi.org/10.1103/PhysRevD.104.115029}{{\em Phys. Rev. D}
  {\bfseries 104} no.~11, (2021) 115029},
  \href{http://arxiv.org/abs/2106.15602}{{\ttfamily arXiv:2106.15602
  [hep-ph]}}.

\bibitem{Huang:2022vkf}
P.~Huang and K.-P. Xie, ``{Leptogenesis triggered by a first-order phase
  transition},'' \href{http://dx.doi.org/10.1007/JHEP09(2022)052}{{\em JHEP}
  {\bfseries 09} (2022) 052}, \href{http://arxiv.org/abs/2206.04691}{{\ttfamily
  arXiv:2206.04691 [hep-ph]}}.

\bibitem{Chun:2023ezg}
E.~J. Chun, T.~P. Dutka, T.~H. Jung, X.~Nagels, and M.~Vanvlasselaer,
  ``{Bubble-assisted leptogenesis},''
  \href{http://dx.doi.org/10.1007/JHEP09(2023)164}{{\em JHEP} {\bfseries 09}
  (2023) 164}, \href{http://arxiv.org/abs/2305.10759}{{\ttfamily
  arXiv:2305.10759 [hep-ph]}}.

\bibitem{Dichtl:2023xqd}
M.~Dichtl, J.~Nava, S.~Pascoli, and F.~Sala, ``{Baryogenesis and leptogenesis
  from supercooled confinement},''
  \href{http://dx.doi.org/10.1007/JHEP02(2024)059}{{\em JHEP} {\bfseries 02}
  (2024) 059}, \href{http://arxiv.org/abs/2312.09282}{{\ttfamily
  arXiv:2312.09282 [hep-ph]}}.

\bibitem{Cataldi:2024pgt}
M.~Cataldi and B.~Shakya, ``{Leptogenesis via bubble collisions},''
  \href{http://dx.doi.org/10.1088/1475-7516/2024/11/047}{{\em JCAP} {\bfseries
  11} (2024) 047}, \href{http://arxiv.org/abs/2407.16747}{{\ttfamily
  arXiv:2407.16747 [hep-ph]}}.

\bibitem{Vachaspati:1991nm}
T.~Vachaspati, ``{Magnetic fields from cosmological phase transitions},''
  \href{http://dx.doi.org/10.1016/0370-2693(91)90051-Q}{{\em Phys. Lett. B}
  {\bfseries 265} (1991) 258--261}.

\bibitem{Ahonen:1997wh}
J.~Ahonen and K.~Enqvist, ``{Magnetic field generation in first order phase
  transition bubble collisions},''
  \href{http://dx.doi.org/10.1103/PhysRevD.57.664}{{\em Phys. Rev. D}
  {\bfseries 57} (1998) 664--673},
  \href{http://arxiv.org/abs/hep-ph/9704334}{{\ttfamily arXiv:hep-ph/9704334}}.

\bibitem{Vachaspati:2001nb}
T.~Vachaspati, ``{Estimate of the primordial magnetic field helicity},''
  \href{http://dx.doi.org/10.1103/PhysRevLett.87.251302}{{\em Phys. Rev. Lett.}
  {\bfseries 87} (2001) 251302},
  \href{http://arxiv.org/abs/astro-ph/0101261}{{\ttfamily
  arXiv:astro-ph/0101261}}.

\bibitem{Ellis:2019tjf}
J.~Ellis, M.~Fairbairn, M.~Lewicki, V.~Vaskonen, and A.~Wickens,
  ``{Intergalactic Magnetic Fields from First-Order Phase Transitions},''
  \href{http://dx.doi.org/10.1088/1475-7516/2019/09/019}{{\em JCAP} {\bfseries
  1909} no.~09, (2019) 019},
\href{http://arxiv.org/abs/1907.04315}{{\ttfamily arXiv:1907.04315
  [astro-ph.CO]}}.

\bibitem{Di:2020kbw}
Y.~Di, J.~Wang, R.~Zhou, L.~Bian, R.-G. Cai, and J.~Liu, ``{Magnetic Field and
  Gravitational Waves from the First-Order Phase Transition},''
  \href{http://dx.doi.org/10.1103/PhysRevLett.126.251102}{{\em Phys. Rev.
  Lett.} {\bfseries 126} no.~25, (2021) 251102},
  \href{http://arxiv.org/abs/2012.15625}{{\ttfamily arXiv:2012.15625
  [astro-ph.CO]}}.

\bibitem{Olea-Romacho:2023rhh}
M.~O. Olea-Romacho, ``{Primordial magnetogenesis in the two-Higgs-doublet
  model},'' \href{http://dx.doi.org/10.1103/PhysRevD.109.015023}{{\em Phys.
  Rev. D} {\bfseries 109} no.~1, (2024) 015023},
  \href{http://arxiv.org/abs/2310.19948}{{\ttfamily arXiv:2310.19948
  [hep-ph]}}.

\bibitem{Balaji:2024rvo}
S.~Balaji, M.~Fairbairn, and M.~O. Olea-Romacho, ``{Magnetogenesis with
  gravitational waves and primordial black hole dark matter},''
  \href{http://dx.doi.org/10.1103/PhysRevD.109.075048}{{\em Phys. Rev. D}
  {\bfseries 109} no.~7, (2024) 075048},
  \href{http://arxiv.org/abs/2402.05179}{{\ttfamily arXiv:2402.05179
  [hep-ph]}}.

\bibitem{Falkowski:2012fb}
A.~Falkowski and J.~M. No, ``{Non-thermal Dark Matter Production from the
  Electroweak Phase Transition: Multi-TeV WIMPs and 'Baby-Zillas'},''
  \href{http://dx.doi.org/10.1007/JHEP02(2013)034}{{\em JHEP} {\bfseries 02}
  (2013) 034}, \href{http://arxiv.org/abs/1211.5615}{{\ttfamily arXiv:1211.5615
  [hep-ph]}}.

\bibitem{Baldes:2020kam}
I.~Baldes, Y.~Gouttenoire, and F.~Sala, ``{String Fragmentation in Supercooled
  Confinement and Implications for Dark Matter},''
  \href{http://dx.doi.org/10.1007/JHEP04(2021)278}{{\em JHEP} {\bfseries 04}
  (2021) 278}, \href{http://arxiv.org/abs/2007.08440}{{\ttfamily
  arXiv:2007.08440 [hep-ph]}}.

\bibitem{Hong:2020est}
J.-P. Hong, S.~Jung, and K.-P. Xie, ``{Fermi-ball dark matter from a
  first-order phase transition},''
  \href{http://dx.doi.org/10.1103/PhysRevD.102.075028}{{\em Phys. Rev. D}
  {\bfseries 102} no.~7, (2020) 075028},
  \href{http://arxiv.org/abs/2008.04430}{{\ttfamily arXiv:2008.04430
  [hep-ph]}}.

\bibitem{Azatov:2021ifm}
A.~Azatov, M.~Vanvlasselaer, and W.~Yin, ``{Dark Matter production from
  relativistic bubble walls},''
  \href{http://dx.doi.org/10.1007/JHEP03(2021)288}{{\em JHEP} {\bfseries 03}
  (2021) 288}, \href{http://arxiv.org/abs/2101.05721}{{\ttfamily
  arXiv:2101.05721 [hep-ph]}}.

\bibitem{Baldes:2021aph}
I.~Baldes, Y.~Gouttenoire, F.~Sala, and G.~Servant, ``{Supercool composite Dark
  Matter beyond 100 TeV},''
  \href{http://dx.doi.org/10.1007/JHEP07(2022)084}{{\em JHEP} {\bfseries 07}
  (2022) 084}, \href{http://arxiv.org/abs/2110.13926}{{\ttfamily
  arXiv:2110.13926 [hep-ph]}}.

\bibitem{Asadi:2021pwo}
P.~Asadi, E.~D. Kramer, E.~Kuflik, G.~W. Ridgway, T.~R. Slatyer, and
  J.~Smirnov, ``{Thermal squeezeout of dark matter},''
  \href{http://dx.doi.org/10.1103/PhysRevD.104.095013}{{\em Phys. Rev. D}
  {\bfseries 104} no.~9, (2021) 095013},
  \href{http://arxiv.org/abs/2103.09827}{{\ttfamily arXiv:2103.09827
  [hep-ph]}}.

\bibitem{Baldes:2022oev}
I.~Baldes, Y.~Gouttenoire, and F.~Sala, ``{Hot and heavy dark matter from a
  weak scale phase transition},''
  \href{http://dx.doi.org/10.21468/SciPostPhys.14.3.033}{{\em SciPost Phys.}
  {\bfseries 14} (2023) 033}, \href{http://arxiv.org/abs/2207.05096}{{\ttfamily
  arXiv:2207.05096 [hep-ph]}}.

\bibitem{Azatov:2022tii}
A.~Azatov, G.~Barni, S.~Chakraborty, M.~Vanvlasselaer, and W.~Yin,
  ``{Ultra-relativistic bubbles from the simplest Higgs portal and their
  cosmological consequences},''
  \href{http://dx.doi.org/10.1007/JHEP10(2022)017}{{\em JHEP} {\bfseries 10}
  (2022) 017}, \href{http://arxiv.org/abs/2207.02230}{{\ttfamily
  arXiv:2207.02230 [hep-ph]}}.

\bibitem{Baldes:2023cih}
I.~Baldes, M.~Dichtl, Y.~Gouttenoire, and F.~Sala, ``{Ultrahigh-Energy Particle
  Collisions and Heavy Dark Matter at Phase Transitions},''
  \href{http://dx.doi.org/10.1103/PhysRevLett.134.061001}{{\em Phys. Rev.
  Lett.} {\bfseries 134} no.~6, (2025) 061001},
  \href{http://arxiv.org/abs/2306.15555}{{\ttfamily arXiv:2306.15555
  [hep-ph]}}.

\bibitem{Kierkla:2022odc}
M.~Kierkla, A.~Karam, and B.~Swiezewska, ``{Conformal model for gravitational
  waves and dark matter: a status update},''
  \href{http://dx.doi.org/10.1007/JHEP03(2023)007}{{\em JHEP} {\bfseries 03}
  (2023) 007}, \href{http://arxiv.org/abs/2210.07075}{{\ttfamily
  arXiv:2210.07075 [astro-ph.CO]}}.

\bibitem{Gehrman:2023qjn}
T.~C. Gehrman, B.~Shams Es~Haghi, K.~Sinha, and T.~Xu, ``{Recycled dark
  matter},'' \href{http://dx.doi.org/10.1088/1475-7516/2024/03/044}{{\em JCAP}
  {\bfseries 03} (2024) 044}, \href{http://arxiv.org/abs/2310.08526}{{\ttfamily
  arXiv:2310.08526 [hep-ph]}}.

\bibitem{Giudice:2024tcp}
G.~F. Giudice, H.~M. Lee, A.~Pomarol, and B.~Shakya, ``{Nonthermal heavy dark
  matter from a first-order phase transition},''
  \href{http://dx.doi.org/10.1007/JHEP12(2024)190}{{\em JHEP} {\bfseries 12}
  (2024) 190}, \href{http://arxiv.org/abs/2403.03252}{{\ttfamily
  arXiv:2403.03252 [hep-ph]}}.

\bibitem{Ai:2024ikj}
W.-Y. Ai, M.~Fairbairn, K.~Mimasu, and T.~You, ``{Non-thermal production of
  heavy vector dark matter from relativistic bubble walls},''
  \href{http://arxiv.org/abs/2406.20051}{{\ttfamily arXiv:2406.20051
  [hep-ph]}}.

\bibitem{Cembranos:2024pvy}
J.~A.~R. Cembranos, J.~Luque, and J.~Rubio, ``{Scalar dark matter production
  through the bubble expansion mechanism: the role of the Lorentz factor and
  non-renormalizable interactions},''
  \href{http://dx.doi.org/10.1140/epjc/s10052-025-14064-6}{{\em Eur. Phys. J.
  C} {\bfseries 85} no.~4, (2025) 368},
  \href{http://arxiv.org/abs/2407.14592}{{\ttfamily arXiv:2407.14592
  [hep-ph]}}.

\bibitem{Benso:2025vgm}
C.~Benso, F.~Kahlhoefer, and H.~Mansour, ``{Dark matter phase-in: producing
  feebly-interacting particles after a first-order phase transition},''
  \href{http://arxiv.org/abs/2504.10593}{{\ttfamily arXiv:2504.10593
  [hep-ph]}}.

\bibitem{Kodama:1982sf}
H.~Kodama, M.~Sasaki, and K.~Sato, ``{Abundance of Primordial Holes Produced by
  Cosmological First Order Phase Transition},''
  \href{http://dx.doi.org/10.1143/PTP.68.1979}{{\em Prog. Theor. Phys.}
  {\bfseries 68} (1982) 1979}.

\bibitem{Kawana:2021tde}
K.~Kawana and K.-P. Xie, ``{Primordial black holes from a cosmic phase
  transition: The collapse of Fermi-balls},''
  \href{http://dx.doi.org/10.1016/j.physletb.2021.136791}{{\em Phys. Lett. B}
  {\bfseries 824} (2022) 136791},
  \href{http://arxiv.org/abs/2106.00111}{{\ttfamily arXiv:2106.00111
  [astro-ph.CO]}}.

\bibitem{Liu:2021svg}
J.~Liu, L.~Bian, R.-G. Cai, Z.-K. Guo, and S.-J. Wang, ``{Primordial black hole
  production during first-order phase transitions},''
  \href{http://dx.doi.org/10.1103/PhysRevD.105.L021303}{{\em Phys. Rev. D}
  {\bfseries 105} no.~2, (2022) L021303},
  \href{http://arxiv.org/abs/2106.05637}{{\ttfamily arXiv:2106.05637
  [astro-ph.CO]}}.

\bibitem{Jung:2021mku}
T.~H. Jung and T.~Okui, ``{Primordial black holes from bubble collisions during
  a first-order phase transition},''
  \href{http://dx.doi.org/10.1103/PhysRevD.110.115014}{{\em Phys. Rev. D}
  {\bfseries 110} no.~11, (2024) 115014},
  \href{http://arxiv.org/abs/2110.04271}{{\ttfamily arXiv:2110.04271
  [hep-ph]}}.

\bibitem{Gouttenoire:2023naa}
Y.~Gouttenoire and T.~Volansky, ``{Primordial black holes from supercooled
  phase transitions},''
  \href{http://dx.doi.org/10.1103/PhysRevD.110.043514}{{\em Phys. Rev. D}
  {\bfseries 110} no.~4, (2024) 043514},
  \href{http://arxiv.org/abs/2305.04942}{{\ttfamily arXiv:2305.04942
  [hep-ph]}}.

\bibitem{Lewicki:2024ghw}
M.~Lewicki, P.~Toczek, and V.~Vaskonen, ``{Black Holes and Gravitational Waves
  from Slow First-Order Phase Transitions},''
  \href{http://dx.doi.org/10.1103/PhysRevLett.133.221003}{{\em Phys. Rev.
  Lett.} {\bfseries 133} no.~22, (2024) 221003},
  \href{http://arxiv.org/abs/2402.04158}{{\ttfamily arXiv:2402.04158
  [astro-ph.CO]}}.

\bibitem{Ai:2024cka}
W.-Y. Ai, L.~Heurtier, and T.~H. Jung, ``{Primordial black holes from an
  interrupted phase transition},''
  \href{http://arxiv.org/abs/2409.02175}{{\ttfamily arXiv:2409.02175
  [astro-ph.CO]}}.

\bibitem{Murai:2025hse}
K.~Murai, K.~Sakurai, and F.~Takahashi, ``{Primordial Black Hole Formation via
  Inverted Bubble Collapse},''
  \href{http://arxiv.org/abs/2502.02291}{{\ttfamily arXiv:2502.02291
  [astro-ph.CO]}}.

\bibitem{Gowling:2021gcy}
C.~Gowling and M.~Hindmarsh, ``{Observational prospects for phase transitions
  at LISA: Fisher matrix analysis},''
  \href{http://dx.doi.org/10.1088/1475-7516/2021/10/039}{{\em JCAP} {\bfseries
  10} (2021) 039}, \href{http://arxiv.org/abs/2106.05984}{{\ttfamily
  arXiv:2106.05984 [astro-ph.CO]}}.

\bibitem{Bea:2021zsu}
Y.~Bea, J.~Casalderrey-Solana, T.~Giannakopoulos, D.~Mateos,
  M.~Sanchez-Garitaonandia, and M.~Zilh\~ao, ``{Bubble wall velocity from
  holography},'' \href{http://dx.doi.org/10.1103/PhysRevD.104.L121903}{{\em
  Phys. Rev. D} {\bfseries 104} no.~12, (2021) L121903},
  \href{http://arxiv.org/abs/2104.05708}{{\ttfamily arXiv:2104.05708
  [hep-th]}}.

\bibitem{Bigazzi:2021ucw}
F.~Bigazzi, A.~Caddeo, T.~Canneti, and A.~L. Cotrone, ``{Bubble wall velocity
  at strong coupling},'' \href{http://dx.doi.org/10.1007/JHEP08(2021)090}{{\em
  JHEP} {\bfseries 08} (2021) 090},
  \href{http://arxiv.org/abs/2104.12817}{{\ttfamily arXiv:2104.12817
  [hep-ph]}}.

\bibitem{Janik:2022wsx}
R.~A. Janik, M.~Jarvinen, H.~Soltanpanahi, and J.~Sonnenschein, ``{Perfect
  Fluid Hydrodynamic Picture of Domain Wall Velocities at Strong Coupling},''
  \href{http://dx.doi.org/10.1103/PhysRevLett.129.081601}{{\em Phys. Rev.
  Lett.} {\bfseries 129} no.~8, (2022) 081601},
  \href{http://arxiv.org/abs/2205.06274}{{\ttfamily arXiv:2205.06274
  [hep-th]}}.

\bibitem{Li:2023xto}
L.~Li, S.-J. Wang, and Z.-Y. Yuwen, ``{Bubble expansion at strong coupling},''
  \href{http://dx.doi.org/10.1103/PhysRevD.108.096033}{{\em Phys. Rev. D}
  {\bfseries 108} no.~9, (2023) 096033},
  \href{http://arxiv.org/abs/2302.10042}{{\ttfamily arXiv:2302.10042
  [hep-th]}}.

\bibitem{Bea:2024bls}
Y.~Bea, M.~Giliberti, D.~Mateos, M.~Sanchez-Garitaonandia, A.~Serantes, and
  M.~Zilh\~ao, ``{Bubble dynamics in a QCD-like phase diagram},''
  \href{http://arxiv.org/abs/2412.09588}{{\ttfamily arXiv:2412.09588
  [hep-th]}}.

\bibitem{Moore:1995ua}
G.~D. Moore and T.~Prokopec, ``{Bubble wall velocity in a first order
  electroweak phase transition},''
  \href{http://dx.doi.org/10.1103/PhysRevLett.75.777}{{\em Phys. Rev. Lett.}
  {\bfseries 75} (1995) 777--780},
  \href{http://arxiv.org/abs/hep-ph/9503296}{{\ttfamily arXiv:hep-ph/9503296}}.

\bibitem{Moore:1995si}
G.~D. Moore and T.~Prokopec, ``{How fast can the wall move? A Study of the
  electroweak phase transition dynamics},''
  \href{http://dx.doi.org/10.1103/PhysRevD.52.7182}{{\em Phys. Rev. D}
  {\bfseries 52} (1995) 7182--7204},
  \href{http://arxiv.org/abs/hep-ph/9506475}{{\ttfamily arXiv:hep-ph/9506475}}.

\bibitem{Espinosa:2010hh}
J.~R. Espinosa, T.~Konstandin, J.~M. No, and G.~Servant, ``{Energy Budget of
  Cosmological First-order Phase Transitions},''
  \href{http://dx.doi.org/10.1088/1475-7516/2010/06/028}{{\em JCAP} {\bfseries
  1006} (2010) 028},
\href{http://arxiv.org/abs/1004.4187}{{\ttfamily arXiv:1004.4187 [hep-ph]}}.

\bibitem{Liu:1992tn}
B.-H. Liu, L.~D. McLerran, and N.~Turok, ``{Bubble nucleation and growth at a
  baryon number producing electroweak phase transition},''
  \href{http://dx.doi.org/10.1103/PhysRevD.46.2668}{{\em Phys. Rev. D}
  {\bfseries 46} (1992) 2668--2688}.

\bibitem{Konstandin:2010dm}
T.~Konstandin and J.~M. No, ``{Hydrodynamic obstruction to bubble expansion},''
  \href{http://dx.doi.org/10.1088/1475-7516/2011/02/008}{{\em JCAP} {\bfseries
  02} (2011) 008}, \href{http://arxiv.org/abs/1011.3735}{{\ttfamily
  arXiv:1011.3735 [hep-ph]}}.

\bibitem{Huber:2011aa}
S.~J. Huber and M.~Sopena, ``{The bubble wall velocity in the minimal
  supersymmetric light stop scenario},''
  \href{http://dx.doi.org/10.1103/PhysRevD.85.103507}{{\em Phys. Rev. D}
  {\bfseries 85} (2012) 103507},
  \href{http://arxiv.org/abs/1112.1888}{{\ttfamily arXiv:1112.1888 [hep-ph]}}.

\bibitem{Kozaczuk:2015owa}
J.~Kozaczuk, ``{Bubble Expansion and the Viability of Singlet-Driven
  Electroweak Baryogenesis},''
  \href{http://dx.doi.org/10.1007/JHEP10(2015)135}{{\em JHEP} {\bfseries 10}
  (2015) 135}, \href{http://arxiv.org/abs/1506.04741}{{\ttfamily
  arXiv:1506.04741 [hep-ph]}}.

\bibitem{Dorsch:2018pat}
G.~C. Dorsch, S.~J. Huber, and T.~Konstandin, ``{Bubble wall velocities in the
  Standard Model and beyond},''
  \href{http://dx.doi.org/10.1088/1475-7516/2018/12/034}{{\em JCAP} {\bfseries
  1812} no.~12, (2018) 034},
\href{http://arxiv.org/abs/1809.04907}{{\ttfamily arXiv:1809.04907 [hep-ph]}}.

\bibitem{Friedlander:2020tnq}
A.~Friedlander, I.~Banta, J.~M. Cline, and D.~Tucker-Smith, ``{Wall speed and
  shape in singlet-assisted strong electroweak phase transitions},''
  \href{http://dx.doi.org/10.1103/PhysRevD.103.055020}{{\em Phys. Rev. D}
  {\bfseries 103} no.~5, (2021) 055020},
  \href{http://arxiv.org/abs/2009.14295}{{\ttfamily arXiv:2009.14295
  [hep-ph]}}.

\bibitem{Balaji:2020yrx}
S.~Balaji, M.~Spannowsky, and C.~Tamarit, ``{Cosmological bubble friction in
  local equilibrium},''
  \href{http://dx.doi.org/10.1088/1475-7516/2021/03/051}{{\em JCAP} {\bfseries
  03} (2021) 051}, \href{http://arxiv.org/abs/2010.08013}{{\ttfamily
  arXiv:2010.08013 [hep-ph]}}.

\bibitem{Cline:2021iff}
J.~M. Cline, A.~Friedlander, D.-M. He, K.~Kainulainen, B.~Laurent, and
  D.~Tucker-Smith, ``{Baryogenesis and gravity waves from a UV-completed
  electroweak phase transition},''
  \href{http://dx.doi.org/10.1103/PhysRevD.103.123529}{{\em Phys. Rev. D}
  {\bfseries 103} no.~12, (2021) 123529},
  \href{http://arxiv.org/abs/2102.12490}{{\ttfamily arXiv:2102.12490
  [hep-ph]}}.

\bibitem{Ai:2021kak}
W.-Y. Ai, B.~Garbrecht, and C.~Tamarit, ``{Bubble wall velocities in local
  equilibrium},'' \href{http://dx.doi.org/10.1088/1475-7516/2022/03/015}{{\em
  JCAP} {\bfseries 03} no.~03, (2022) 015},
  \href{http://arxiv.org/abs/2109.13710}{{\ttfamily arXiv:2109.13710
  [hep-ph]}}.

\bibitem{Lewicki:2021pgr}
M.~Lewicki, M.~Merchand, and M.~Zych, ``{Electroweak bubble wall expansion:
  gravitational waves and baryogenesis in Standard Model-like thermal
  plasma},'' \href{http://dx.doi.org/10.1007/JHEP02(2022)017}{{\em JHEP}
  {\bfseries 02} (2022) 017}, \href{http://arxiv.org/abs/2111.02393}{{\ttfamily
  arXiv:2111.02393 [astro-ph.CO]}}.

\bibitem{Dorsch:2021nje}
G.~C. Dorsch, S.~J. Huber, and T.~Konstandin, ``{A sonic boom in bubble wall
  friction},'' \href{http://dx.doi.org/10.1088/1475-7516/2022/04/010}{{\em
  JCAP} {\bfseries 04} no.~04, (2022) 010},
  \href{http://arxiv.org/abs/2112.12548}{{\ttfamily arXiv:2112.12548
  [hep-ph]}}.

\bibitem{Jiang:2022btc}
S.~Jiang, F.~P. Huang, and X.~Wang, ``{Bubble wall velocity during electroweak
  phase transition in the inert doublet model},''
  \href{http://dx.doi.org/10.1103/PhysRevD.107.095005}{{\em Phys. Rev. D}
  {\bfseries 107} no.~9, (2023) 095005},
  \href{http://arxiv.org/abs/2211.13142}{{\ttfamily arXiv:2211.13142
  [hep-ph]}}.

\bibitem{Laurent:2022jrs}
B.~Laurent and J.~M. Cline, ``{First principles determination of bubble wall
  velocity},'' \href{http://dx.doi.org/10.1103/PhysRevD.106.023501}{{\em Phys.
  Rev. D} {\bfseries 106} no.~2, (2022) 023501},
  \href{http://arxiv.org/abs/2204.13120}{{\ttfamily arXiv:2204.13120
  [hep-ph]}}.

\bibitem{Wang:2022txy}
S.-J. Wang and Z.-Y. Yuwen, ``{Hydrodynamic backreaction force of cosmological
  bubble expansion},''
  \href{http://dx.doi.org/10.1103/PhysRevD.107.023501}{{\em Phys. Rev. D}
  {\bfseries 107} no.~2, (2023) 023501},
  \href{http://arxiv.org/abs/2205.02492}{{\ttfamily arXiv:2205.02492
  [hep-ph]}}.

\bibitem{Ai:2023see}
W.-Y. Ai, B.~Laurent, and J.~van~de Vis, ``{Model-independent bubble wall
  velocities in local thermal equilibrium},''
  \href{http://dx.doi.org/10.1088/1475-7516/2023/07/002}{{\em JCAP} {\bfseries
  07} (2023) 002}, \href{http://arxiv.org/abs/2303.10171}{{\ttfamily
  arXiv:2303.10171 [astro-ph.CO]}}.

\bibitem{Krajewski:2023clt}
T.~Krajewski, M.~Lewicki, and M.~Zych, ``{Hydrodynamical constraints on the
  bubble wall velocity},''
  \href{http://dx.doi.org/10.1103/PhysRevD.108.103523}{{\em Phys. Rev. D}
  {\bfseries 108} no.~10, (2023) 103523},
  \href{http://arxiv.org/abs/2303.18216}{{\ttfamily arXiv:2303.18216
  [astro-ph.CO]}}.

\bibitem{Wang:2023kux}
J.-C. Wang, Z.-Y. Yuwen, Y.-S. Hao, and S.-J. Wang, ``{General backreaction
  force of cosmological bubble expansion},''
  \href{http://dx.doi.org/10.1103/PhysRevD.110.016031}{{\em Phys. Rev. D}
  {\bfseries 110} no.~1, (2024) 016031},
  \href{http://arxiv.org/abs/2310.07691}{{\ttfamily arXiv:2310.07691
  [hep-ph]}}.

\bibitem{Dorsch:2023tss}
G.~C. Dorsch and D.~A. Pinto, ``{Bubble wall velocities with an extended fluid
  Ansatz},'' \href{http://dx.doi.org/10.1088/1475-7516/2024/04/027}{{\em JCAP}
  {\bfseries 04} (2024) 027}, \href{http://arxiv.org/abs/2312.02354}{{\ttfamily
  arXiv:2312.02354 [hep-ph]}}.

\bibitem{Kang:2024xqk}
Z.~Kang and J.~Zhu, ``{Confinement Bubble Wall Velocity via Quasiparticle
  Determination},'' \href{http://arxiv.org/abs/2401.03849}{{\ttfamily
  arXiv:2401.03849 [hep-ph]}}.

\bibitem{Ai:2024shx}
W.-Y. Ai, X.~Nagels, and M.~Vanvlasselaer, ``{Criterion for ultra-fast bubble
  walls: the impact of hydrodynamic obstruction},''
  \href{http://dx.doi.org/10.1088/1475-7516/2024/03/037}{{\em JCAP} {\bfseries
  03} (2024) 037}, \href{http://arxiv.org/abs/2401.05911}{{\ttfamily
  arXiv:2401.05911 [hep-ph]}}.

\bibitem{Krajewski:2024gma}
T.~Krajewski, M.~Lewicki, and M.~Zych, ``{Bubble-wall velocity in local thermal
  equilibrium: hydrodynamical simulations vs analytical treatment},''
  \href{http://dx.doi.org/10.1007/JHEP05(2024)011}{{\em JHEP} {\bfseries 05}
  (2024) 011}, \href{http://arxiv.org/abs/2402.15408}{{\ttfamily
  arXiv:2402.15408 [astro-ph.CO]}}.

\bibitem{Wang:2024wcs}
D.-W. Wang, Q.-S. Yan, and M.~Huang, ``{Bubble wall velocity and gravitational
  wave in the minimal left-right symmetric model},''
  \href{http://dx.doi.org/10.1103/PhysRevD.110.076011}{{\em Phys. Rev. D}
  {\bfseries 110} no.~7, (2024) 076011},
  \href{http://arxiv.org/abs/2405.01949}{{\ttfamily arXiv:2405.01949 [gr-qc]}}.

\bibitem{Barni:2024lkj}
G.~Barni, S.~Blasi, and M.~Vanvlasselaer, ``{The hydrodynamics of inverse phase
  transitions},'' \href{http://dx.doi.org/10.1088/1475-7516/2024/10/042}{{\em
  JCAP} {\bfseries 10} (2024) 042},
  \href{http://arxiv.org/abs/2406.01596}{{\ttfamily arXiv:2406.01596
  [hep-ph]}}.

\bibitem{Ekstedt:2024fyq}
A.~Ekstedt, O.~Gould, J.~Hirvonen, B.~Laurent, L.~Niemi, P.~Schicho, and
  J.~van~de Vis, ``{How fast does the WallGo? A package for computing wall
  velocities in first-order phase transitions},''
  \href{http://arxiv.org/abs/2411.04970}{{\ttfamily arXiv:2411.04970
  [hep-ph]}}.

\bibitem{Ai:2024btx}
W.-Y. Ai, B.~Laurent, and J.~van~de Vis, ``{Bounds on the bubble wall
  velocity},'' \href{http://dx.doi.org/10.1007/JHEP02(2025)119}{{\em JHEP}
  {\bfseries 02} (2025) 119}, \href{http://arxiv.org/abs/2411.13641}{{\ttfamily
  arXiv:2411.13641 [hep-ph]}}.

\bibitem{Krajewski:2024xuz}
T.~Krajewski, M.~Lewicki, M.~Vasar, V.~Vaskonen, H.~Veerm\"ae, and M.~Zych,
  ``{Thermalization effects on the dynamics of growing vacuum bubbles},''
  \href{http://arxiv.org/abs/2411.15094}{{\ttfamily arXiv:2411.15094
  [hep-ph]}}.

\bibitem{Dorsch:2024jjl}
G.~C. Dorsch, T.~Konstandin, E.~Perboni, and D.~A. Pinto, ``{Non-singular
  solutions to the Boltzmann equation with a fluid Ansatz},''
  \href{http://arxiv.org/abs/2412.09266}{{\ttfamily arXiv:2412.09266
  [hep-ph]}}.

\bibitem{Carena:2025flp}
M.~Carena, A.~Ireland, T.~Ou, and I.~R. Wang, ``{The Discriminant Power of
  Bubble Wall Velocities: Gravitational Waves and Electroweak Baryogenesis},''
  \href{http://arxiv.org/abs/2504.17841}{{\ttfamily arXiv:2504.17841
  [hep-ph]}}.

\bibitem{Dine:1992wr}
M.~Dine, R.~G. Leigh, P.~Y. Huet, A.~D. Linde, and D.~A. Linde, ``{Towards the
  theory of the electroweak phase transition},''
  \href{http://dx.doi.org/10.1103/PhysRevD.46.550}{{\em Phys. Rev. D}
  {\bfseries 46} (1992) 550--571},
  \href{http://arxiv.org/abs/hep-ph/9203203}{{\ttfamily arXiv:hep-ph/9203203}}.

\bibitem{Bodeker:2009qy}
D.~Bodeker and G.~D. Moore, ``{Can electroweak bubble walls run away?},''
  \href{http://dx.doi.org/10.1088/1475-7516/2009/05/009}{{\em JCAP} {\bfseries
  05} (2009) 009}, \href{http://arxiv.org/abs/0903.4099}{{\ttfamily
  arXiv:0903.4099 [hep-ph]}}.

\bibitem{Bodeker:2017cim}
D.~Bodeker and G.~D. Moore, ``{Electroweak Bubble Wall Speed Limit},''
  \href{http://dx.doi.org/10.1088/1475-7516/2017/05/025}{{\em JCAP} {\bfseries
  05} (2017) 025}, \href{http://arxiv.org/abs/1703.08215}{{\ttfamily
  arXiv:1703.08215 [hep-ph]}}.

\bibitem{BarrosoMancha:2020fay}
M.~Barroso~Mancha, T.~Prokopec, and B.~Swiezewska, ``{Field-theoretic
  derivation of bubble-wall force},''
  \href{http://dx.doi.org/10.1007/JHEP01(2021)070}{{\em JHEP} {\bfseries 01}
  (2021) 070}, \href{http://arxiv.org/abs/2005.10875}{{\ttfamily
  arXiv:2005.10875 [hep-th]}}.

\bibitem{Hoche:2020ysm}
S.~H\"oche, J.~Kozaczuk, A.~J. Long, J.~Turner, and Y.~Wang, ``{Towards an
  all-orders calculation of the electroweak bubble wall velocity},''
  \href{http://dx.doi.org/10.1088/1475-7516/2021/03/009}{{\em JCAP} {\bfseries
  03} (2021) 009}, \href{http://arxiv.org/abs/2007.10343}{{\ttfamily
  arXiv:2007.10343 [hep-ph]}}.

\bibitem{Azatov:2020ufh}
A.~Azatov and M.~Vanvlasselaer, ``{Bubble wall velocity: heavy physics
  effects},'' \href{http://dx.doi.org/10.1088/1475-7516/2021/01/058}{{\em JCAP}
  {\bfseries 01} (2021) 058}, \href{http://arxiv.org/abs/2010.02590}{{\ttfamily
  arXiv:2010.02590 [hep-ph]}}.

\bibitem{Gouttenoire:2021kjv}
Y.~Gouttenoire, R.~Jinno, and F.~Sala, ``{Friction pressure on relativistic
  bubble walls},'' \href{http://dx.doi.org/10.1007/JHEP05(2022)004}{{\em JHEP}
  {\bfseries 05} (2022) 004}, \href{http://arxiv.org/abs/2112.07686}{{\ttfamily
  arXiv:2112.07686 [hep-ph]}}.

\bibitem{GarciaGarcia:2022yqb}
I.~Garcia~Garcia, G.~Koszegi, and R.~Petrossian-Byrne, ``{Reflections on Bubble
  Walls},'' \href{http://arxiv.org/abs/2212.10572}{{\ttfamily arXiv:2212.10572
  [hep-ph]}}.

\bibitem{Ai:2023suz}
W.-Y. Ai, ``{Logarithmically divergent friction on ultrarelativistic bubble
  walls},'' \href{http://dx.doi.org/10.1088/1475-7516/2023/10/052}{{\em JCAP}
  {\bfseries 10} (2023) 052}, \href{http://arxiv.org/abs/2308.10679}{{\ttfamily
  arXiv:2308.10679 [hep-ph]}}.

\bibitem{Azatov:2023xem}
A.~Azatov, G.~Barni, R.~Petrossian-Byrne, and M.~Vanvlasselaer, ``{Quantisation
  across bubble walls and friction},''
  \href{http://dx.doi.org/10.1007/JHEP05(2024)294}{{\em JHEP} {\bfseries 05}
  (2024) 294}, \href{http://arxiv.org/abs/2310.06972}{{\ttfamily
  arXiv:2310.06972 [hep-ph]}}.

\bibitem{Baldes:2024wuz}
I.~Baldes, M.~Dichtl, Y.~Gouttenoire, and F.~Sala, ``{Particle shells from
  relativistic bubble walls},''
  \href{http://dx.doi.org/10.1007/JHEP07(2024)231}{{\em JHEP} {\bfseries 07}
  (2024) 231}, \href{http://arxiv.org/abs/2403.05615}{{\ttfamily
  arXiv:2403.05615 [hep-ph]}}.

\bibitem{Azatov:2024auq}
A.~Azatov, G.~Barni, and R.~Petrossian-Byrne, ``{NLO friction in symmetry
  restoring phase transitions},''
  \href{http://dx.doi.org/10.1007/JHEP12(2024)056}{{\em JHEP} {\bfseries 12}
  (2024) 056}, \href{http://arxiv.org/abs/2405.19447}{{\ttfamily
  arXiv:2405.19447 [hep-ph]}}.

\bibitem{Schwinger:1960qe}
J.~S. Schwinger, ``{Brownian motion of a quantum oscillator},''
  \href{http://dx.doi.org/10.1063/1.1703727}{{\em J. Math. Phys.} {\bfseries 2}
  (1961) 407--432}.

\bibitem{Keldysh:1964ud}
L.~V. Keldysh, ``{Diagram technique for nonequilibrium processes},'' {\em Sov.
  Phys. JETP} {\bfseries 20} (1965) 1018.

\bibitem{Chou:1984es}
K.-C. Chou, Z.-B. Su, B.-L. Hao, and L.~Yu, ``{Equilibrium and Nonequilibrium
  Formalisms Made Unified},''
  \href{http://dx.doi.org/10.1016/0370-1573(85)90136-X}{{\em Phys. Rept.}
  {\bfseries 118} (1985) 1--131}.

\bibitem{Cornwall:1974vz}
J.~M. Cornwall, R.~Jackiw, and E.~Tomboulis, ``{Effective Action for Composite
  Operators},'' \href{http://dx.doi.org/10.1103/PhysRevD.10.2428}{{\em Phys.
  Rev. D} {\bfseries 10} (1974) 2428--2445}.

\bibitem{Carosi:2024lop}
M.~Carosi and B.~Garbrecht, ``{False vacuum decay beyond the quadratic
  approximation: summation of non-local self-energies},''
  \href{http://arxiv.org/abs/2411.18421}{{\ttfamily arXiv:2411.18421
  [hep-th]}}.

\bibitem{Ignatius:1993qn}
J.~Ignatius, K.~Kajantie, H.~Kurki-Suonio, and M.~Laine, ``{The growth of
  bubbles in cosmological phase transitions},''
  \href{http://dx.doi.org/10.1103/PhysRevD.49.3854}{{\em Phys. Rev. D}
  {\bfseries 49} (1994) 3854--3868},
  \href{http://arxiv.org/abs/astro-ph/9309059}{{\ttfamily
  arXiv:astro-ph/9309059}}.

\bibitem{Konstandin:2014zta}
T.~Konstandin, G.~Nardini, and I.~Rues, ``{From Boltzmann equations to steady
  wall velocities},''
  \href{http://dx.doi.org/10.1088/1475-7516/2014/09/028}{{\em JCAP} {\bfseries
  09} (2014) 028}, \href{http://arxiv.org/abs/1407.3132}{{\ttfamily
  arXiv:1407.3132 [hep-ph]}}.

\bibitem{Ramsey-Musolf:2025jyk}
M.~J. Ramsey-Musolf and J.~Zhu, ``{Bubble wall velocity from Kadanoff-Baym
  equations: fluid dynamics and microscopic interactions},''
  \href{http://arxiv.org/abs/2504.13724}{{\ttfamily arXiv:2504.13724
  [hep-ph]}}.

\bibitem{Calzetta:1986cq}
E.~Calzetta and B.~L. Hu, ``{Nonequilibrium Quantum Fields: Closed Time Path
  Effective Action, Wigner Function and Boltzmann Equation},''
  \href{http://dx.doi.org/10.1103/PhysRevD.37.2878}{{\em Phys. Rev. D}
  {\bfseries 37} (1988) 2878}.

\bibitem{Berges:2004yj}
J.~Berges, ``{Introduction to nonequilibrium quantum field theory},''
  \href{http://dx.doi.org/10.1063/1.1843591}{{\em AIP Conf. Proc.} {\bfseries
  739} no.~1, (2004) 3--62},
  \href{http://arxiv.org/abs/hep-ph/0409233}{{\ttfamily arXiv:hep-ph/0409233}}.

\bibitem{Ai:2023qnr}
W.-Y. Ai, A.~Beniwal, A.~Maggi, and D.~J.~E. Marsh, ``{From QFT to Boltzmann:
  freeze-in in the presence of oscillating condensates},''
  \href{http://dx.doi.org/10.1007/JHEP02(2024)122}{{\em JHEP} {\bfseries 02}
  (2024) 122}, \href{http://arxiv.org/abs/2310.08272}{{\ttfamily
  arXiv:2310.08272 [hep-ph]}}.

\bibitem{Cheung:2015iqa}
Y.-K.~E. Cheung, M.~Drewes, J.~U. Kang, and J.~C. Kim, ``{Effective Action for
  Cosmological Scalar Fields at Finite Temperature},''
  \href{http://dx.doi.org/10.1007/JHEP08(2015)059}{{\em JHEP} {\bfseries 08}
  (2015) 059}, \href{http://arxiv.org/abs/1504.04444}{{\ttfamily
  arXiv:1504.04444 [hep-ph]}}.

\bibitem{Ai:2021gtg}
W.-Y. Ai, M.~Drewes, D.~Glavan, and J.~Hajer, ``{Oscillating scalar dissipating
  in a medium},'' \href{http://dx.doi.org/10.1007/JHEP11(2021)160}{{\em JHEP}
  {\bfseries 11} (2021) 160}, \href{http://arxiv.org/abs/2108.00254}{{\ttfamily
  arXiv:2108.00254 [hep-ph]}}.

\bibitem{Wang:2022mvv}
Z.-L. Wang and W.-Y. Ai, ``{Dissipation of oscillating scalar backgrounds in an
  FLRW universe},'' \href{http://dx.doi.org/10.1007/JHEP11(2022)075}{{\em JHEP}
  {\bfseries 11} (2022) 075}, \href{http://arxiv.org/abs/2202.08218}{{\ttfamily
  arXiv:2202.08218 [hep-ph]}}.

\bibitem{Ai:2023ahr}
W.-Y. Ai and Z.-L. Wang, ``{Fate of oscillating homogeneous
  \ensuremath{\mathbb{Z}}$_{2}$-symmetric scalar condensates in the early
  Universe},'' \href{http://dx.doi.org/10.1088/1475-7516/2024/06/075}{{\em
  JCAP} {\bfseries 06} (2024) 075},
  \href{http://arxiv.org/abs/2307.14811}{{\ttfamily arXiv:2307.14811
  [hep-ph]}}.

\bibitem{Cutkosky:1960sp}
R.~E. Cutkosky, ``{Singularities and discontinuities of Feynman amplitudes},''
  \href{http://dx.doi.org/10.1063/1.1703676}{{\em J. Math. Phys.} {\bfseries 1}
  (1960) 429--433}.

\bibitem{Weldon:1983jn}
H.~A. Weldon, ``{Simple Rules for Discontinuities in Finite Temperature Field
  Theory},'' \href{http://dx.doi.org/10.1103/PhysRevD.28.2007}{{\em Phys. Rev.
  D} {\bfseries 28} (1983) 2007}.

\bibitem{Kobes:1985kc}
R.~L. Kobes and G.~W. Semenoff, ``{Discontinuities of Green Functions in Field
  Theory at Finite Temperature and Density},''
  \href{http://dx.doi.org/10.1016/0550-3213(85)90056-2}{{\em Nucl. Phys. B}
  {\bfseries 260} (1985) 714--746}.

\bibitem{Kobes:1986za}
R.~L. Kobes and G.~W. Semenoff, ``{Discontinuities of Green Functions in Field
  Theory at Finite Temperature and Density. 2},''
  \href{http://dx.doi.org/10.1016/0550-3213(86)90006-4}{{\em Nucl. Phys. B}
  {\bfseries 272} (1986) 329--364}.

\bibitem{Landshoff:1996ta}
P.~V. Landshoff, ``{Simple physical approach to thermal cutting rules},''
  \href{http://dx.doi.org/10.1016/0370-2693(96)00919-7}{{\em Phys. Lett. B}
  {\bfseries 386} (1996) 291--296},
  \href{http://arxiv.org/abs/hep-ph/9606426}{{\ttfamily arXiv:hep-ph/9606426}}.

\bibitem{Gelis:1997zv}
F.~Gelis, ``{Cutting rules in the real time formalisms at finite
  temperature},'' \href{http://dx.doi.org/10.1016/S0550-3213(97)00511-7}{{\em
  Nucl. Phys. B} {\bfseries 508} (1997) 483--505},
  \href{http://arxiv.org/abs/hep-ph/9701410}{{\ttfamily arXiv:hep-ph/9701410}}.

\bibitem{Bedaque:1996af}
P.~F. Bedaque, A.~K. Das, and S.~Naik, ``{Cutting rules at finite
  temperature},'' \href{http://dx.doi.org/10.1142/S0217732397002612}{{\em Mod.
  Phys. Lett. A} {\bfseries 12} (1997) 2481--2496},
  \href{http://arxiv.org/abs/hep-ph/9603325}{{\ttfamily arXiv:hep-ph/9603325}}.

\bibitem{Prokopec:2003pj}
T.~Prokopec, M.~G. Schmidt, and S.~Weinstock, ``{Transport equations for chiral
  fermions to order h bar and electroweak baryogenesis. Part 1},''
  \href{http://dx.doi.org/10.1016/j.aop.2004.06.002}{{\em Annals Phys.}
  {\bfseries 314} (2004) 208--265},
  \href{http://arxiv.org/abs/hep-ph/0312110}{{\ttfamily arXiv:hep-ph/0312110}}.

\bibitem{Garbrecht:2018mrp}
B.~Garbrecht, ``{Why is there more matter than antimatter? Calculational
  methods for leptogenesis and electroweak baryogenesis},''
  \href{http://dx.doi.org/10.1016/j.ppnp.2019.103727}{{\em Prog. Part. Nucl.
  Phys.} {\bfseries 110} (2020) 103727},
  \href{http://arxiv.org/abs/1812.02651}{{\ttfamily arXiv:1812.02651
  [hep-ph]}}.

\bibitem{Dashko:2020qwy}
A.~Dashko and A.~Ekstedt, ``{Bubble-wall speed with loop corrections},''
  \href{http://dx.doi.org/10.1007/JHEP03(2025)024}{{\em JHEP} {\bfseries 25}
  (2020) 024}, \href{http://arxiv.org/abs/2411.05075}{{\ttfamily
  arXiv:2411.05075 [hep-ph]}}.

\bibitem{DeCurtis:2022hlx}
S.~De~Curtis, L.~D. Rose, A.~Guiggiani, A.~G. Muyor, and G.~Panico, ``{Bubble
  wall dynamics at the electroweak phase transition},''
  \href{http://arxiv.org/abs/2201.08220}{{\ttfamily arXiv:2201.08220
  [hep-ph]}}.

\bibitem{DeCurtis:2023hil}
S.~De~Curtis, L.~Delle~Rose, A.~Guiggiani, A.~Gil~Muyor, and G.~Panico,
  ``{Collision integrals for cosmological phase transitions},''
  \href{http://dx.doi.org/10.1007/JHEP05(2023)194}{{\em JHEP} {\bfseries 05}
  (2023) 194}, \href{http://arxiv.org/abs/2303.05846}{{\ttfamily
  arXiv:2303.05846 [hep-ph]}}.

\bibitem{DeCurtis:2024hvh}
S.~De~Curtis, L.~Delle~Rose, A.~Guiggiani, A.~Gil~Muyor, and G.~Panico,
  ``{Non-linearities in cosmological bubble wall dynamics},''
  \href{http://dx.doi.org/10.1007/JHEP05(2024)009}{{\em JHEP} {\bfseries 05}
  (2024) 009}, \href{http://arxiv.org/abs/2401.13522}{{\ttfamily
  arXiv:2401.13522 [hep-ph]}}.

\bibitem{Azatov:2024crd}
A.~Azatov, X.~Nagels, M.~Vanvlasselaer, and W.~Yin, ``{Populating secluded dark
  sector with ultra-relativistic bubbles},''
  \href{http://dx.doi.org/10.1007/JHEP11(2024)129}{{\em JHEP} {\bfseries 11}
  (2024) 129}, \href{http://arxiv.org/abs/2406.12554}{{\ttfamily
  arXiv:2406.12554 [hep-ph]}}.

\bibitem{Bellac:2011kqa}
M.~L. Bellac, \href{http://dx.doi.org/10.1017/CBO9780511721700}{{\em {Thermal
  Field Theory}}}.
\newblock Cambridge Monographs on Mathematical Physics. Cambridge University
  Press, 3, 2011.

\bibitem{Liu:2011jh}
T.~Liu, M.~J. Ramsey-Musolf, and J.~Shu, ``{Electroweak Beautygenesis: From b
  {\textbackslash{}to} s CP-violation to the Cosmic Baryon Asymmetry},''
  \href{http://dx.doi.org/10.1103/PhysRevLett.108.221301}{{\em Phys. Rev.
  Lett.} {\bfseries 108} (2012) 221301},
  \href{http://arxiv.org/abs/1109.4145}{{\ttfamily arXiv:1109.4145 [hep-ph]}}.

\bibitem{Cline:2021dkf}
J.~M. Cline and B.~Laurent, ``{Electroweak baryogenesis from light fermion
  sources: A critical study},''
  \href{http://dx.doi.org/10.1103/PhysRevD.104.083507}{{\em Phys. Rev. D}
  {\bfseries 104} no.~8, (2021) 083507},
  \href{http://arxiv.org/abs/2108.04249}{{\ttfamily arXiv:2108.04249
  [hep-ph]}}.

\bibitem{Chadha-Day:2022inf}
F.~Chadha-Day, B.~Garbrecht, and J.~McDonald, ``{Superradiance in stars:
  non-equilibrium approach to damping of fields in stellar media},''
  \href{http://dx.doi.org/10.1088/1475-7516/2022/12/008}{{\em JCAP} {\bfseries
  12} (2022) 008}, \href{http://arxiv.org/abs/2207.07662}{{\ttfamily
  arXiv:2207.07662 [hep-ph]}}.

\bibitem{Croon:2020cgk}
D.~Croon, O.~Gould, P.~Schicho, T.~V.~I. Tenkanen, and G.~White, ``{Theoretical
  uncertainties for cosmological first-order phase transitions},''
  \href{http://dx.doi.org/10.1007/JHEP04(2021)055}{{\em JHEP} {\bfseries 04}
  (2021) 055}, \href{http://arxiv.org/abs/2009.10080}{{\ttfamily
  arXiv:2009.10080 [hep-ph]}}.

\bibitem{Chala:2024xll}
M.~Chala, J.~C. Criado, L.~Gil, and J.~L. Miras, ``{Higher-order-operator
  corrections to phase-transition parameters in dimensional reduction},''
  \href{http://dx.doi.org/10.1007/JHEP10(2024)025}{{\em JHEP} {\bfseries 10}
  (2024) 025}, \href{http://arxiv.org/abs/2406.02667}{{\ttfamily
  arXiv:2406.02667 [hep-ph]}}.

\bibitem{Kierkla:2023von}
M.~Kierkla, B.~Swiezewska, T.~V.~I. Tenkanen, and J.~van~de Vis,
  ``{Gravitational waves from supercooled phase transitions: dimensional
  transmutation meets dimensional reduction},''
  \href{http://dx.doi.org/10.1007/JHEP02(2024)234}{{\em JHEP} {\bfseries 02}
  (2024) 234}, \href{http://arxiv.org/abs/2312.12413}{{\ttfamily
  arXiv:2312.12413 [hep-ph]}}.

\bibitem{Kierkla:2025qyz}
M.~Kierkla, P.~Schicho, B.~Swiezewska, T.~V.~I. Tenkanen, and J.~van~de Vis,
  ``{Finite-temperature bubble nucleation with shifting scale hierarchies},''
  \href{http://arxiv.org/abs/2503.13597}{{\ttfamily arXiv:2503.13597
  [hep-ph]}}.

\bibitem{Bernardo:2025vkz}
F.~Bernardo, P.~Klose, P.~Schicho, and T.~V.~I. Tenkanen, ``{Higher-dimensional
  operators at finite-temperature affect gravitational-wave predictions},''
  \href{http://arxiv.org/abs/2503.18904}{{\ttfamily arXiv:2503.18904
  [hep-ph]}}.

\bibitem{Gould:2024chm}
O.~Gould, A.~Kormu, and D.~J. Weir, ``{Nonperturbative test of nucleation
  calculations for strong phase transitions},''
  \href{http://dx.doi.org/10.1103/PhysRevD.111.L051901}{{\em Phys. Rev. D}
  {\bfseries 111} no.~5, (2025) L051901},
  \href{http://arxiv.org/abs/2404.01876}{{\ttfamily arXiv:2404.01876
  [hep-th]}}.

\bibitem{Hirvonen:2024rfg}
J.~Hirvonen, ``{Real-Time Nucleation and Off-Equilibrium Effects in
  High-Temperature Quantum Field Theories},''
  \href{http://arxiv.org/abs/2403.07987}{{\ttfamily arXiv:2403.07987
  [hep-ph]}}.

\end{thebibliography}\endgroup

\end{document}